\documentclass[preprint,12pt]{elsarticle}

\usepackage{amssymb}
\usepackage{amsmath}
\usepackage{graphicx}% Include figure files
\usepackage{hyperref}% add hypertext capabilities
\usepackage{physics}
\usepackage{booktabs}
\usepackage{esint}
\usepackage{soul}
\usepackage[rgb,dvipsnames]{xcolor}
\usepackage{cancel}
\usepackage{kotex}
\usepackage{subcaption}  % subfigure 대체 패키지
\usepackage[margin=1in]{geometry}
\usepackage{array}
\usepackage{siunitx}

\makeatletter
\def\ps@pprintTitle{%
 \let\@oddhead\@empty
 \let\@evenhead\@empty
 \let\@oddfoot\@empty
 \let\@evenfoot\@empty
}
\makeatother
\usepackage{lineno}

\begin{document}

\begin{frontmatter}

%% Title, authors and addresses

%% use the tnoteref command within \title for footnotes;
%% use the tnotetext command for theassociated footnote;
%% use the fnref command within \author or \affiliation for footnotes;
%% use the fntext command for theassociated footnote;
%% use the corref command within \author for corresponding author footnotes;
%% use the cortext command for theassociated footnote;
%% use the ead command for the email address,
%% and the form \ead[url] for the home page:
%% \title{Title\tnoteref{label1}}
%% \tnotetext[label1]{}
%% \author{Name\corref{cor1}\fnref{label2}}
%% \ead{email address}
%% \ead[url]{home page}
%% \fntext[label2]{}
%% \cortext[cor1]{}
%% \affiliation{organization={},
%%             addressline={},
%%             city={},
%%             postcode={},
%%             state={},
%%             country={}}
%% \fntext[label3]{}

\title{
Numerical Framework for Multimode Jaynes– and Tavis–Cummings Models Incorporating the Modified Langevin Noise Formalism: Non-Markovian Analysis of Atom–Field Interactions in Dissipative Electromagnetic Environments}

%% use optional labels to link authors explicitly to addresses:
%% \author[label1,label2]{}
%% \affiliation[label1]{organization={},
%%             addressline={},
%%             city={},
%%             postcode={},
%%             state={},
%%             country={}}
%%
%% \affiliation[label2]{organization={},
%%             addressline={},
%%             city={},
%%             postcode={},
%%             state={},
%%             country={}}

\author[label1]{Hyunwoo Choi} %% Author name

%% Author affiliation
\affiliation[label1]{organization={Department of Electrical Engineering, Pohang University of Science and Technology},%Department and Organization
            addressline={77 Cheongam-ro, Nam-gu}, 
            city={Pohang},
            postcode={37673}, 
            state={Gyeongsangbuk-do},
            country={Republic of Korea}
            }

\author[label2]{Weng C. Chew} %% Author name

%% Author affiliation
\affiliation[label2]{organization={Elmore Family School of Electrical and Computer Engineering, Purdue University},%Department and Organization
            addressline={610 Purdue Mall}, 
            city={West Lafayette},
            postcode={47907}, 
            state={Indiana},
            country={United States of America}
            }

\author{Dong-Yeop Na\corref{cor1}\fnref{label1}}
\ead{dyna22@postech.ac.kr}
\ead[url]{https://sites.google.com/view/acem-postech/home}
\cortext[cor1]{corresponding author}

%\author[label1]{Dong-Yeop Na \corref{cor1}} %% Author name

%% Abstract
\begin{abstract}
We present a novel numerical framework that integrates the modified Langevin noise formalism into the multimode Jaynes– and Tavis–Cummings models, enabling a first-principles, non-Markovian analysis of atom–field interactions in dissipative electromagnetic (EM) environments that account for both radiative losses and absorptive dissipation in lossy dielectric media (or satisfying general inhomogeneous causal media exhibiting both dispersion and absorption effects).
In the modified Langevin noise formalism \cite{na2023numerical}, the boundary- and medium-assisted (BA and MA) fields, which constitute a continuum set of EM modes in dissipative EM environments, are numerically obtained using the finite-element method (FEM). 
Specifically, BA field modes are extracted by solving plane-wave scattering problems, while MA field modes are determined through point-source radiation problems. 
To reduce the computational load, we employ (i) an adaptive frequency refinement technique that selectively identifies the BA and MA modes contributing most significantly to the atom–field interactions, while also (ii) enabling coarse-graining over the degeneracy space of the BA and MA fields.
These numerically obtained BA and MA field modes are then incorporated into the multimode Jaynes– and Tavis–Cummings models such that the coupling strength between atoms and BA-MA field modes can be calculated for the study of atom–field interactions in dissipative EM environments.
For a specific application, we appropriately truncate the Hilbert space and define quantum states within the resulting subspace, as the Jaynes– and Tavis–Cummings models preserve the total number of initially excited quanta.
The numerical integrator is then employed to solve the matrix representation of the Schrödinger equation, yielding the time-dependent probability amplitudes of quantum states.
These amplitudes are subsequently used to compute the expectation values of various observables in the Schrödinger picture, such as atomic populations, emitted EM fields, and related quantities.
The proposed methodology captures non-Markovian atomic dynamics that cannot be described by traditional quantum master equations under the Markovian approximation. 
More importantly, the present numerical framework enables the estimation of expectation values of EM field operators—such as the single-photon EM field amplitude emitted from atoms and the second-order correlation function $g^{(2)}(\tau)$—which are also not accessible within the traditional quantum master equation formalism. 
This capability is particularly important for evaluating the radiation patterns of designed single-photon sources, the out-coupling efficiency to a posterior optical fiber, and tracking the entanglement transfer from atomic systems into EM fields.
To validate the accuracy of the proposed numerical framework, we present four numerical examples: (i) a two-level system (TLS) in a perfect electric conductor (PEC) half-space; (ii) dissipative cavity electrodynamics with two limiting cases approaching free space (spontaneous emission in free space) and a perfect mirror (ideal Rabi oscillations); (iii) super-radiance in TLS arrays; and (iv) entanglement sudden death of two TLSs initially prepared in an entangled mixed state inside dissipative cavities. 
The proposed methodology can serve as a ground-truth numerical simulator for studying atom–field interactions in general dissipative EM environments.
\end{abstract}

%%Graphical abstract
%\begin{graphicalabstract}
%\includegraphics{grabs}
%\end{graphicalabstract}

%%Research highlights
%\begin{highlights}
%\item Research highlight 1
%\item Research highlight 2
%\end{highlights}

%% Keywords
\begin{keyword}
quantum optics, dissipative cavity electrodynamics, modified Langevin noise formalism, boundary-assisted fields, medium-assisted fields, spontaneous emission, Rabi oscillation, entanglement sudden death, finite-element method
\end{keyword}

\end{frontmatter}

%% Add \usepackage{lineno} before \begin{document} and uncomment 
%% following line to enable line numbers
%% \linenumbers

%% main text
%%

%% Use \section commands to start a section
\section{Introduction}\label{SEC_INTRO}

In quantum optics, the quantization of electromagnetic (EM) fields is typically performed by identifying the normal modes (or eigenmodes) of the fields within a given system and treating each normal mode as a quantum harmonic oscillator\cite{Gerry2004introductory,fox2006quantum,loudon2000quantum} so that the subsequent quantization process is mathematically equivalent to the quantization of a mechanical harmonic oscillator \cite{miller2008quantum,Chew2015-vo}.
The key point here is that the given system must be {\it energy-conserving} or {\it Hermitian}. 
Under this condition, the normal modes have real eigenfrequencies and form a complete set, allowing the resulting Hamiltonian operator to be exactly diagonalized. 
For example, a periodic box filled with a homogeneous medium can be treated analytically, as its normal modes correspond to trigonometric functions \cite{Gerry2004introductory,fox2006quantum,loudon2000quantum}. 
In cases involving inhomogeneous media, numerical mode-decomposition methods \cite{Chew2016Quantum_1,Chew2016Quantum_2,Chew2021Quantum} based on Computational Electromagnetics (CEM) can be employed \cite{Na2020quantum}.

However, when a system includes absorbing dielectric objects of finite size in free space—introducing both radiation and medium losses—it is no longer Hermitian.
As a result, normal modes do not form a complete set, and their eigenfrequencies become complex. 
This causes the standard bosonic commutator relation
\begin{flalign}
\left[\hat{a}(t),\hat{a}^{\dag}(t)\right] \neq \left[\hat{a}(0),\hat{a}^{\dag}(0)\right]
\end{flalign}
to no longer be preserved over time, violating the Heisenberg uncertainty principle and thereby losing fundamental quantum physical properties \cite{Gruner1996Green}.
Therefore, the quantization of EM fields in lossy environments cannot be achieved using the conventional mode-decomposition approach, which has remained a significant challenge over the past decades.

To address this issue, numerous studies have been conducted, broadly exploring four major approaches in previous research.
The first approach is to use the so-called {\it quasi-normal modes} (QNM) for the quantization of EM fields \cite{franke2019quantization,medina2021few,ren2020near,franke2020quantized,monica2022few}.
The QNM quantization extends traditional mode decomposition by incorporating dissipation and radiation effects through complex eigenfrequencies. 
Unlike normal modes, QNMs satisfy a bi-orthogonal relation rather than standard orthonormality, requiring careful normalization to maintain mode quality \cite{sauvan2022normalization}. 
However, finding QNMs numerically is challenging, and ensuring bi-orthogonality can be problematic in practical implementations. 
Additionally, QNMs are typically defined near the cavity regions and do not fully describe the radiated field propagating into free space or interactions with other devices at intermediate distances.
As a result, QNM-based quantization struggles to capture out-coupling efficiency and far-field radiation, necessitating additional methods such as near-to-far-field transformation \cite{ren2020near}.
The second approach, which is the most rigorous, is the microscopic model introduced by Huttner and Barnett \cite{huttner1992quantization} and later extended by subsequent researchers \cite{Philbin2010canonical,Dorier2019Canonical,Na2021Diagonalization}.
The microscopic model describes the polarization density induced in absorbing dielectric media by representing matter oscillators coupled with infinite bath oscillators at every spatial point. 
Then, the loss of electromagnetic energy can be modeled as the transfer of energy from the EM fields to the bath oscillators through the matter oscillators.
The resulting composite system, where the EM fields interact with the matter and infinite bath oscillators, remains strictly energy-conserving, allowing, in principle, for exact diagonalization of the Hamiltonian operator for the composite system \cite{huttner1992quantization,Philbin2010canonical,Na2021Diagonalization}. 
The normal modes required to describe the EM fields form a subset of the normal modes of the composite system. 
By adjusting the coupling strength, one can control the extent of dispersion and absorption \cite{Philbin2010canonical}.
Despite the strong theoretical foundation of this approach, its practical implementation remains infeasible. 
This is because modeling an infinite set of bath oscillators requires enormous computational resources, especially when the absorbing medium is large.
Without sufficient coarse graining, the effects of the actual bath are not accurately captured, and Poincaré recurrence may lead to unphysical artifacts that deviate from real physical behavior.
The third approach, a strong alternative to the microscopic model, is the Green's function approach (also known as the Langevin noise formalism), which is based on the fluctuation-dissipation theorem and was proposed by Welsch and co-workers \cite{Gruner1996Green,Dung1998three}.
The fundamental idea behind this method is to introduce noise current sources within the absorbing dielectrics to compensate for the energy loss due to the medium loss, ensuring quasi-energy conservation.
Specifically, the resulting electric fields are determined by the linear superposition of the so-called {\it medium-assisted fields}, which correspond to the radiation from the noise source currents.
Due to its computational efficiency, this approach has been widely used in quantum optics to estimate the Purcell factor of an atom embedded in plasmonic structures \cite{yao2009ultrahigh,dzsotjan2010quantum}.
However, it was found that the Green's function approach is incomplete in the sense that it does not account for radiation loss due to the openness of the system \cite{Stefano2001Mode,Drezet2017Equivalence,Drezet2017quantizing,na2023numerical,Dorier2020Critical}.

The fourth approach addresses this issue and is referred to as the modified Langevin noise (M-LN) formalism \cite{Stefano2001Mode,Drezet2017Equivalence,na2023numerical}. 
This formalism introduces an additional term—referred to as the ``boundary-assisted (BA)'' field—which accounts for thermal radiation propagating toward the system's infinite boundary. 
As the name suggests (i.e., fields assisted by the boundaries of materials), when a bounded lossy dielectric is present, the effect of this missing term in the original Langevin noise (O-LN) formalism becomes more pronounced. 
Therefore, incorporating the BA field in the M-LN formalism is particularly important in such cases. 
With this modification, the EM field dynamics is governed by two distinct types of fluctuations—those arising from radiation and those due to medium losses—treated on equal footing, in full accordance with the fluctuation-dissipation theorem (FDT). 
This ensures that the system remains (quasi)Hermitian even in the presence of finite-sized lossy dielectric media, which corresponds to most practical scenarios. 
As a result, the quantization of EM fields in lossy environments can be achieved in a consistent and physically accurate manner.
In recent years, the modified Langevin approach has been actively investigated for both its validation and practical applications \cite{Ciattoni2024quantum,Semin2024canonical,Miano2025quantum}.

Meanwhile, to describe the atom-field interactions, additional two-level-system(s) (TLSs) can be introduced into the system. For this purpose, the Jaynes- and Tavis-Cummings models are most widely used \cite{Shore1993JC,Fink2009Dressed}. 
With the dipole approximation under the radiation gauge, this model compactly implements the atom-field interaction using simplified interaction Hamiltonian term, also known as $\mathbf{d\cdot E}$ Hamiltonian. 
This makes the equation analytically solvable, but only for very simple cases (e.g., free space). 
However, when the environment becomes realistic (e.g., lossy environment), the situation becomes significantly more challenging. 
There are two main approaches to dealing with this difficulty. 
The first approach is using the quantum master equation. 
In this method, referred to as Lindblad master equation, the system we observed is reduced, and the remaining complex reservoir is addressed through a phenomenological treatment of dynamics under Markovian approximation. 
The other approach traces the total system by considering dissipative effects during the field quantization process, as introduced through the four representative methods mentioned earlier. 
Because this approach does not reduce the system to the atom(s) and reservoir, there is no need to rely on the Markovian approximations. 
In doing so, It can capture realistic effects that phenomenological treatments cannot. 
In addition, the quantum state explicitly incorporates the field’s Fock states (photon number states), which offer significant advantages, such as capturing not only the dynamics of atomic states but also time-dependent field properties (e.g., photon statistics and entanglement).
However, this approach is more complex as it does not statistically trace out the environment DoFs but instead considers the quantization of the lossy field. Therefore, to accurately capture the dynamics of quantum state within this framework, a numerical treatment is required.

Rather than applying well-established quantum optics theories to simplified structures, there is a growing trend toward applying them to complex scenarios with practical relevance in real-world environments. In such settings, accurately determining the mode profiles in complicated electromagnetic (EM) environments becomes essential. To achieve this, integrating Computational Electromagnetics (CEM) methodologies with quantum optics theory—known as Computational Quantum Electromagnetics (CQEM)—is critically important. Reflecting this trend, recent years have seen a surge in research related to CQEM \cite{Roth2021Macroscopic, Khan2024Field, Moon2024Analytical, Forestiere2023Integral, Forestiere2022Operative, Zhu2024Quantum, Ryu2023Efficient, Ryu2023Matrix}.
This study is also part of the broader CQEM effort, aiming to describe atom–field interactions from first principles in the presence of loss. The ultimate goal is to develop the analytical capability to rigorously understand atom–field interactions and out-coupling fields in arbitrary non-Markovian environments.

This article formulates the numerical recipe to address these difficulties. Namely, we present a numerical framework that incorporates the M-LN with multimode Jaynes- and Tavis-Cummings model. For this purpose, we use coarse-graining strategy based on the finite element method. To be specific, We obtain each degenerated BA field by solving plane-wave scattering problem, while obtain each degenerate MA field through point-source radiation problem. We also introduce a unified ladder operator to account for all BA and MA fields at once. With the use of the rotating wave approximation and the dipole approximation, the Hamiltonian of given system is described within the framework of the Jaynes- and Tavis-Cummings models. Especially for the Tavis-Cummings model, we consider truncated Hilbert space to refine the expression of the general quantum state. Then we can calculate the Schrödinger equation in the matrix representation by numerical integration.  we validate our methodology through four examples: (i) a two-level system (TLS) in a perfect electric conductor
(PEC) half-space; (ii) dissipative cavity electrodynamics with two limiting cases approaching
free space (spontaneous emission in free space) and a perfect mirror (ideal Rabi oscillations);
(iii) super-radiance in TLS arrays; and (iv) entanglement sudden death of two TLSs initially
prepared in an entangled mixed state inside dissipative cavities. 

This work adopts the time convention $e^{-i\omega t}$ and natural units where $\hbar=c=\epsilon_0=\mu_0=1$.

\section{Modified Langevin noise formalism}
This section briefly explains the modified Langevin noise (M-LN) formalism \cite{Drezet2017Equivalence,Stefano2001Mode,na2023numerical}.
Extending the original Langevin noise formalism (O-LN) \cite{Gruner1996Green,Dung1998three}, the M-LN formalism provides the complete quantization of electromagnetic (EM) fields in the presence of inhomogeneous, dispersive, and absorbing (IDA) dielectric objects of finite size in free space (i.e., are involved in open systems).
The primary modification of the O-LN in the M-LN was the incorporation of additional noise fluctuations in response to radiation loss in the electric field operator, arising from the system's openness.
Note that such dispersive and absorbing dielectric media should satisfy the Kramers-Kr\"{o}nig relation to preserve the causality of the system.
Here, we assume that dielectric media are assumed to be non-magnetic, i.e., the permeability of the dielectric medium is $\mu_0$.

\begin{figure}[t]
\centering
\includegraphics[width=3in]{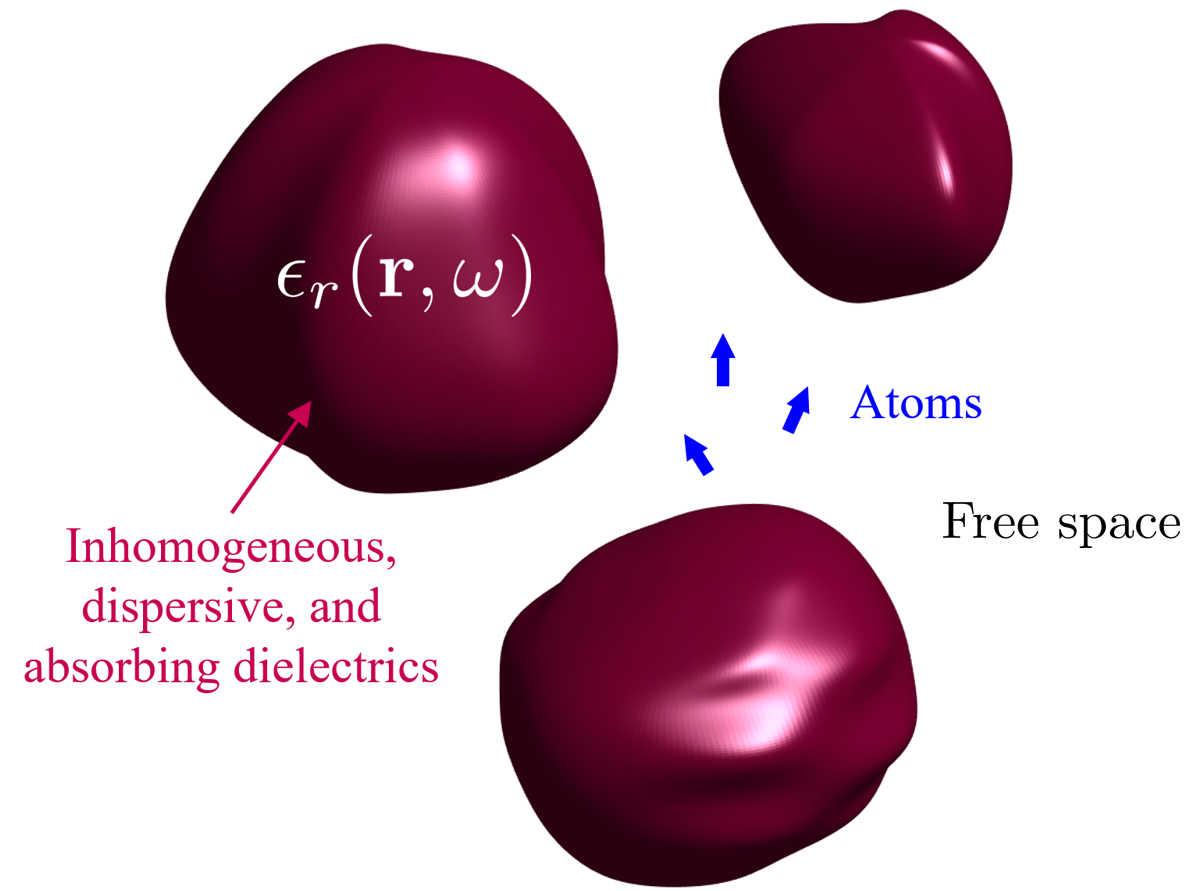}
\caption{Illustration of atoms in the vicinity of inhomogeneous, dispersive, and absorbing dielectric objects of finite size and arbitrary shape in free space.}
\label{fig:IDA_geometry}
\end{figure}

Consider IDA dielectric objects of finite size and arbitrary shape, with relative permittivity denoted as $\epsilon_r(\mathbf{r},\omega)$, that occupy the volume $V_m$ in free space, as illustrated in Fig. \ref{fig:IDA_geometry}.
According to the M-LN formalism, the positive-frequency part of the monochromatic electric field operator arises from contributions of both boundary-assisted (BA) and medium-assisted (MA) monochromatic fields \cite{na2023numerical}, as follows:
\begin{flalign}
\hat{\mathbf{E}}^{(+)}(\mathbf{r},\omega)
=
\hat{\mathbf{E}}_{(\text{BA})}^{(+)}(\mathbf{r},\omega)
+
\hat{\mathbf{E}}_{(\text{MA})}^{(+)}(\mathbf{r},\omega)
\label{eqn:BA/MA_fields}
\end{flalign}
where the monochromatic BA and MA fields, denoted as $\hat{\mathbf{E}}_{(\text{BA})}^{(+)}(\mathbf{r},\omega)$ and $\hat{\mathbf{E}}_{(\text{MA})}^{(+)}(\mathbf{r},\omega)$, respectively, have the infinite degeneracy and can be represented by
\begin{flalign}
\hat{\mathbf{E}}_{(\text{BA})}^{(+)}(\mathbf{r},\omega)
&=
\int_{\mathbb{S}^2}
d\Omega
\sum_{s\in \left\{H,V\right\}}
\mathbf{E}^{\text{(BA)}}\left(\mathbf{r},\omega,\left\{\Omega,s\right\}\right)
\hat{a}\left(\omega,\left\{\Omega,s\right\}\right),
\label{eqn:BA_field}
\\
\hat{\mathbf{E}}_{(\text{MA})}^{(+)}(\mathbf{r},\omega)
&=
\int_{V_{m}}d\mathbf{r}'
\sum_{\xi\in\left\{x,y,z\right\}}
\mathbf{E}^{\text{(MA)}}\left(\mathbf{r},\omega,\left\{\mathbf{r}',\xi\right\}\right)
\hat{f}\left(\omega,\left\{\mathbf{r}',\xi\right\}\right).
\label{eqn:MA_field}
\end{flalign}
Here, $\mathbb{S}^2$ denotes represents the surface of a unit sphere, $\Omega$ represents a solid angle, $s$ denotes the polarization index of a uniform plane wave (either horizontal or vertical), $\xi$ represents the orientation of a point current source, $\mathbf{E}^{\text{(BA)}}$ and $\mathbf{E}^{\text{(MA)}}$ are the degenerate BA and MA fields, respectively, and $\hat{a}$ and $\hat{f}$ are annihilation operators for the degenerate BA and MA fields, respectively.
As observed in \eqref{eqn:BA_field} and \eqref{eqn:MA_field}, monochromatic BA fields exhibit degeneracy in terms of uniform plane waves, namely polarization and propagation direction, while monochromatic MA fields exhibit degeneracy in terms of position within $V_m$ and orientation of a point current source.
This result has been extensively discussed in \cite{Drezet2017Equivalence,Stefano2001Mode,na2023numerical}.

Physically speaking, each degenerate BA field correspond to a total field composed of (i) an incident uniform plane wave with a wavevector of 
\begin{flalign}
\mathbf{k}_{\text{inc}}\left(\omega,\left\{\Omega,s\right\}\right)
=
\frac{\omega}{c}
\left(\sin\theta\cos\phi\hat{x}+\sin\theta\sin\phi\hat{y}+\cos\theta\hat{z}\right)
\end{flalign} 
in three-dimensional space where wavenumber in free space $k=\omega/c$ and polarization $s$ and (ii) scattered fields due to the IDA dielectric objects.
On the other hand, each degenerate MA field corresponds to the radiation field produced by a point current source at every individual point within the IDA dielectric objects.
More specifically, degenerate BA and MA fields can be written by
\begin{flalign}
\mathbf{E}^{\text{(BA)}}\left(\mathbf{r},\omega,\left\{\Omega,s\right\}\right) 
&= 
\frac{i}{\left(\sqrt{2\pi}\right)^{3}}
\mathbf{E}_{\text{tot}}\left(\mathbf{r},\omega,\left\{\Omega,s\right\}\right)
\sqrt{\frac{\hbar\omega}{2}},
\label{eqn:BA_mode}
\\
\mathbf{E}^{(\text{MA})}\left(\mathbf{r},\omega,\left\{\mathbf{r}',\xi\right\}\right) 
& = 
i k^2 
\left[
\overline{\mathbf{G}}_{\text{E}}(\mathbf{r},\mathbf{r}',\omega)
\cdot
\hat{\xi}
\right]
\sqrt{\frac{\hbar\chi_{I}(\mathbf{r}',\omega)}{\pi\epsilon_0}}
\label{eqn:MA_mode}
\end{flalign}
where $\chi_{I}$ represents the imaginary part of the electric susceptibility of the IDA dielectric objects, $\mathbf{E}_{\text{tot}}\left(\mathbf{r},\omega,\left\{\Omega,s\right\}\right)$ denotes the total field including the incident uniform plane wave and resulting scattered fields, and $\overline{\mathbf{G}}_{\text{E}}(\mathbf{r},\mathbf{r}'',\omega)$ is the electric-field dyadic Green's function governed by
\begin{flalign}
\nabla\times\nabla\times \overline{\mathbf{G}}_{\text{E}}(\mathbf{r},\mathbf{r}',\omega)
-\frac{\omega^2}{c^2}\epsilon_r(\mathbf{r},\omega)
\overline{\mathbf{G}}_{\text{E}}(\mathbf{r},\mathbf{r}',\omega)
=
\overline{\mathbf{I}}\delta(\mathbf{r}-\mathbf{r}').
\label{eqn:DGF}
\end{flalign}

We shall discuss later in detail how to obtain these fields numerically.

The monochromatic Hamiltonian operator is diagonalized in terms of ladder operators of degenerate BA and MA fields, and can be expressed by
\begin{flalign}
\hat{H}_{field}(\omega)
&=
\hbar\omega
\int_{\mathbb{S}^2} d\Omega
\sum_{s\in\left\{H,V\right\}}
\hat{a}^{\dag}\left(\omega,\left\{\Omega,s\right\}\right)
\hat{a}\left(\omega,\left\{\Omega,s\right\}\right)
\nonumber \\
&+
\hbar\omega
\int_{V_{m}}d\mathbf{r}'
\sum_{\xi\in\left\{x,y,z\right\}}
\hat{f}^{\dag}\left(\omega,\left\{\mathbf{r}',\xi\right\}\right)\hat{f}\left(\omega,\left\{\mathbf{r}',\xi\right\}\right).
\label{eqn:Ham_BAMA_freq}
\end{flalign}
Note that the ladder operators associated with the degenerate BA and MA fields satisfy the standard bosonic commutation relation and take Fock states as their eigenstates.

The total Hamiltonian and the positive-frequency part of the electric field operator are then obtained by integrating the monochromatic components over frequency, as follows:
\begin{flalign}
\hat{H}_{field}
&=
\int_{0}^{\infty} d\omega \hat{H}_{field}(\omega),
\label{eqn:c_Ham}
\\
\hat{\mathbf{E}}^{(+)}(\mathbf{r},t)
&=
\int_{0}^{\infty}d\omega
\hat{\mathbf{E}}^{(+)}(\mathbf{r},\omega)e^{-i\omega t}.
\label{eqn:full_E_operator}
\end{flalign}

\subsection{Degenerate boundary-assisted (BA) field}
The total field used for modeling degenerate BA fields is given by  
\begin{flalign}
\mathbf{E}_{\text{tot}}\left(\mathbf{r},\omega,\left\{\Omega,s\right\}\right)
&=
\mathbf{E}_{\text{inc}}\left(\mathbf{r},\omega,\left\{\Omega,s\right\}\right)
+
\mathbf{E}_{\text{sca}}\left(\mathbf{r},\omega,\left\{\Omega,s\right\}\right),
\label{eqn:BA_total_field}
\end{flalign}
where \( \mathbf{E}_{\text{tot}} \), \( \mathbf{E}_{\text{inc}} \), and \( \mathbf{E}_{\text{sca}} \) denote the total, incident, and scattered fields, respectively. 
The total electric field satisfies the homogeneous wave equation:
\begin{flalign}
    \nabla \times \nabla \times \mathbf{E}_{\text{tot}}\left(\mathbf{r},\omega,\left\{\Omega,s\right\}\right)
    - k^2 \epsilon_r(\mathbf{r},\omega)
    \mathbf{E}_{\text{tot}}\left(\mathbf{r},\omega,\left\{\Omega,s\right\}\right)
    = 0,
    \label{eqn:homogen_sol_eqn}
\end{flalign}
subject to open boundary conditions.
Since 
\begin{flalign}
\mathbf{E}_{\text{inc}}\left(\mathbf{r},\omega,\left\{\Omega,s\right\}\right)
=
\hat{e}_{s} e^{i \mathbf{k}_{\text{inc}}(\omega,\left\{\Omega,s\right\})\cdot\mathbf{r}}
\end{flalign}
where $\hat{e}_{s}$ is the polarization unit vector of the plane wave, satisfying
\begin{flalign}
\nabla\times\nabla\times
\mathbf{E}_{\text{inc}}\left(\mathbf{r},\omega,\left\{\Omega,s\right\}\right)
-k^2
\mathbf{E}_{\text{inc}}\left(\mathbf{r},\omega,\left\{\Omega,s\right\}\right)
=
0,
\label{eqn:free space_homogen_sol_eqn}
\end{flalign}
one can derive a plane-wave scattering problem by substituting \eqref{eqn:BA_total_field} into \eqref{eqn:homogen_sol_eqn}, as follows:  
\begin{flalign}
&\nabla\times\nabla\times\mathbf{E}_{\text{sca}}\left(\mathbf{r},\omega,\left\{\Omega,s\right\}\right)
-k^2\epsilon_r(\mathbf{r},\omega)
\mathbf{E}_{\text{sca}}\left(\mathbf{r},\omega,\left\{\Omega,s\right\}\right)
\nonumber \\
&\quad\quad\quad\quad\quad\quad\quad\quad\quad\quad\quad\quad\quad\quad\quad
=
k^2\chi(\mathbf{r},\omega)
\mathbf{E}_{\text{inc}}\left(\mathbf{r},\omega,\left\{\Omega,s\right\}\right)
\label{eqn:plane_wave_scattering_problem}
\end{flalign}
where $\chi$ denotes the electric susceptibility of the IDA dielectric objects.

From the above, one can interpret an incident uniform plane wave field within the IDA dielectric objects serve as a current source to produce the resulting scattered fields, i.e.,
\begin{flalign}
\mathbf{J}_{\text{BA}}\left(\mathbf{r},\omega,\left\{\Omega,s\right\}\right) = -i\omega \epsilon_0 \chi(\mathbf{r},\omega)\mathbf{E}_{\text{inc}}\left(\mathbf{r},\omega,\left\{\Omega,s\right\}\right)
\end{flalign}
Hence, Eq. \eqref{eqn:plane_wave_scattering_problem} can be rewritten by
\begin{flalign}
&
\nabla\times\nabla\times\mathbf{E}_{\text{sca}}\left(\mathbf{r},\omega,\left\{\Omega,s\right\}\right)
-k^2\epsilon_r(\mathbf{r},\omega)
\mathbf{E}_{\text{sca}}\left(\mathbf{r},\omega,\left\{\Omega,s\right\}\right)
\nonumber \\
&\quad\quad\quad\quad\quad\quad\quad\quad\quad\quad\quad\quad\quad\quad\quad\quad\quad
=
i\omega\mu_0\mathbf{J}_{\text{BA}}\left(\mathbf{r},\omega,\left\{\Omega,s\right\}\right).
\label{eqn:plane_wave_scattering_problem2}
\end{flalign}
Thus, the scattered fields can be found by the radiation integral: 
\begin{flalign}
\mathbf{E}_{\text{sca}}\left(\mathbf{r},\omega,\left\{\Omega,s\right\}\right)
=
i\omega\mu_0
\int_{V_{m}} 
d\mathbf{r}'
\overline{\mathbf{G}}_{\text{E}}(\mathbf{r},\mathbf{r}',\omega)
\cdot
\mathbf{J}_{\text{BA}}\left(\mathbf{r}',\left\{\Omega,s\right\}\right).
\end{flalign}
Consequently, degenerate BA fields can be expressed by
\begin{flalign}
\mathbf{E}^{\text{(BA)}}\left(\mathbf{r},\omega,\left\{\Omega,s\right\}\right) 
&= 
\frac{i}{\left(\sqrt{2\pi}\right)^{3}}
\Bigl[
\hat{e}_{s} e^{i \mathbf{k}_{\text{inc}}(\omega,\left\{\Omega,s\right\})\cdot\mathbf{r}}
\nonumber \\
&+
i\omega\mu_0
\int_{V_{m}} 
d\mathbf{r}'
\overline{\mathbf{G}}_{\text{E}}(\mathbf{r},\mathbf{r}',\omega)
\cdot
\mathbf{J}_{\text{BA}}\left(\mathbf{r}',\left\{\Omega,s\right\}\right)
\Bigr]
\sqrt{\frac{\hbar\omega}{2}},
\end{flalign}

\subsection{Degenerate medium-assisted (MA) field}
Unlike degenerate BA fields, each degenerate MA field in \eqref{eqn:MA_mode} physically represents a radiation field generated by a scaled point current source at every single point $\mathbf{r}'$ within the IDA dielectric objects.
When defining the scaled point current source by
\begin{flalign}
\mathbf{J}_{\text{MA}}(\mathbf{r}'',\omega,\left\{\mathbf{r}',\xi\right\})
=
\hat{\xi}
\omega \epsilon_0
\sqrt{\frac{\hbar\chi_{I}(\mathbf{r}',\omega)}{\pi\epsilon_0}}
\delta(\mathbf{r}''-\mathbf{r}'),
\label{eqb:MA_source}
\end{flalign}
one can derive the governing equation for each degenerate MA field as
\begin{flalign}
&\nabla\times\nabla\times\mathbf{E}^{\text{(MA)}}\left(\mathbf{r},\omega,\left\{\mathbf{r}',\xi\right\}\right)
-k^2\epsilon_r(\mathbf{r},\omega)
\mathbf{E}^{\text{(MA)}}\left(\mathbf{r},\omega,\left\{\mathbf{r}',\xi\right\}\right)
\nonumber \\
&\quad\quad\quad\quad\quad\quad\quad\quad\quad\quad\quad\quad\quad\quad\quad\quad\quad\quad
=
i\omega \mu_0
\mathbf{J}_{\text{MA}}\left(\mathbf{r},\omega,\left\{\mathbf{r}',\xi\right\}\right).
\label{eqn:MA_mode_governing_eqn}
\end{flalign}
As a result, degenerate MA fields can be expressed by
\begin{flalign}
\mathbf{E}^{(\text{MA})}\left(\mathbf{r},\omega,\left\{\mathbf{r}',\xi\right\}\right) 
& = 
i\omega \mu_0
\int_{V_{m}}d\mathbf{r}''
\overline{\mathbf{G}}_{\text{E}}(\mathbf{r},\mathbf{r}'',\omega)
\cdot
\mathbf{J}_{\text{MA}}(\mathbf{r}'',\omega,\left\{\mathbf{r}',\xi\right\}).
\label{eqn:degen_MA_field}
\end{flalign}

\section{Numerical evaluation of degenerate BA and MA fields}

\subsection{Degenerate BA fields}
Each degenerate BA field physically represents a total field composed of an incident uniform plane wave and scattered fields from IDA dielectric objects. 
In other words, solving a {\it plane-wave scattering problem} is necessary to obtain each degenerate BA field.
If the dielectric object has a simple shape, such as a sphere, an analytical solution can be obtained using the Mie scattering formula. 
Otherwise, numerical methods in computational electromagnetics, such as the Finite Element Method (FEM) \cite{jin2002finite} or the Method of Moments \cite{gibson2024method}, may be used.
Here, we demonstrate the use of the FEM to obtain numerical solutions for degenerate BA fields.

For IDA dielectric objects of finite size and arbitrary shape in free space, we solve \eqref{eqn:plane_wave_scattering_problem} to find scattered fields.
For an incident uniform plane wave with wavevector $\mathbf{k}_{\text{inc}}(\omega,\left\{\Omega,s\right\})$ and polarization $s$ (either horizontal or vertical), the FEM formulation of \eqref{eqn:plane_wave_scattering_problem} can be written by
\begin{flalign}
\overline{\mathbf{S}}\cdot \mathbf{e}_{sca}\left(\omega,\left\{\Omega,s\right\}\right) -  k^2  \overline{\mathbf{M}}\cdot \mathbf{e}_{sca}\left(\omega,\left\{\Omega,s\right\}\right) =i\omega\mu_0 \mathbf{j}_{BA}\left(\omega,\left\{\Omega,s\right\}\right).
\end{flalign}
Here, 
$\overline{\mathbf{S}}$ and $\overline{\mathbf{M}}$ are stiffness and mass matrices, respectively, whose elements can be computed by
\begin{flalign}
\left[\overline{\mathbf{S}}\right]_{p,q}&=
-\int_{V} 
d\mathbf{r}
\Bigl[\nabla\times
\mathbf{W}^{(1)}_{p}(\mathbf{r})
\Bigr]
\cdot
\Bigl[
\nabla\times
\mathbf{W}^{(1)}_{q}(\mathbf{r})
\Bigr],
\\
\left[\overline{\mathbf{M}}\right]_{p,q}&=
\int_{V} 
d\mathbf{r}
\epsilon_r(\mathbf{r})
\Bigl[
\mathbf{W}^{(1)}_{p}(\mathbf{r})
\cdot
\mathbf{W}^{(1)}_{q}(\mathbf{r})
\Bigr],
\end{flalign}
where indices $p$ and $q$ refer to the index of edges in a given simplicial mesh, $\mathbf{W}^{(1)}_{p}(\mathbf{r})$ denotes Whitney-1 form (also known as curl-conforming vector basis function or edge element) associated with $p$-th edge, the column vector $\mathbf{e}_{sca}$ represents Degrees of Freedom (DoFs) for the scattered field $\mathbf{E}_{\text{sca}}$ with its $p$-th element corresponding to a co-chain associated with $p$-th edge, and $\mathbf{j}_{BA}$ is a column vector collecting DoFs for the current density $\mathbf{J}_{\text{BA}}$ whose elements can be computed by
\begin{flalign}
\Bigl[
\mathbf{j}_{BA}\left(\omega,\left\{\Omega,s\right\}\right)
\Bigr]_{p} 
= 
\int_{V_{m}}
d\mathbf{r}
\mathbf{W}^{(1)}_{p}(\mathbf{r})\cdot
\mathbf{J}_{\text{BA}}\left(\mathbf{r},\omega,\left\{\Omega,s\right\}\right).
\end{flalign}
Solving the FEM sparse linear system, one can have the DoFs for the scattered fields and reconstruct the scattered fields via
\begin{flalign}
\mathbf{E}_{\text{sca}}\left(\mathbf{r},\omega,\left\{\Omega,s\right\}\right)
\approx
\sum_{p=1}^{N_1}
\Bigl[\mathbf{e}_{sca}\left(\omega,\left\{\Omega,s\right\}\right)\Bigr]_{p}
\mathbf{W}^{(1)}_p(\mathbf{r})
\end{flalign}
where $N_1$ denotes the total number of edges.
Then, the total field, which corresponds to each discrete BA mode, can be represented by
\begin{flalign}
\mathbf{E}_{\text{tot}}\left(\mathbf{r},\omega,\left\{\Omega,s\right\}\right)
\approx
\hat{e}_{s} e^{i \mathbf{k}_{\text{inc}}\left(\omega,\left\{\Omega,s\right\}\right)\cdot\mathbf{r}}
+
\sum_{p=1}^{N_1}
\Bigl[\mathbf{e}_{sca}\left(\omega,\left\{\Omega,s\right\}\right)\Bigr]_{p}
\mathbf{W}^{(1)}_p(\mathbf{r}).
\end{flalign}
Finally, one can obtain a numerical solution to each degenerate BA field as
\begin{flalign}
\mathbf{E}^{\text{(BA)}}\left(\mathbf{r},\omega,\left\{\Omega,s\right\}\right) 
&\approx
\frac{i}{\left(\sqrt{2\pi}\right)^{3}}
\Biggl[
\hat{e}_{s} e^{i \mathbf{k}_{\text{inc}}(\omega,\left\{\Omega,s\right\})\cdot\mathbf{r}}
\nonumber \\
&+
\sum_{p=1}^{N_1}
\Bigl[\mathbf{e}_{sca}\left(\omega,\left\{\Omega,s\right\}\right)\Bigr]_{p}
\mathbf{W}^{(1)}_p(\mathbf{r})
\Biggr]
\sqrt{\frac{\hbar\omega}{2}},
\label{eqn:discrete_BA_mode}
\end{flalign}

\subsection{Degenerate MA fields}
Similarly, we use the FEM formulation to search for a numerical solution of each degenerate MA field.
The FEM counterpart of \eqref{eqn:MA_mode_governing_eqn} can be written as
\begin{flalign}
\overline{\mathbf{S}}\cdot \mathbf{e}_{MA}\left(\omega,\left\{\mathbf{r}',\xi\right\}\right) -  k^2  \overline{\mathbf{M}}\cdot \mathbf{e}_{MA}\left(\omega,\left\{\mathbf{r}',\xi\right\}\right) =i\omega\mu_0 \mathbf{j}_{MA}\left(\omega,\left\{\mathbf{r}',\xi\right\}\right).
\end{flalign}
Here, the column vector $\mathbf{e}_{MA}$ represents DoFs for each degenerate MA field $\mathbf{E}_{\text{MA}}$ with its $p$-th element corresponding to a cochain associated with the $p$-th edge, and $\mathbf{j}_{MA}$ is a column vector that collects DoFs for the current density $\mathbf{J}_{\text{MA}}$ whose elements can be computed by
\begin{flalign}
\Bigl[
\mathbf{j}_{MA}\left(\omega,\left\{\mathbf{r}',\xi\right\}\right)
\Bigr]_{p} 
= 
\int_{V_{m}}
d\mathbf{r}
\mathbf{W}^{(1)}_{p}(\mathbf{r})\cdot
\mathbf{J}_{\text{MA}}\left(\mathbf{r},\omega,\left\{\mathbf{r}',\xi\right\}\right)
=
\omega \epsilon_0
\sqrt{\frac{\hbar\chi_{I}(\mathbf{r}',\omega)}{\pi\epsilon_0}}
\mathbf{W}_p^{(1)}(\mathbf{r}')\cdot \hat{\xi}.
\end{flalign}
That is, the MA field can be obtained from a {\it point current-source radiation problem} since $\mathbf{J}_{\text{MA}}$ in \eqref{eqb:MA_source} represents the current density of a noise point current source.
Thus, each discrete MA mode can be calculated by
\begin{flalign}
\mathbf{E}^{(\text{MA})}\left(\mathbf{r},\omega,\left\{\mathbf{r}',\xi\right\}\right)
\approx
\sum_{p=1}^{N_1}
\left[\hat{e}_{M}\left(\omega,\left\{\mathbf{r}',\xi\right\}\right)\right]_p
\mathbf{W}^{(1)}_{p}(\mathbf{r}).
\end{flalign}

\subsection{Coarse-graining of continuum BA and MA fields into discrete field modes}
The expressions in \eqref{eqn:BA_field} and \eqref{eqn:MA_field} represent the continuum, considering infinitesimal frequency differentials, viz., being expanded by an uncountably infinite BA and MA fields.
However, the number of the BA and MA fields to be numerically obtained for simulations must form a countably finite set, that is, coarse-grained.
Thus, integrals in the continuum expressions should be changed by summations.
When coarse graining is performed much finely, the numerical solutions for the BA and MA fields should converge to the continuum solution.

When replacing the infinite degeneracy indices of BA and MA fields with discrete ones such as 
\begin{flalign}
&\left(\omega,\left\{\Omega,s\right\}\right) \rightarrow m,\\
&\left(\omega,\left\{\mathbf{r}',\xi\right\}\right) \rightarrow n,
\end{flalign}
one can express the resulting electric field operator in \eqref{eqn:BA/MA_fields}, \eqref{eqn:BA_field}, and \eqref{eqn:MA_field} with the countably finite set of numerical BA and MA fields:
\begin{flalign}
\mathbf{E}^{(+)}(\mathbf{r},t)
\approx
\sum_{m=1}^{N_{BA}}
\mathbf{E}^{(\text{BA})}_{m}(\mathbf{r})\hat{a}_{m}e^{-i\omega_m t}
\mathcal{D}^{(\text{BA})}_{m}
+
\sum_{n=1}^{N_{MA}}
\mathbf{E}^{(\text{MA})}_{n}(\mathbf{r})\hat{f}_{n}e^{-i\omega_n t}
\mathcal{D}^{(\text{MA})}_{m}.
\end{flalign}
Here, $\mathcal{D}_{m}^{(\text{BA})}$ and $\mathcal{D}_{n}^{(\text{MA})}$ denote the volume of the mode for the numerically obtained degenerate BA and MA fields, taking the form of
\begin{flalign}
\mathcal{D}_{m}^{(\text{BA})} &= \Delta \omega \Delta \Omega,\\
\mathcal{D}_{m}^{(\text{BA})} &= \Delta \omega \Delta \mathbf{r}'.
\end{flalign}
For efficient computations, one can employ (1) adaptive frequency refinement, (2) adaptive angular refinement on the unit sphere, and (3) adaptive mesh refinement; hence, $\mathcal{D}_{m}^{(\text{BA})}$ and $\mathcal{D}_{m}^{(\text{MA})}$ may not be a constant but varying for each discrete BA and MA field mode.

To further simplify the notation, we introduce a unified discrete mode index \( l \), representing the discrete modal index of BA or MA field modes, denoted as \( \mathbf{E}_l(\mathbf{r}) \).
And we define a unified annihilation operator, denoted as \( \hat{b}_{l} \), for the \( l \)-th discrete field mode.
Consequently, the positive-frequency component of the electric field operator becomes
\begin{flalign}
\mathbf{E}^{(+)}(\mathbf{r},t)
\approx
\sum_{l=1}^{N_f}
\mathbf{E}_{l}(\mathbf{r})\hat{b}_{l}e^{-i\omega_l t}
\mathcal{D}^{(f)}_{l}.
\end{flalign}
where \( \mathcal{D}^{(f)}_{l} \) denotes the mode volume for the unified fields, and \( N_f \) is the total number of discrete field modes, including discrete BA and MA modes.

The continuum Hamiltonian operator in \eqref{eqn:c_Ham} is replaced by 
\begin{flalign}
\hat{H}_{field} \approx \sum_{l=1}^{N_f}
\hbar\omega_l \hat{b}_{l}^{\dag}\hat{b}_{l} \mathcal{D}_{l}^{(f)}.
\label{eqn:d_Ham}
\end{flalign}

Instead of explicitly writing the mode volume, we can absorb the square root of the mode volume into the factors of the mode and annihilation operators such as
\begin{flalign}
\mathbf{E}_{l}(\mathbf{r}) &\rightarrow \sqrt{\mathcal{D}^{(f)}_l}\mathbf{E}_{l}(\mathbf{r}), \\
\hat{b}_l &\rightarrow \sqrt{\mathcal{D}^{(f)}_l}\hat{b}_l, \\
\hat{b}^{\dag}_l &\rightarrow \sqrt{\mathcal{D}^{(f)}_l}\hat{b}^{\dag}_l,
\end{flalign}
allowing for a more simplified expression, which is the traditional expression for the quantized EM fields in a periodic vacuum box, below:
\begin{flalign}
\mathbf{E}^{(+)}(\mathbf{r},t)
&\approx
\sum_{l=1}^{N_f}
\mathbf{E}_{l}(\mathbf{r})\hat{b}_{l}e^{-i\omega_l t},
\label{eqn:d_BAMA_E_FIN}\\
\hat{H}_{field} 
&\approx \sum_{l=1}^{N_f}
\hbar\omega_l \hat{b}_{l}^{\dag}\hat{b}_{l}.
\label{eqn:d_BAMA_H_FIN}
\end{flalign}
In what follows without specifying, we assume that discrete BA and MA modes and their annihilation (and creation) operators possess the square root of the mode volume factor.

\section{Multimode Jaynes- and Tavis-Cummings models incorporating the modified Langevin noise formalism}
Let us incorporate the discrete BA and MA modes obtained in the previous section into the multimode Jayens- and Tavis-Cummings models in order to investigate interactions between atom(s) and fields in which EM systems involve in IDA dielectric objects of finite size and arbitrary shape in free space.

The Schrödinger equation, written by
\begin{flalign}
\hat{H}\ket{\psi(t)} = i\hbar\frac{\partial}{\partial t}\ket{\psi(t)}
\end{flalign}
takes the following Hamiltonian operator consisting of atom(s), field, and interaction parts:
\begin{flalign}
\hat{H} = \hat{H}_{atom} +\hat{H}_{field} + \hat{H}_{int}.
\end{flalign}
The Hamiltonian for atom(s) can be written by
\begin{flalign}
\hat{H}_{atom} 
= 
\sum_{j=1}^{N_a} \hbar\frac{\omega_j^{(a)}}{2}\hat{\sigma}_{j}^{(Z)}
\end{flalign}
where for $j$-th atom $\omega_j^{(a)}$ = ($E^{(e)}_{j}-E^{(g)}_{j})/\hbar$ denotes a transition frequency, and $\hat{\sigma}_{j}^{(Z)} = \ket{e_j}\bra{e_j}-\ket{g_j}\bra{g_j}$ represents the Pauli Z-operator.
The Hamiltonian for the discrete BA and MA modes is expressed by
\begin{flalign}
\hat{H}_{field} = \sum_{l=1}^{N_f} \hbar\omega_{l}\hat{b}_{l}^{\dag}\hat{b}_{l}.
\end{flalign}
With the use of the rotating wave approximation and the dipole approximation, the interaction Hamiltonian can be written by
\begin{flalign}
\hat{H}_{int} = 
-\sum_{j=1}^{N_{a}} \hat{\mathbf{d}}_{j}\cdot \hat{\mathbf{E}}(\mathbf{r}_{j}^{(a)},t=0)
\approx
-\sum_{j=1}^{N_{a}} 
\sum_{l=1}^{N_f}
\hbar
\left(
\gamma_{j,l}
\hat{\sigma}^{(+)}_{j}
\hat{b}_{l}
+
\gamma_{j,l}^{*}
\hat{\sigma}^{(-)}_{j}
\hat{b}_{l}^{\dag}
\right)
\end{flalign}
where $N_a$ denotes the total number of atoms, $\mathbf{r}_j^{(a)}$ refers to the location of $j$-th atom, $\hat{\mathbf{d}}_j=\mathbf{d}_{j}^{(eg)}\hat{\sigma}^{(-)}_j+\mathbf{d}_j^{(ge)}\hat{\sigma}^{(+)}_j$ represents a dipole moment operator with the the transition dipole moment $\mathbf{d}_j^{(eg)}=\mel{e_j}{\hat{\mathbf{d}}_j}{g_j}=\left(\mathbf{d}_j^{(ge)}\right)^*$, $\hat{\sigma}^{(\pm)}_j$ denote raising and lowering operators for $j$-th atom, respectively, and $\gamma_{j,l}$ is a coupling strength between $j$-th atom and $l$-th discrete BA and MA mode given by
\begin{flalign}
\gamma_{j,l} = 
-
\frac{1}{\hbar}
\mel
{e_j}
{\hat{\mathbf{d}}_j}
{g_j}
\cdot
\mathbf{E}_{l}(\mathbf{r}_{j}^{(a)})
=
\frac{1}{\hbar}
\mathbf{d}_{j}^{(eg)}\cdot \mathbf{E}_{l}(\mathbf{r}_{j}^{(a)}).
\end{flalign}

\subsection{Quantum states in the truncated Hilbert space}
The Jaynes- and Tavis-Cummings models always conserve the total number of quanta initially excited.
This study focuses on how initially excited atoms (or two-level systems) emit photons in the presence of IDA dielectric objects of finite size and arbitrary shape in free space.
In other words, the quanta initially stored in atoms are transferred to the BA and MA field modes while preserving the total number of quanta.
More precisely, if the atom is initially excited, it will lose its initial quanta over time, while either the BA or MA field modes acquire the atom’s initial quanta.
In a non-Markovian environment, the photon field scattered back to the atom can return some of the quanta, re-exciting the atom while causing the BA or MA field modes to lose quanta.
However, since the BA and MA fields do not interact with each other, there is no transfer of quanta between the BA and MA field modes.
Thus, the total atom-field system conserves the number of initial quanta, with quanta being exchanged between the atom and the BA or MA field modes depending on the surrounding environment.
Therefore, a truncated Hilbert space can be considered.

For a maximum of $M$ quanta, a general bare quantum state in the truncated Hilbert space can be expressed as
\begin{flalign}
    \ket{\psi(t)}
    &= \sum_{i=1}^{N_{(0)}} \Psi_{i}^{(0)}(t)\ket{\psi_i^{(0)}}
    + \sum_{i=1}^{N_{(1)}} \Psi_{i}^{(1)}(t)\ket{\psi_i^{(1)}}
    + \cdots
    + \sum_{i=1}^{N_{(M)}} \Psi_{i}^{(M)}(t)\ket{\psi_i^{(M)}}.
\end{flalign}

Here, $\ket{\psi_i^{(q)}}$ represents the $i$-th bare eigenstate having $q$ quanta, and $\Psi_{i}^{(q)}(t)$ denotes the corresponding probability amplitude for $\ket{\psi_i^{(q)}}$. The quantity $N_{(q)}$ refers to the total number of possible bare eigenstates containing exactly $q$ quanta.
For example, followings are of single-quanta quantum states for two atoms and three field modes:
\begin{flalign}
\ket{\psi_1^{(1)}} &= \ket{e,g,0,0,0},~\ket{\psi_2^{(1)}} = \ket{g,e,0,0,0},~\ket{\psi_3^{(1)}} = \ket{g,g,1,0,0},
\nonumber \\
\ket{\psi_4^{(1)}} &= \ket{g,g,0,1,0}, \ket{\psi_5^{(1)}} = \ket{g,g,0,0,1}.
\end{flalign}
By replacing the indices $(q,i)$ with a single bare eigenstate index $s$, one can rewrite the general quantum state more compactly as
\begin{flalign}
    \ket{\psi(t)} = \sum_{s=1}^{N_s} \Psi_s(t) \ket{\psi_s},
    \label{eqn:gen_es}
\end{flalign}
where $N_s$ represents the total number of bare eigenstates up to the maximum of $M$ quanta, $\ket{\psi_s}$ denotes the $s$-th bare eigenstate, and $\Psi_s(t)$ corresponds to the probability amplitude for $\ket{\psi_s}$.

\subsection{Matrix representation of quantum states and the Hamiltonian operators}
By collecting all the probability amplitudes of the bare eigenstates and reshaping them into a column vector, one can obtain the matrix representation of a generic quantum state.
\begin{flalign}
\boldsymbol{\Psi}(t)
&=
\left[
\left\{\Psi_{i}^{(0)}(t)\right\},
\left\{\Psi_{i}^{(1)}(t)\right\},
\left\{\Psi_{i}^{(2)}(t)\right\}
\cdots
\right]^{T}
\nonumber \\
&=
\left[
\Psi_{1}(t),
\Psi_{2}(t),
\Psi_{3}(t),
\cdots,
\Psi_{q}(t),
\cdots,
\right]^{T}.
\end{flalign}

In order to solve the Schrödinger equation, that is, to find the probability amplitudes, one needs to find the matrix representations of the Hamiltonian operators with respect to the general quantum state found in \eqref{eqn:gen_es}. 
The matrix representation can be obtained through
\begin{flalign}
\left[\overline{\mathbf{H}}\right]_{p,q}
=
\mel
{\psi_p}
{\hat{H}}
{\psi_q}.
\end{flalign}
%The mathematical procedures are provided in Appendix \ref{} in detail.

\subsection{Numerical integration of the Schrödinger equation}
We have obtained the matrix representations of the quantum states and the Hamiltonian operators for the truncated Hilbert space.
Using these, one needs to introduce a numerical integration in order to solve the Schrödinger equation which includes the first-order time derivatives.
The Schrödinger equation in the matrix representation can be written by
\begin{flalign}
\overline{\mathbf{H}}\cdot \boldsymbol{\Psi}(t)
=
i\hbar
\frac{\partial}{\partial t}
\boldsymbol{\Psi}(t)
\label{eqn:MR_SCE}
\end{flalign}
When putting 
\begin{flalign}
\boldsymbol{\Psi}^{n} \triangleq \boldsymbol{\Psi}(n \Delta t)
\end{flalign}
where $\Delta t$ is time segment.
Here, we apply the finite (central) difference method and time averaging to \eqref{eqn:MR_SCE} such that the following time-update scheme can be obtained:
\begin{flalign}
\overline{\mathbf{H}}\cdot 
\frac{
\boldsymbol{\Psi}^{n+1}+\boldsymbol{\Psi}^{n}
}{2}
=
i\hbar
\frac{
\boldsymbol{\Psi}^{n+1}-\boldsymbol{\Psi}^{n}
}{\Delta t}
\end{flalign}
which updates finally
\begin{flalign}
\boldsymbol{\Psi}^{n+1}
=
\left(\overline{\mathbf{H}}-\frac{2 i\hbar}{\Delta t}\overline{\mathbf{I}}\right)^{-1}
\cdot
\left(-\overline{\mathbf{H}}-\frac{2 i\hbar}{\Delta t}\overline{\mathbf{I}}\right)
\cdot
\boldsymbol{\Psi}^{n}.
\end{flalign}
The time evolution matrix $\overline{\mathbf{A}} = \left(\overline{\mathbf{H}}-\frac{2 i\hbar}{\Delta t}\overline{\mathbf{I}}\right)^{-1}
\cdot
\left(-\overline{\mathbf{H}}-\frac{2 i\hbar}{\Delta t}\overline{\mathbf{I}}\right)$ is a unitary matrix; hence, the magnitude of its spectral radius is always equal to 1.
Thus, the time update will be stable regardless of $\Delta t$.
Note that the Runge-Kutta 4\textsuperscript{th} order numerical integrator may be used alternatively.

\section{Numerical examples}
In this section, we validate our framework with several numerical examples: Two single atom cases use Jaynes-Cummings model and two multi-atom cases using the Tavis-Cummings model. 
These examples are considered to examine the non-Markovian effects, open, and dissipative environment, as well as collective atomic behaviors.  

\subsection{Dynamics of a TLS in PEC half space}
\begin{figure}[t]
\centering
\includegraphics[width=4in]{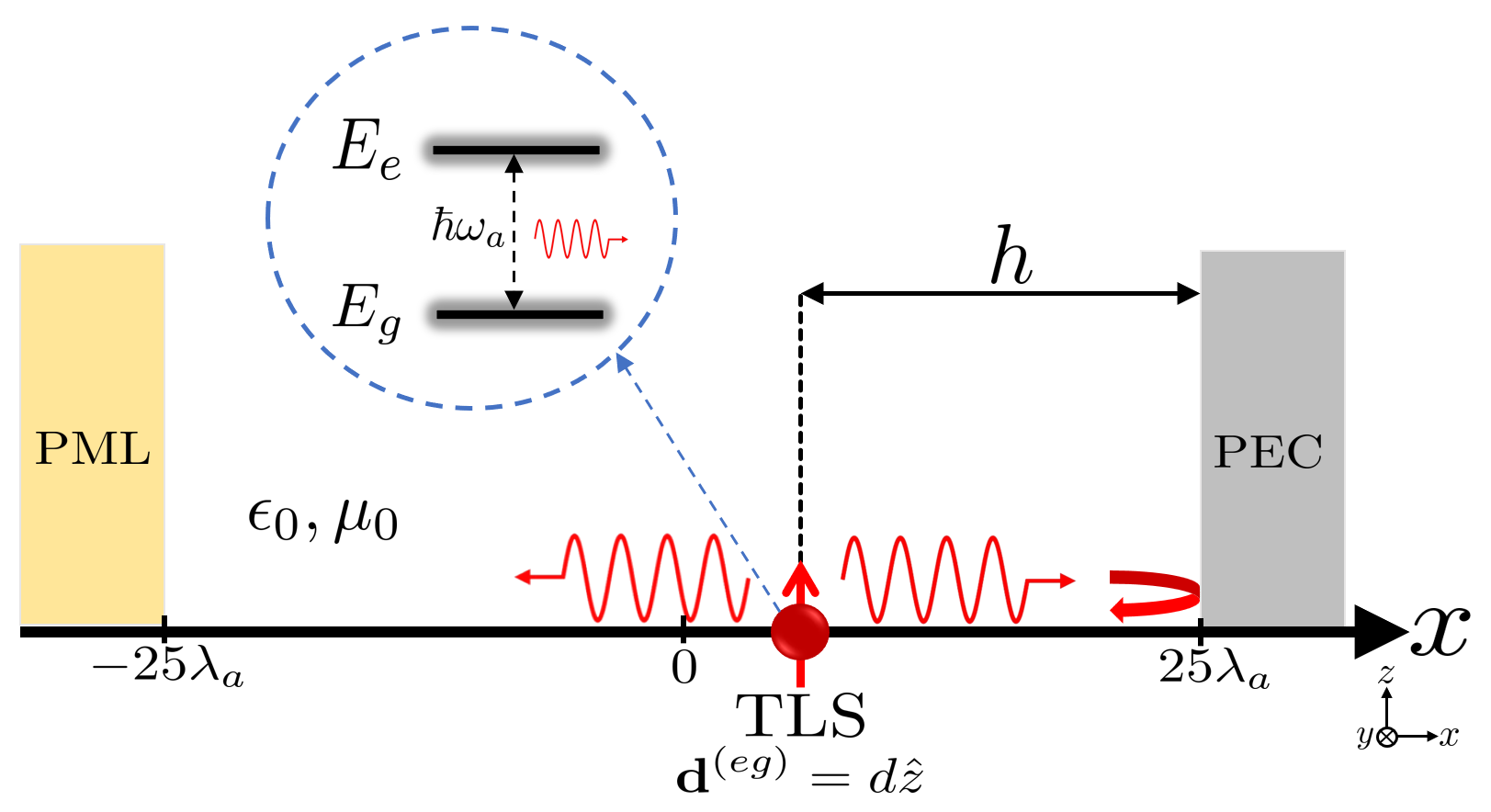}
\caption{Schematic of the numerical simulation depicting the dynamics of a TLS in a PEC half space at different distance ($h$) from the PEC mirror}
\label{fig:half_PEC}
\end{figure}

%% 20250316 by Hyunwoo Choi
First, we consider a simple one-dimensional setup, which involves the dynamics of a two-level atom (TLS) in a perfectly electric conductor (PEC) half-space. Specifically, we observe the atomic population probability and calculate the decaying constant of an atom positioned at a certain distance ($h$) from the PEC mirror. This example primarily focuses on validating the non-Markovian process in open environment. The problem geometry is illustrated in Fig. \ref{fig:half_PEC}. In our setup, the radiated field propagates in the $\pm x$-direction while being polarized in $+z$-direction. The atomic angular transition frequency ($\omega_a$) is set to 50c, while the dipole moment is 0.1 aligns to the polarization direction, i.e., $+z$ direction. To model the open environment, we implement a perfectly matched layer (PML) at the left boundary. The simulation domain without PML is set to 50$\lambda_a$, where $\lambda_a$ is defined as $2\pi/\omega_a$. 
In this simulation, we compare our multi mode Jaynes-Cummings model with BA fields (MMJC-BA) formalism with the Finite-Difference Time-Domain method for quantum emitters (FDTD-QE), a recently developed non-Markovian algorithm by Q. Zhou and colleagues \cite{Zhou2024Simulating}. In our example decribed in figure 2, the change in atomic population probability for different $h$ is based on the Purcell effect which refers to the dependence of the atom's decaying rate on the surrounding environment. In our simulation setup, the TLS is initially in the excited state, and the Purcell effect primarily arises from its interaction with the reflected field—originating from the atom itself—off the PEC surface.

\begin{figure}[htbp]
    \centering

    \begin{subfigure}[b]{0.22\textwidth}
        \includegraphics[width=\linewidth]{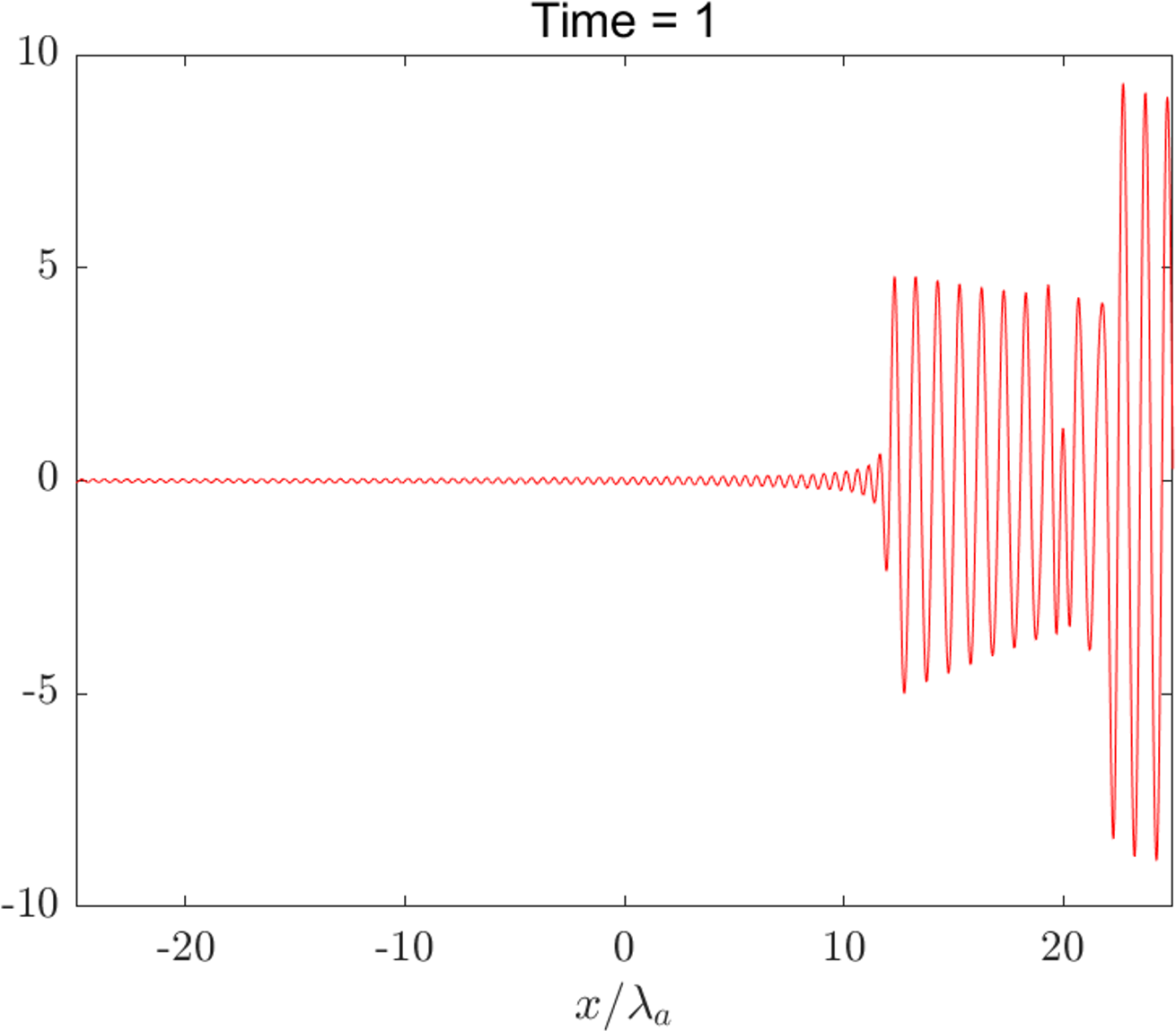}
        \caption{$(h,t) = (5\lambda_a,1)$}
    \end{subfigure}
    \begin{subfigure}[b]{0.22\textwidth}
        \includegraphics[width=\linewidth]{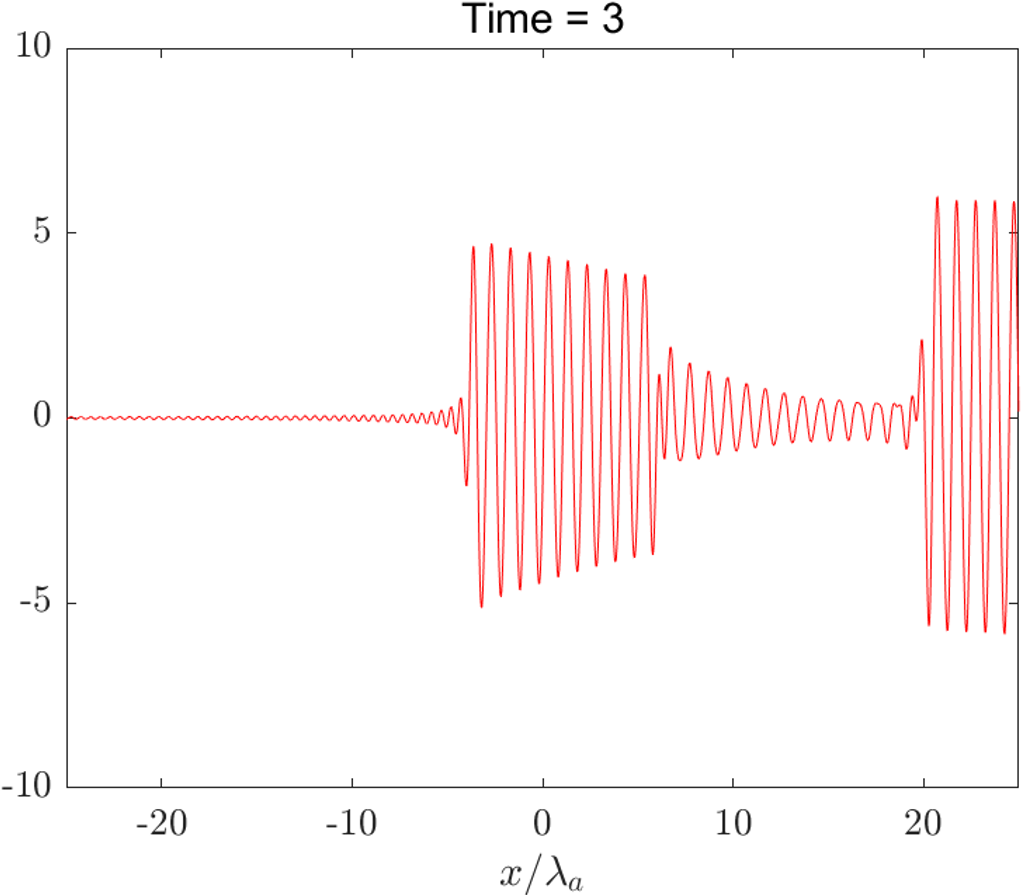}
        \caption{$(h,t) = (5\lambda_a,3)$}
    \end{subfigure}
    \begin{subfigure}[b]{0.22\textwidth}
        \includegraphics[width=\linewidth]{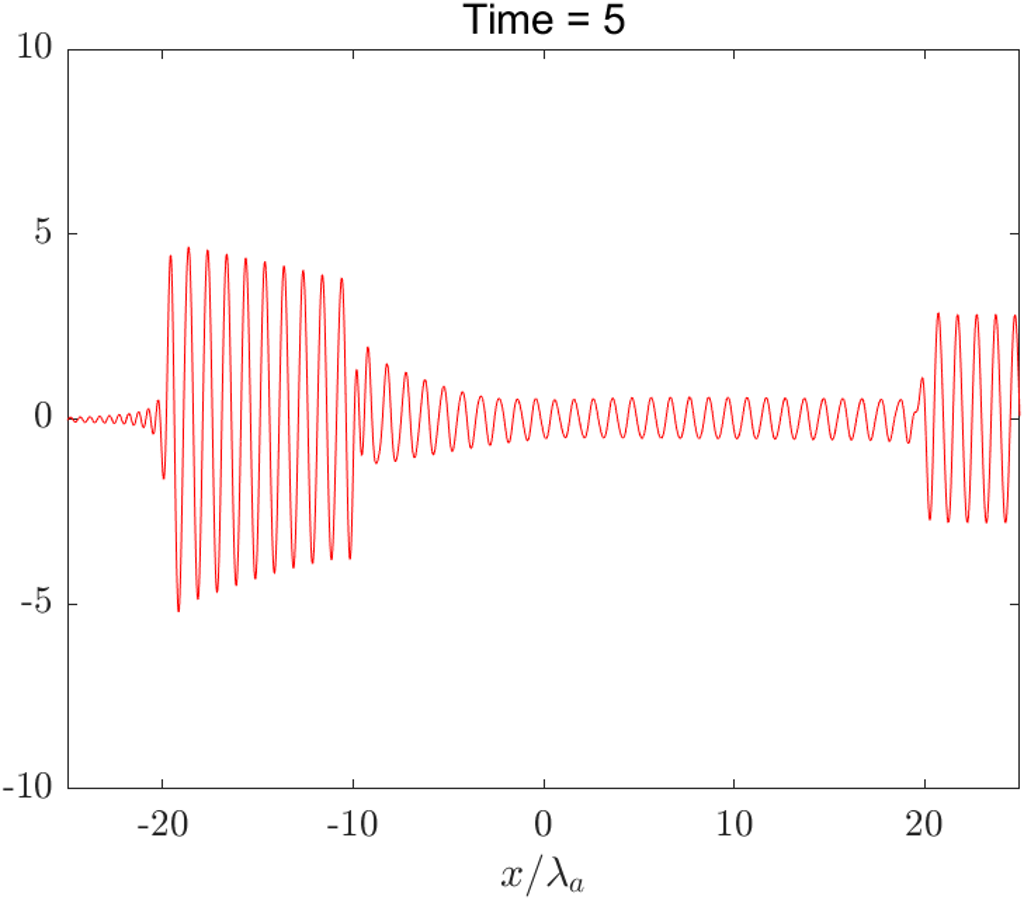}
        \caption{$(h,t) = (5\lambda_a,5)$}
    \end{subfigure}
    \begin{subfigure}[b]{0.22\textwidth}
        \includegraphics[width=\linewidth]{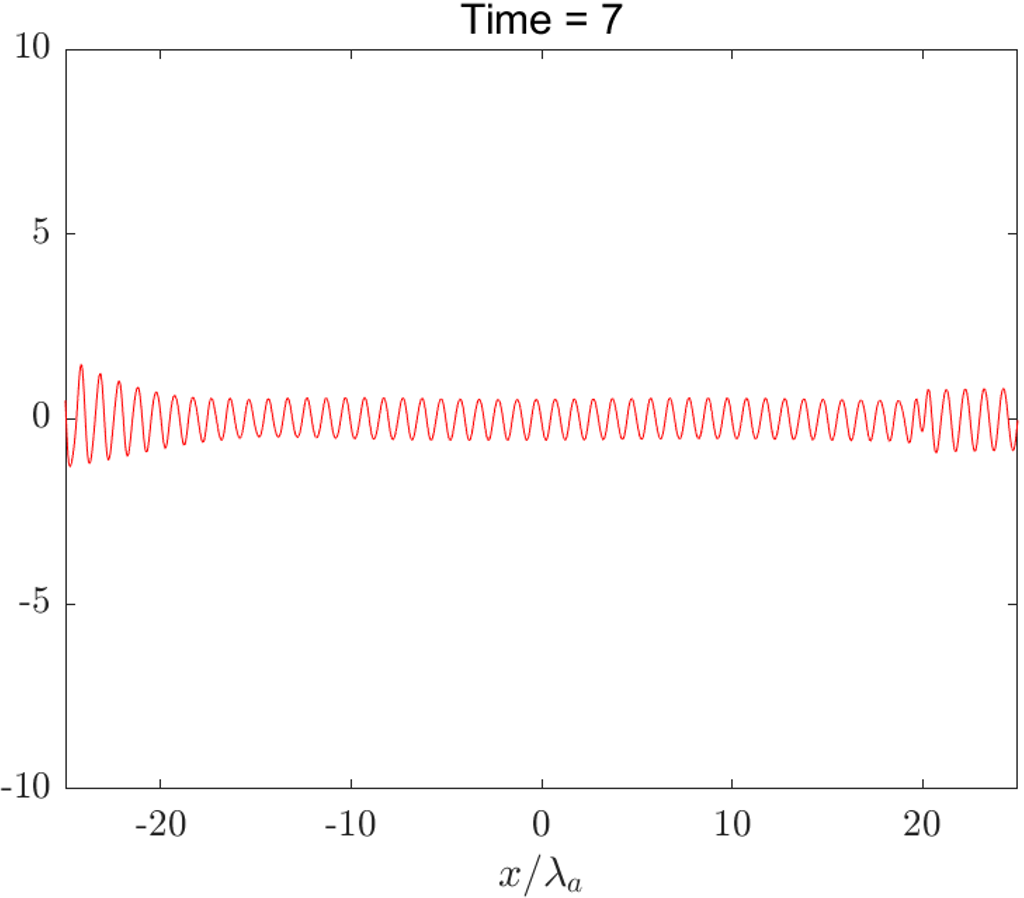}
        \caption{$(h,t) = (5\lambda_a,7)$}
    \end{subfigure}

    \vspace{0.5em}

    \begin{subfigure}[b]{0.22\textwidth}
        \includegraphics[width=\linewidth]{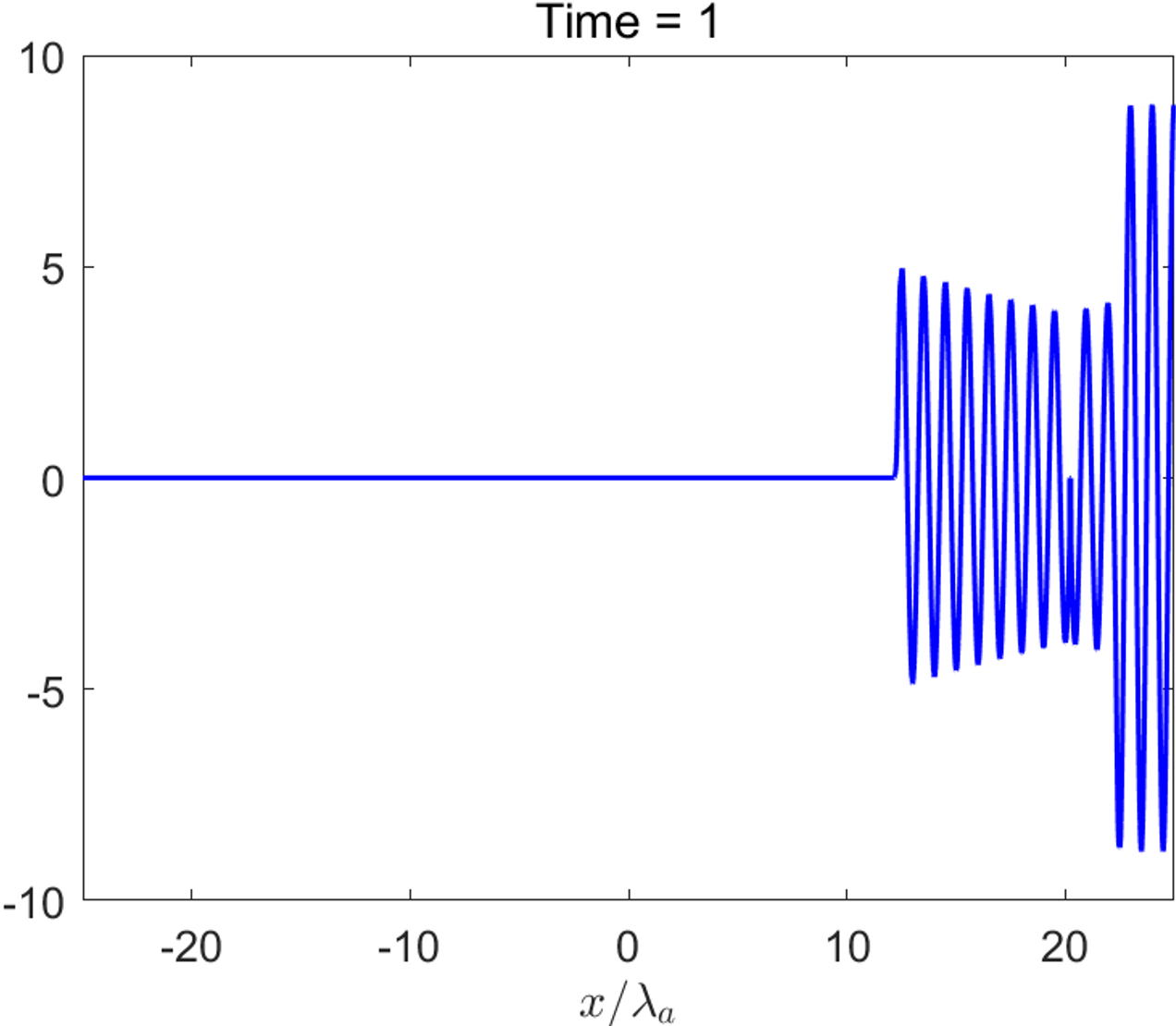}
        \caption{$(h,t) = (5\lambda_a,1)$}
    \end{subfigure}
    \begin{subfigure}[b]{0.22\textwidth}
        \includegraphics[width=\linewidth]{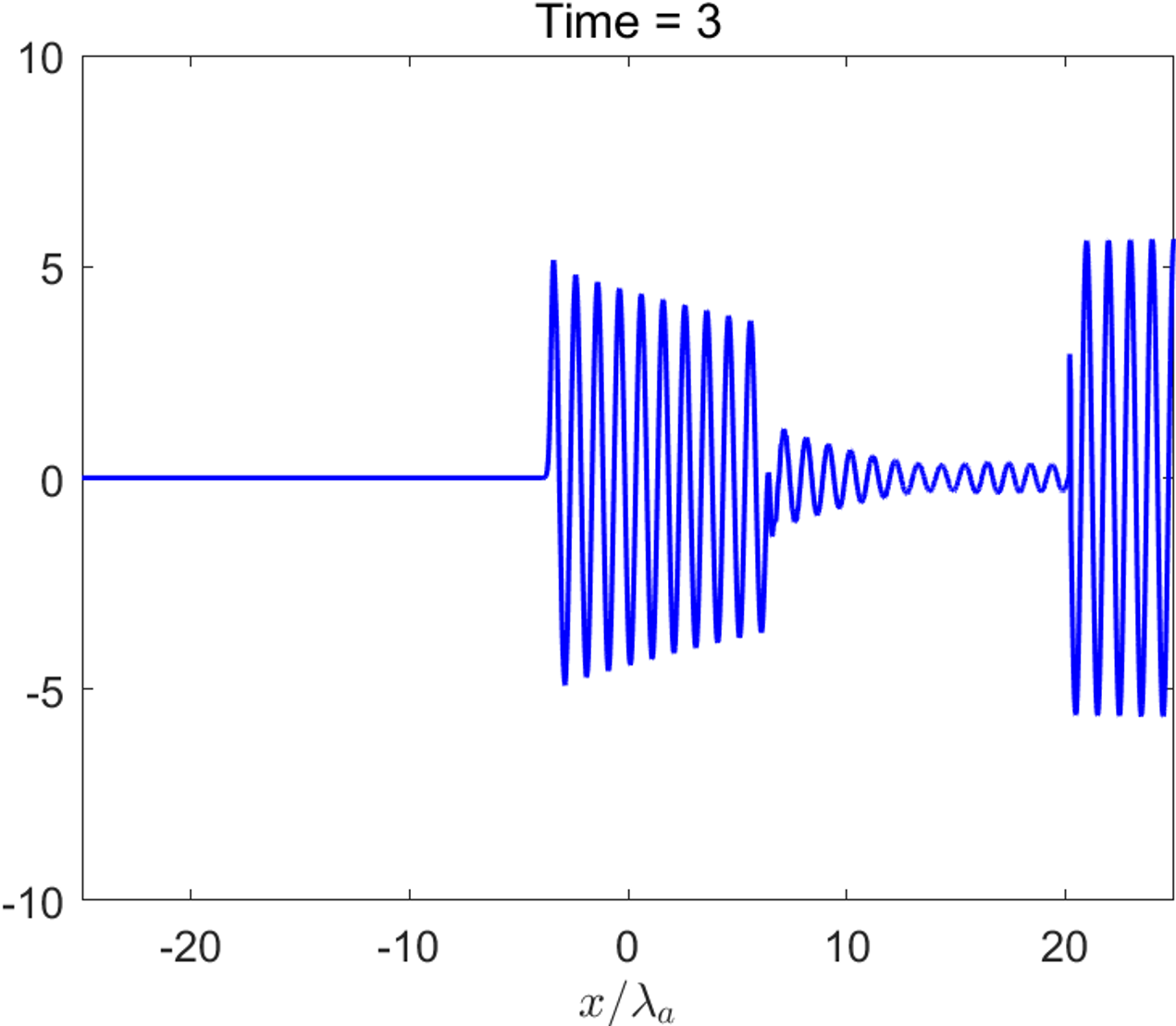}
        \caption{$(h,t) = (5\lambda_a,3)$}
    \end{subfigure}
    \begin{subfigure}[b]{0.22\textwidth}
        \includegraphics[width=\linewidth]{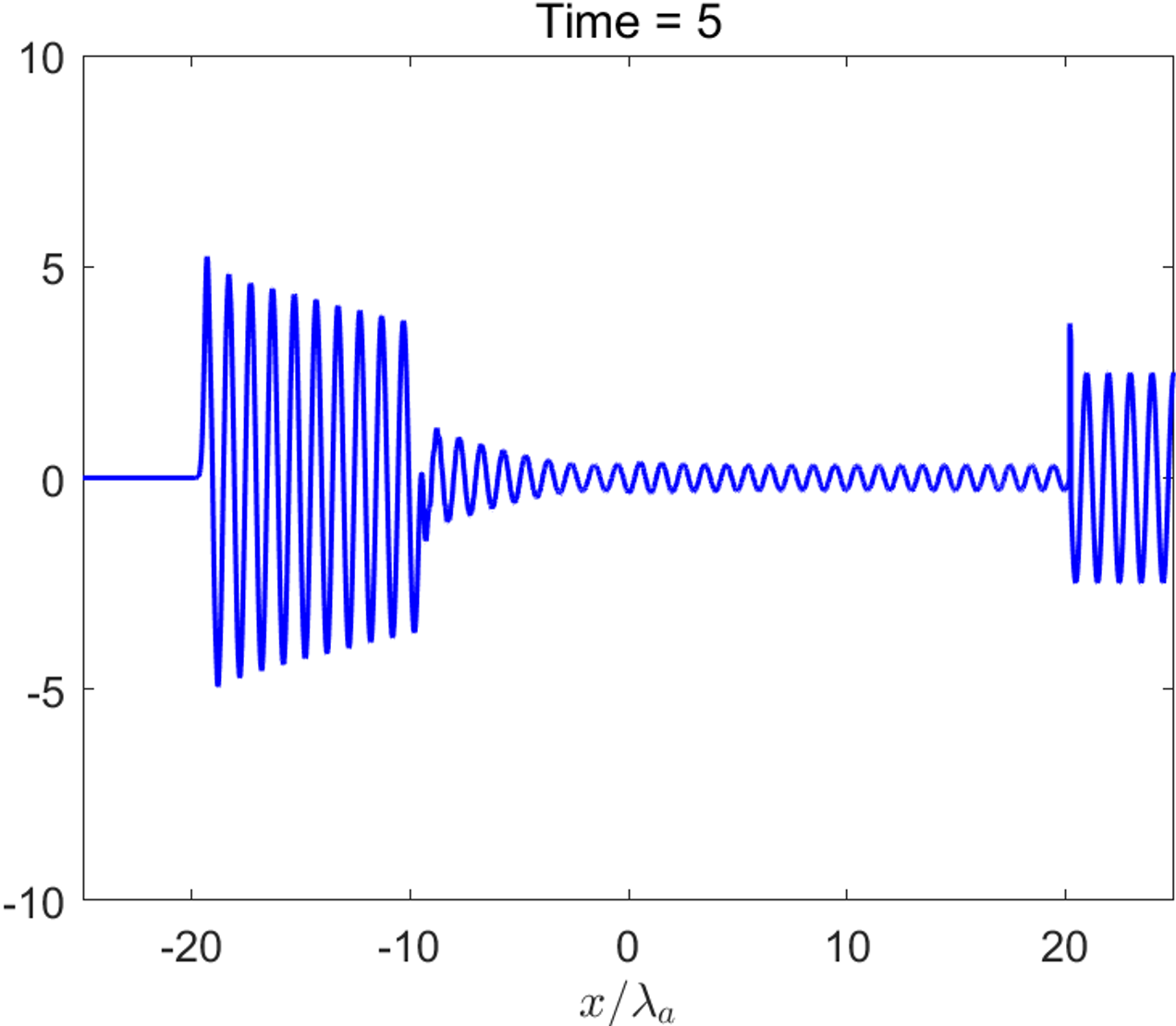}
        \caption{$(h,t) = (5\lambda_a,5)$}
    \end{subfigure}
    \begin{subfigure}[b]{0.22\textwidth}
        \includegraphics[width=\linewidth]{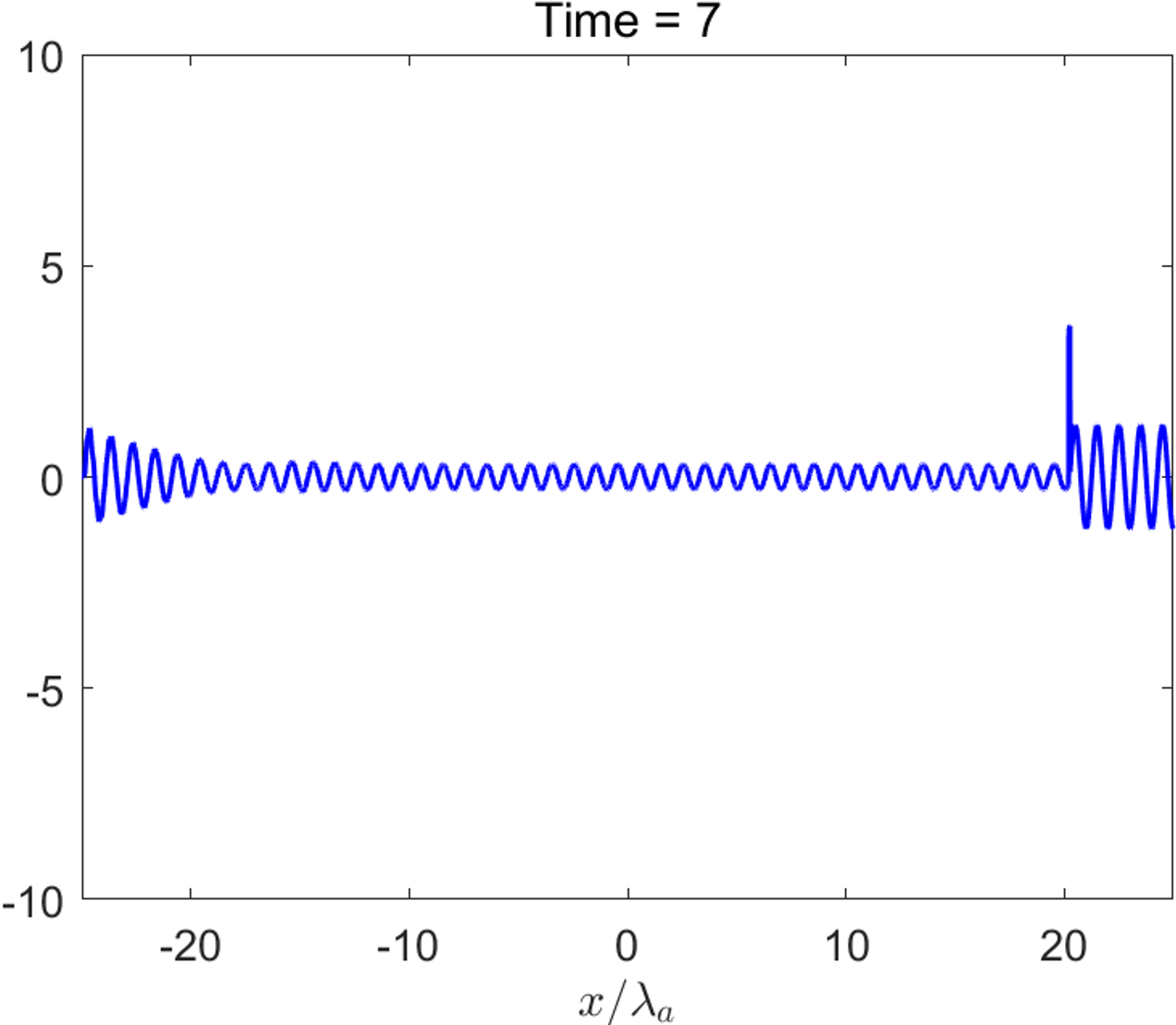}
        \caption{$(h,t) = (5\lambda_a,7)$}
    \end{subfigure}

    \vspace{0.5em}

    \begin{subfigure}[b]{0.22\textwidth}
        \includegraphics[width=\linewidth]{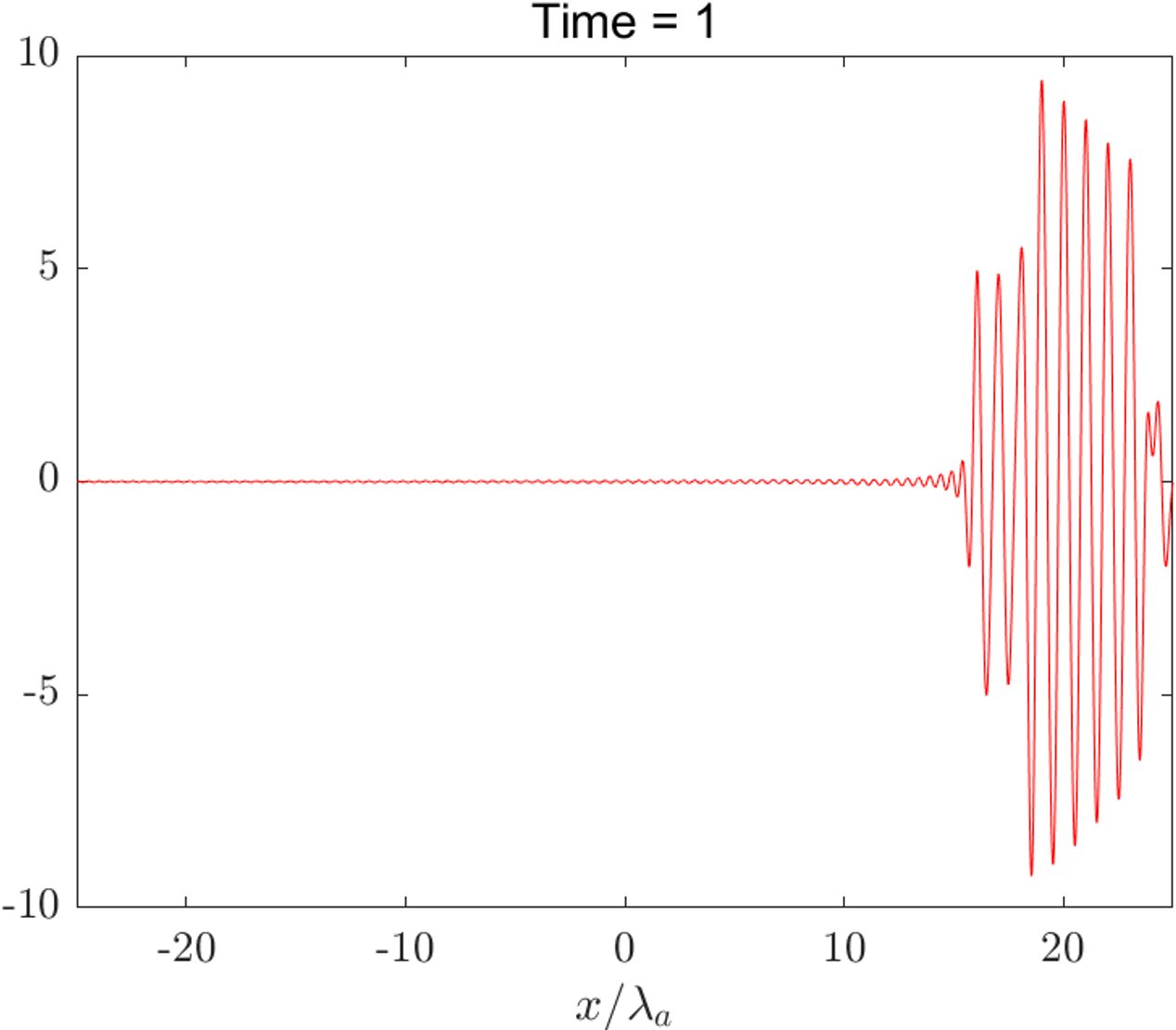}
        \caption{$(h,t) = (1.25\lambda_a,1)$}
    \end{subfigure}
    \begin{subfigure}[b]{0.22\textwidth}
        \includegraphics[width=\linewidth]{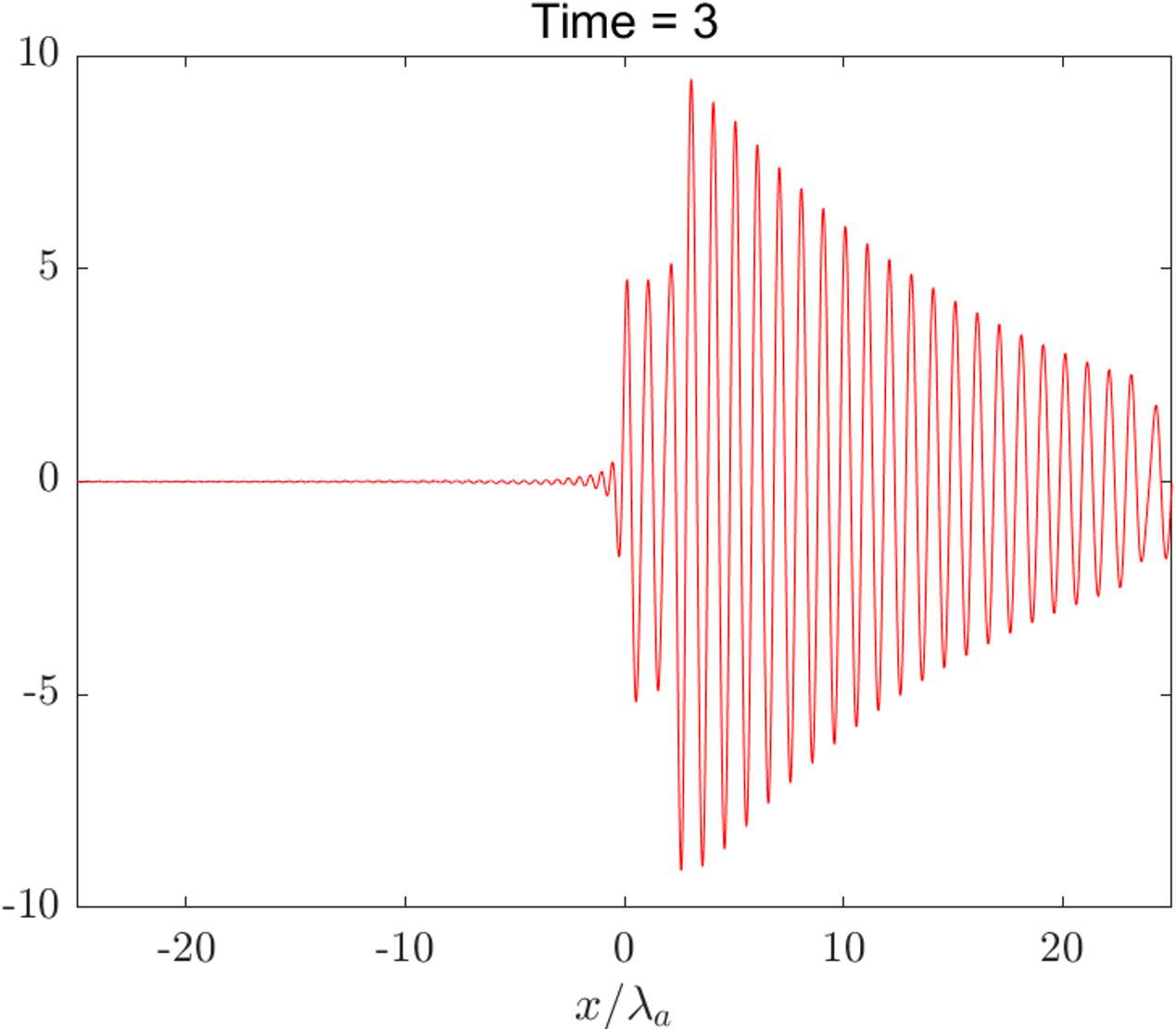}
        \caption{$(h,t) = (1.25\lambda_a,3)$}
    \end{subfigure}
    \begin{subfigure}[b]{0.22\textwidth}
        \includegraphics[width=\linewidth]{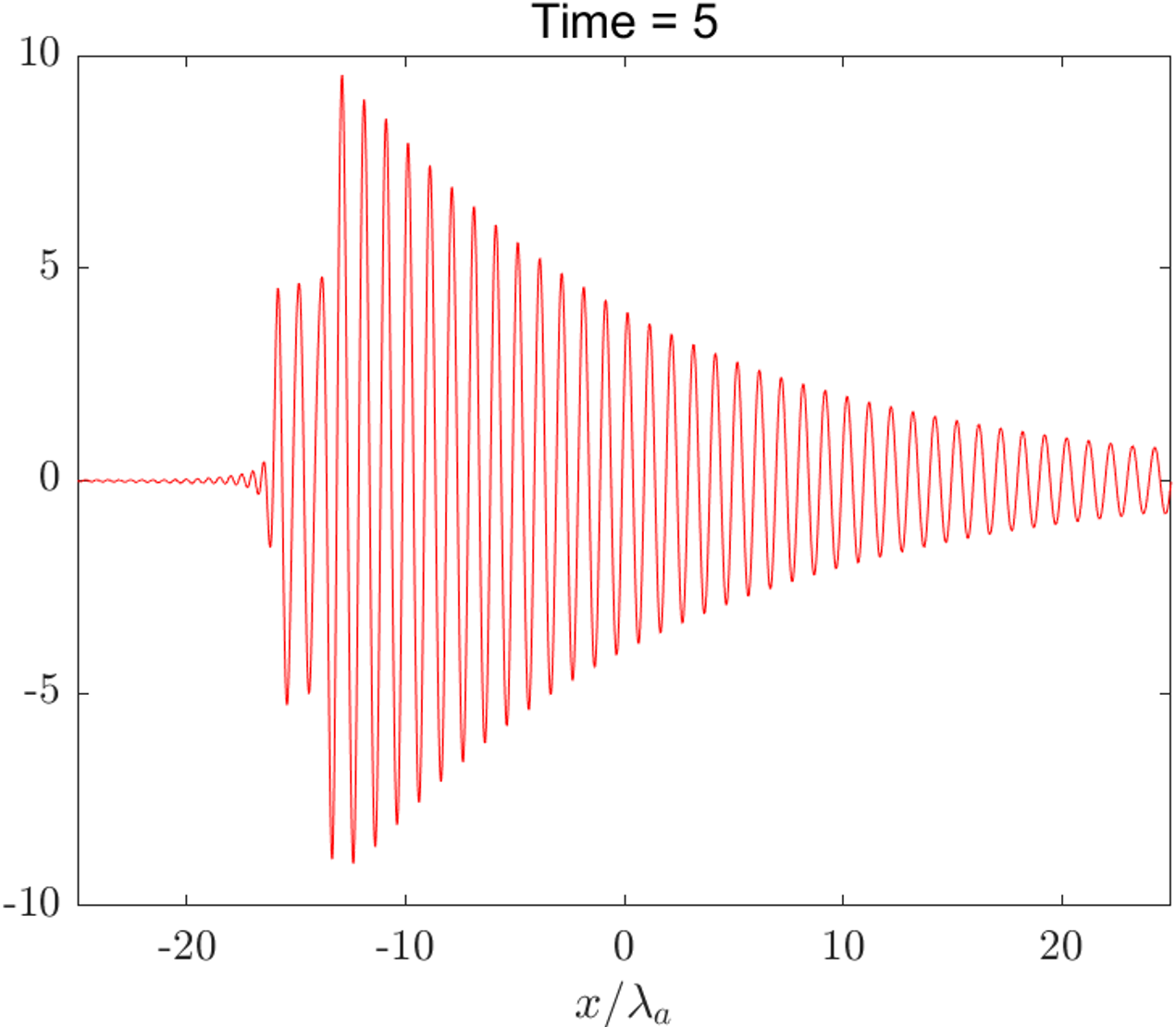}
        \caption{$(h,t) = (1.25\lambda_a,5)$}
    \end{subfigure}
    \begin{subfigure}[b]{0.22\textwidth}
        \includegraphics[width=\linewidth]{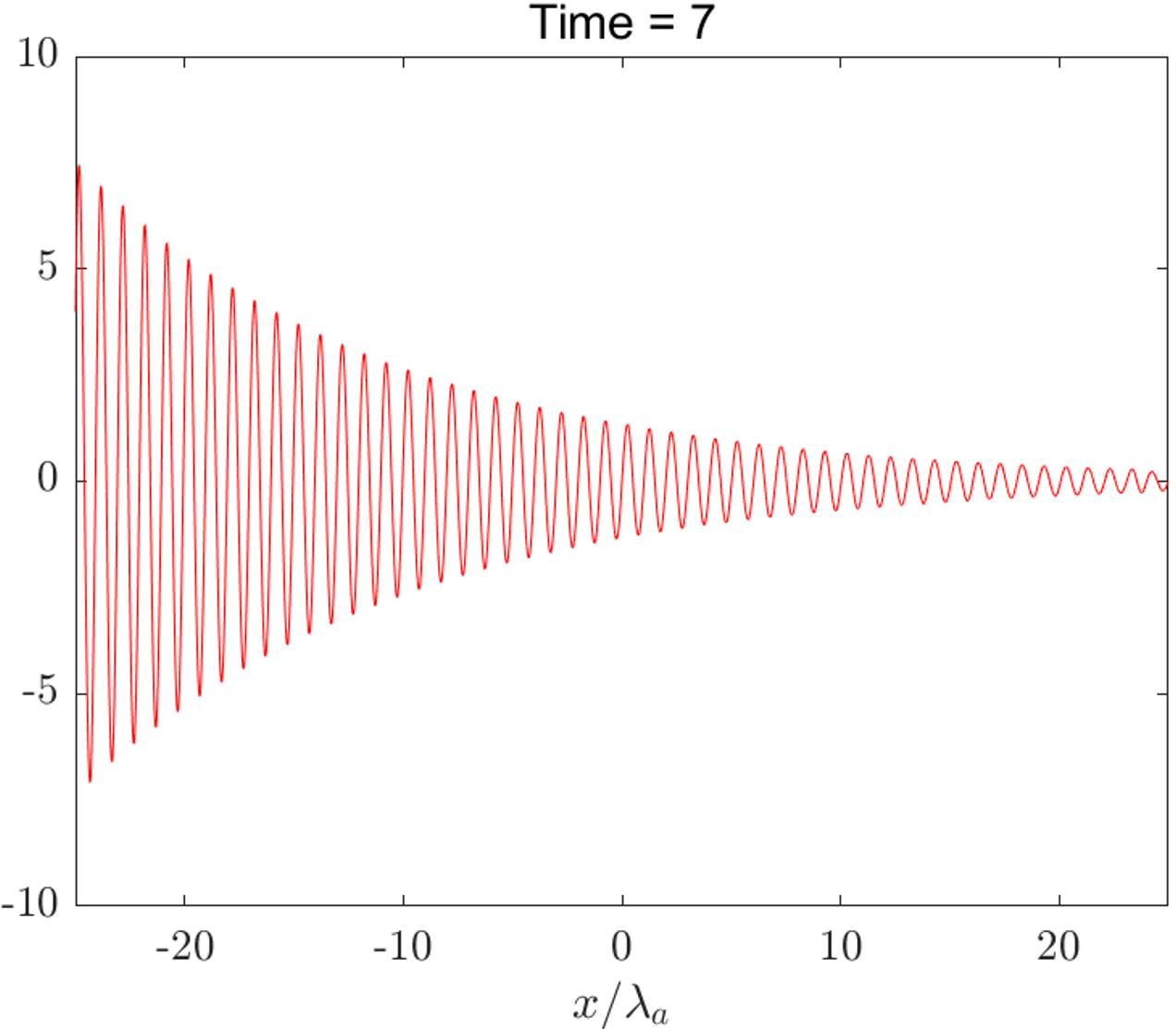}
        \caption{$(h,t) = (1.25\lambda_a,7)$}
    \end{subfigure}

    \vspace{0.5em}

    \begin{subfigure}[b]{0.22\textwidth}
        \includegraphics[width=\linewidth]{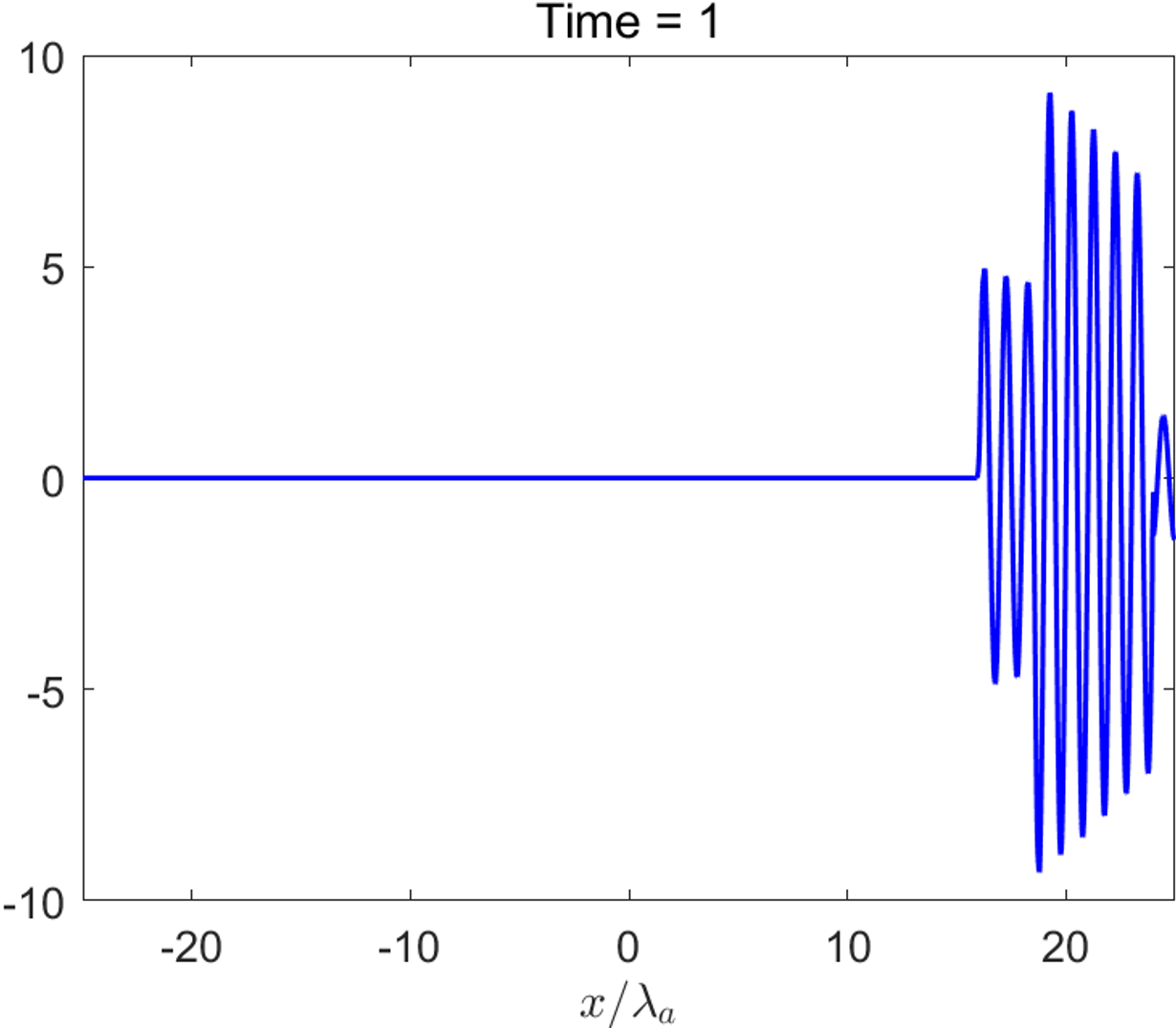}
        \caption{$(h,t) = (1.25\lambda_a,1)$}
    \end{subfigure}
    \begin{subfigure}[b]{0.22\textwidth}
        \includegraphics[width=\linewidth]{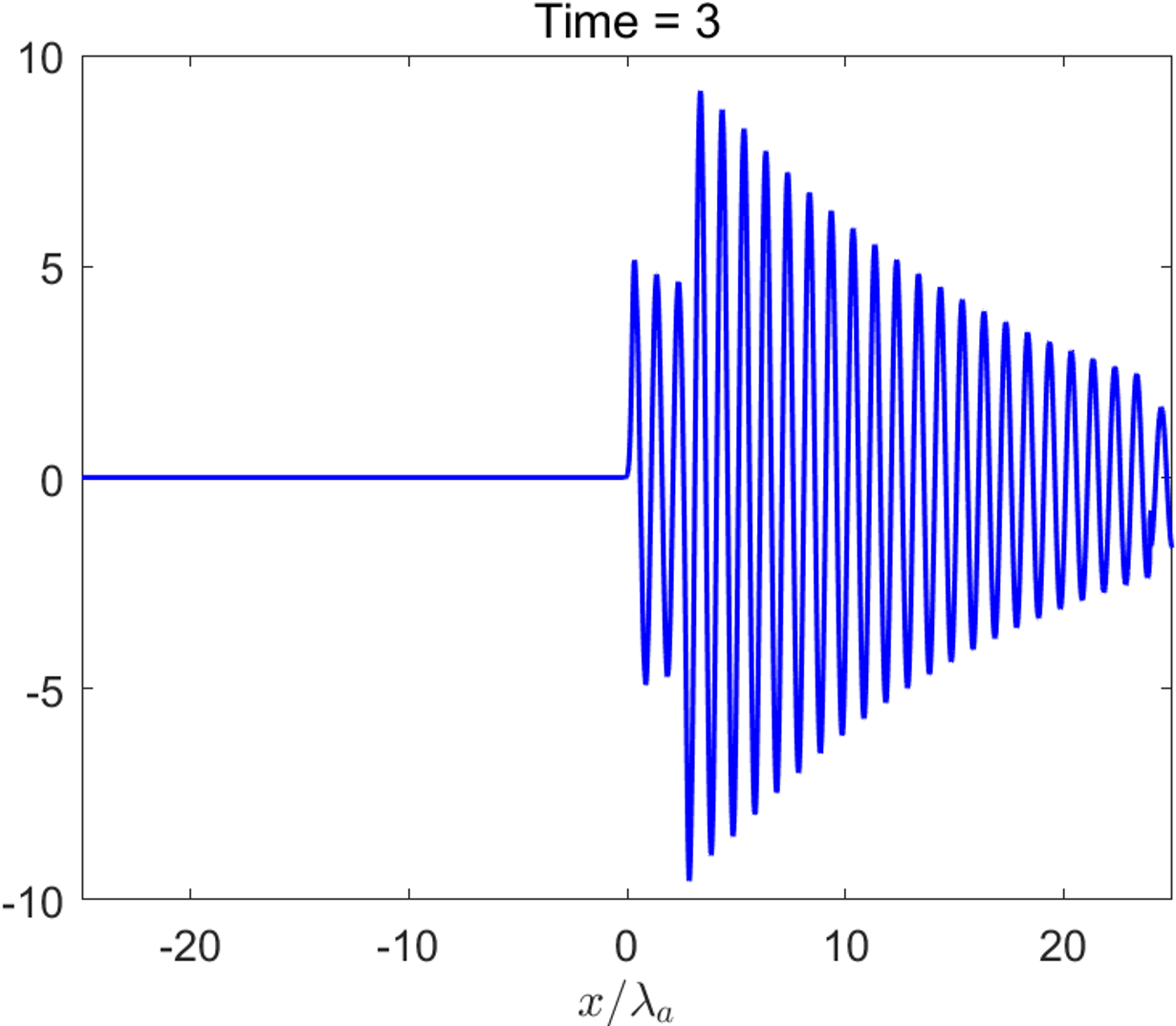}
        \caption{$(h,t) = (1.25\lambda_a,3)$}
    \end{subfigure}
    \begin{subfigure}[b]{0.22\textwidth}
        \includegraphics[width=\linewidth]{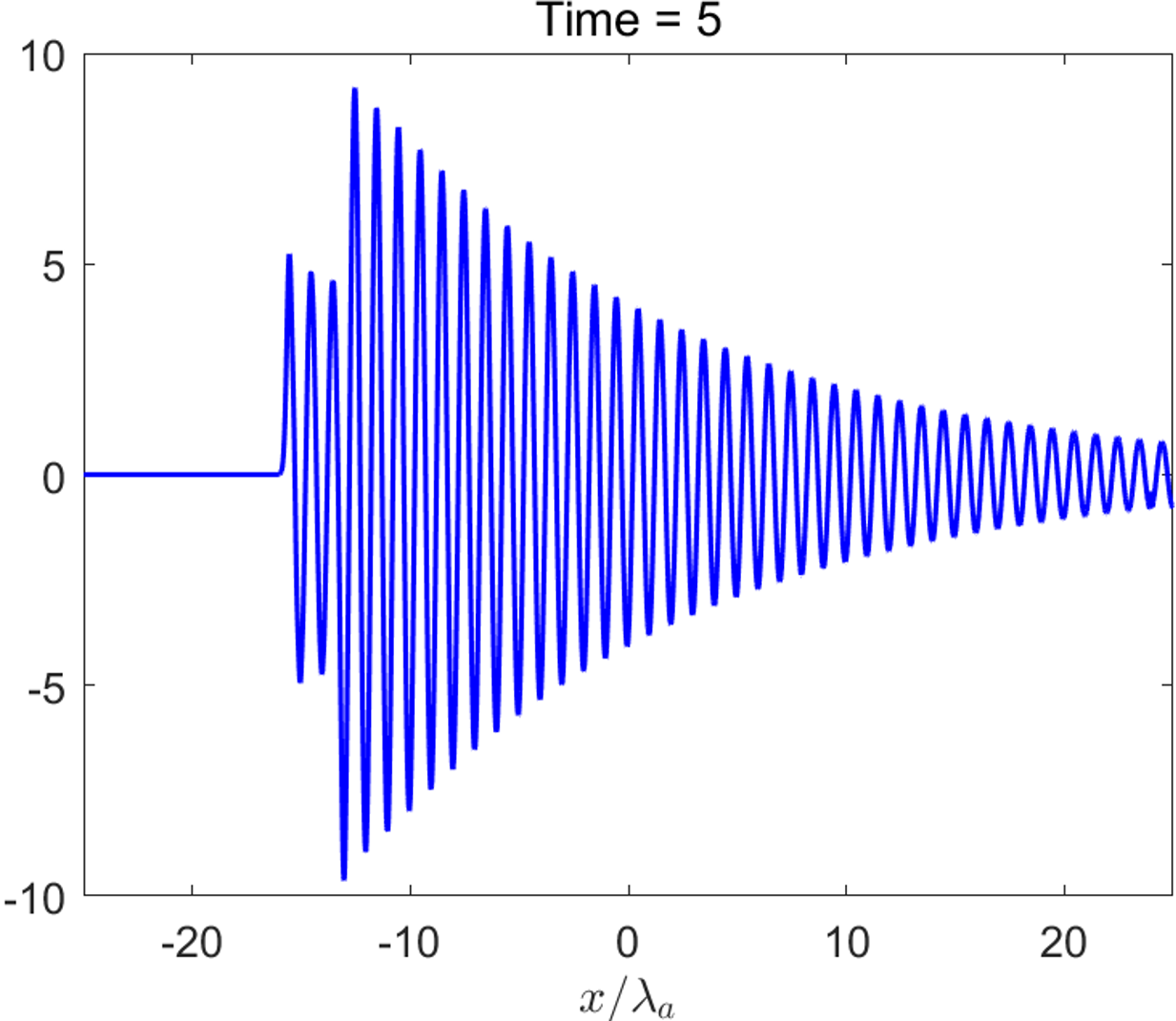}
        \caption{$(h,t) = (1.25\lambda_a,5)$}
    \end{subfigure}
    \begin{subfigure}[b]{0.22\textwidth}
        \includegraphics[width=\linewidth]{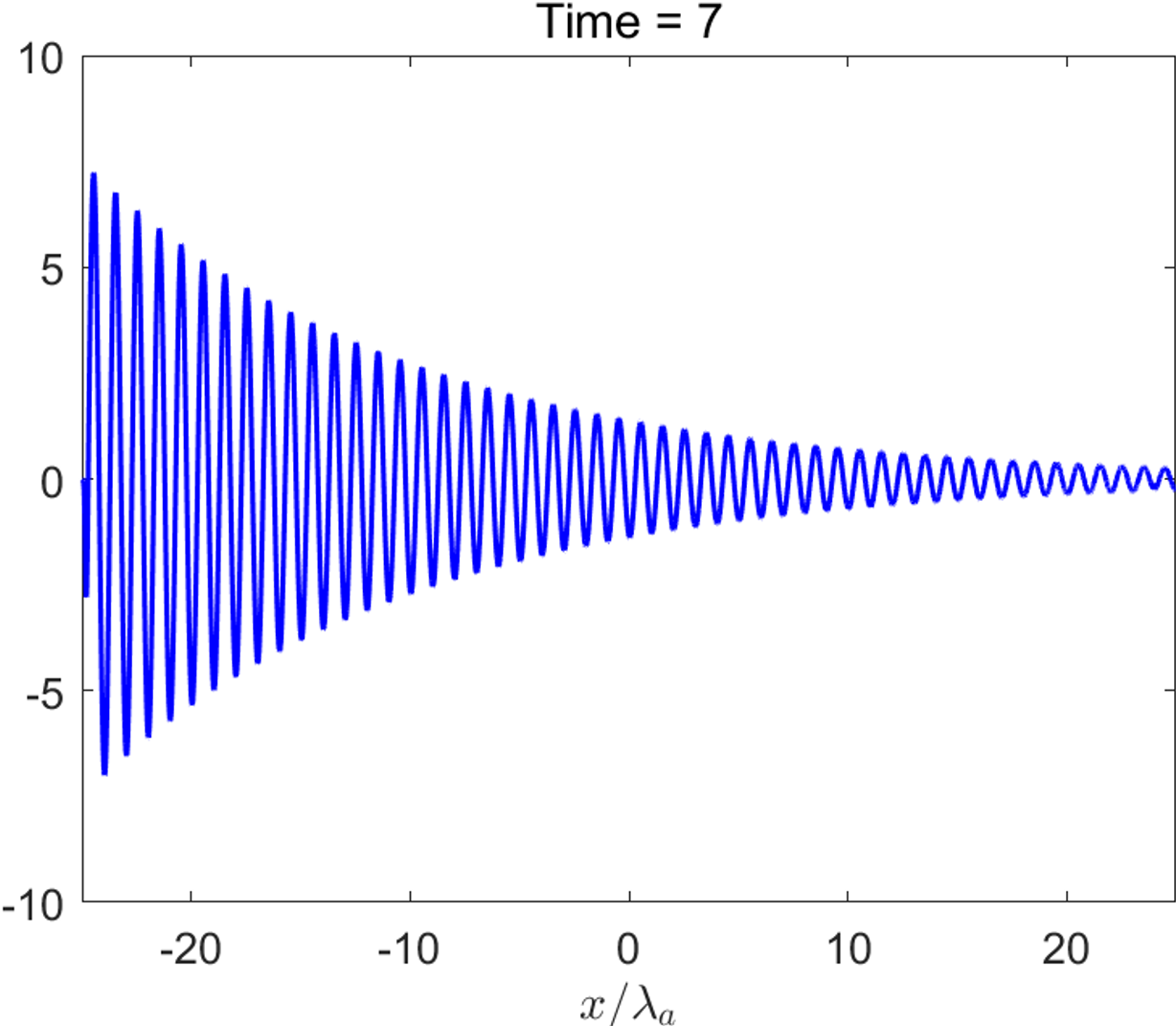}
        \caption{$(h,t) = (1.25\lambda_a,7)$}
    \end{subfigure}

    \caption{Field representation. The red and blue line correspond to BA field and FDTD-QE field, repectively.}
    \label{fig:field}
\end{figure}

In Fig. \ref{fig:field}, we visualize the BA- and FDTD-QE field at different times $(t\in\{1,3,5,7\})$ where $h= 1.25\lambda_a$ and $h= 5\lambda_a$. The main difference between FDTD-QE and our formalism is that, during propagation, our method preserves the information of the time dependent quantum state, whereas FDTD-QE propagates the electric field averaged with the ground state thereby discarding the quantum state information except initial state during time evolving. For this reason, the FDTD-QE method only tracks the dynamics of atomic population and averaged field, whereas our method can capture the full quantum statistical properties (e.g., photon correlations $\langle\hat{E}^{(-)}\hat{E}^{(+)}\rangle$). The fields obtained through two different methods show good agreement at all times. For specific, both methods accurately capture the scattered field and the open boundaries. When 
$h$ is an integer multiple of 0.5$\lambda_a$,the scattered field is trapped between the TLS and PEC; otherwise, it passes by the TLS without significant interaction. Compared to $5\lambda_a$ and $1.25\lambda_a$ cases, both method successfully capture this phenomenon.

\begin{figure}[t]
\centering
\includegraphics[width=3in]{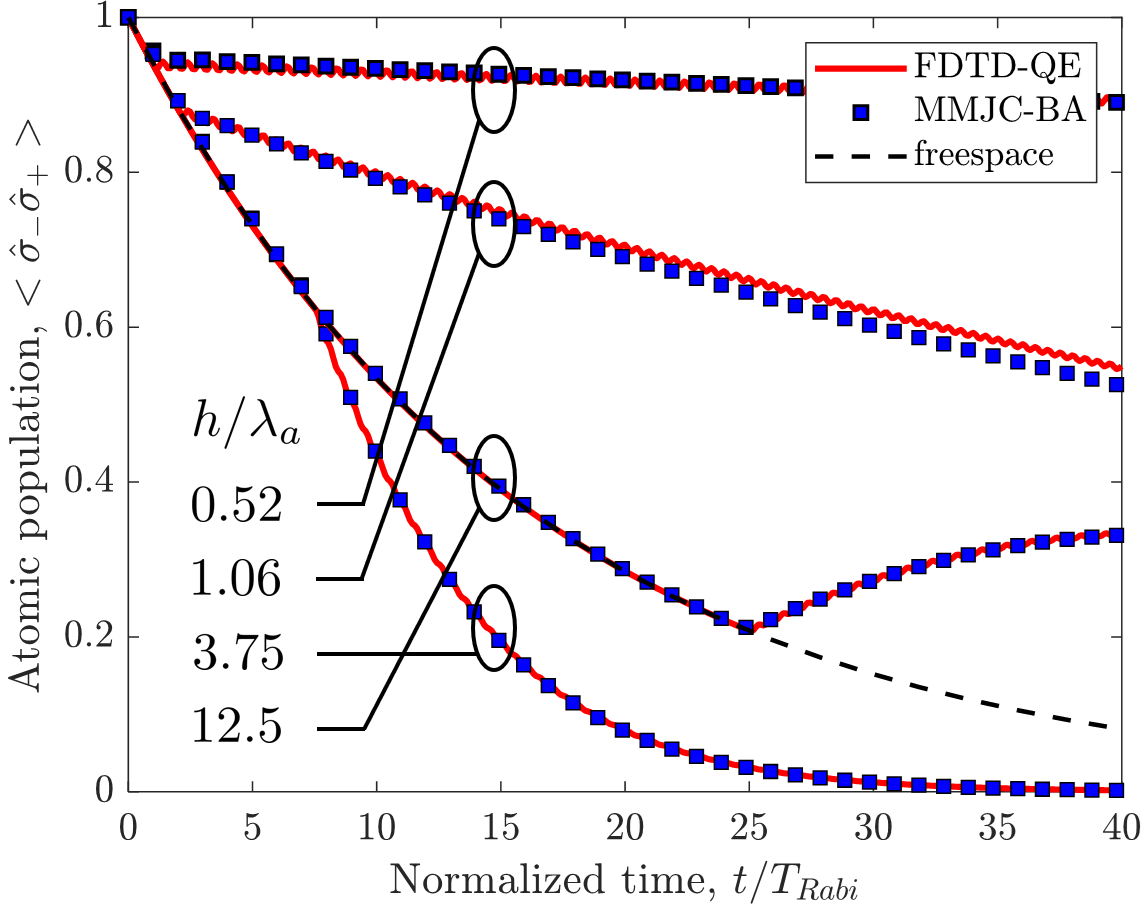}
\caption{Atomic population probability versus time normalized by Rabi period ($T_{Rabi}$) for different $h$ using different method, compared to those in the free space}
\label{fig:PEC_Atomic_Population}
\end{figure}

Fig. \ref{fig:PEC_Atomic_Population} illustrate the atomic population probability for different distance $h$ from PEC using FDTD-QE and our method.
The conventional master equation approach describe this dynamics with different decaying rate ($\Gamma$). This constant is usually determined by the local density of states (LDOS) of the environment, which is proportional to the imaginary part of the Green's function. In this case, the time evolution of the atomic population always follows an exponential decay with calculated $\Gamma$, which apparently violates relativistic causality. Namely, the atom instantaneously perceives changes regardless of how far away it is from a PEC mirror, leading to non-physical artifact of the Markovian approximation. In our approach, however, follow those of free space  until the reflected field from PEC reaches the atom as in the non-Markovian FDTD-QE algorithm, and therefore do not violate causality.

\begin{figure}[t]
\centering
\includegraphics[width=3in]{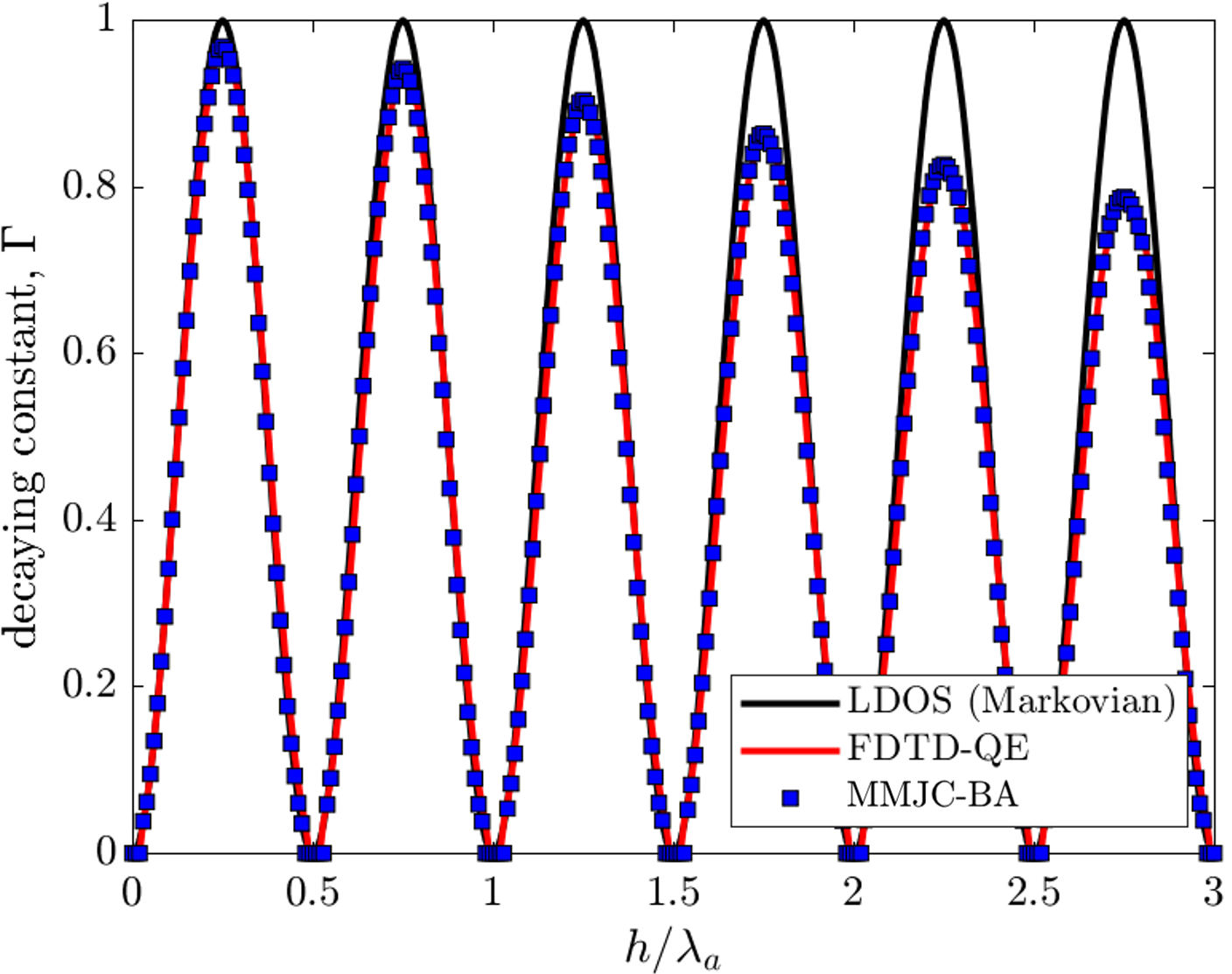}
\caption{Decaying constant versus normalized distance ($h/\lambda_a$) from PEC mirror using LDOS, FDTD-QE and L-MN formalism}
\label{fig:PEC_half_Purcell}
\end{figure}

This becomes more evident in Fig. \ref{fig:PEC_half_Purcell}, which illustrates the decaying constant depend on the normalized distance $h/\lambda_a$ from the PEC mirror. This is the typical illustration of the Purcell effect of atom in PEC mirror half space, which shows local minimum and maximum decaying constant with a period of $\lambda_a/2$ \cite{Drexhage1970Influence,Chance1978Molecular}. The decaying constant is calculated as  the reciprocal of the lifetime ($\tau$) which the atomic population probability decreases to $e^{-1}$ of the initial excited state. Due to computational limitations, cases where $\tau$ exceeds 30 are assumed to have $\Gamma=0$. In our simulations, the results clearly capture the typical $\lambda_a/2$ periodic phenomenon. Besides, unlike the decaying rate derived from LDOS, the two non-Markovian methods show a decreasing  decay constant at local maximum as more periods pass. This is correct because, in cases where $\Gamma$ is local maximum, the receiving field radiated from the TLS takes more time to return as $h/\lambda_a$ increases. During this time, the atom undergoes decay same to that in free space which decays more slowly. As a result, $\Gamma$ at local maximum points should decrease as more period pass.

\subsection{Dissipative cavity quantum electrodynamics}
In the second numerical example, we model one-dimensional dissipative cavity quantum electrodyanmic phenomena.
Let us consider a two-level system (TLS) placed at the center of a cavity made of electrically conducting walls in free space. If the electrically conducting walls have conductivity $\sigma_w$, then in the limit where $\sigma_w$ becomes infinitely large, the walls behave as perfect electric conductors (PEC), effectively isolating the cavity from the external space. In this case, the eigenmodes inside the cavity can be independently defined, and the system can be described by the conventional single-mode Jaynes–Cummings model. The atomic population then exhibits perfect Rabi oscillations over time.
On the other hand, when the conductivity of the conducting walls approaches zero, the cavity becomes completely open to the external space, and the TLS undergoes spontaneous emission in free space, with the atomic population dynamics reflecting this behavior.
In this numerical example, we demonstrate that by applying the MMJC model incorporating the BA–MA formulation, the system accurately converges to the above two limiting cases as the conductivity of the cavity walls is varied.
Consequently, when the wall conductivity becomes very high, the interaction between the interior and exterior of the cavity is significantly suppressed, and the influence of the BA field becomes negligible. However, as the conductivity decreases, the electromagnetic interaction between the inside and outside of the cavity increases, requiring both BA and MA fields to be considered.
We observe that if only the MA field is considered, as in the previous Green’s function approach, the atomic dynamics can be accurately captured when the conductivity is very high. However, as the conductivity decreases, this approach fails to correctly describe the dynamics of the TLS.

The simulation parameters are listed in the table \ref{tab:sim_params_nex2}.
\begin{table}[h!]
\centering
\caption{Problem geometry and simulation parameters}
\label{tab:sim_params_nex2}
\begin{tabular}{|l|c|c|}
\hline
\textbf{Parameter} & \textbf{Symbol} & \textbf{Value} \\
\hline
Transition frequency of TLS & $\omega_a$ & 50 [rad/s]\\
\hline
Wavelength corresponding to $\omega_a$ & $\lambda_a$ & 0.125663706143592 [m]\\
\hline
Dipole moment of TLS & $d_{eg}$ & 0.075 [C$\cdot$m]\\
\hline
Cavity size & $L_{\mathrm{cav}}$ & 0.062831853071796 [m] \\
\hline
Cavity wall thickness & $L_{\mathrm{wall}}$ & 12.56637061435917 [$\mu$m] \\
\hline
Total problem domain size & $L$ & 0.314159265358979 [m] \\
\hline
PML size & $L_{\mathrm{PML}}$ & 0.219911485751286 [m] (1.75 $\lambda_a$) \\
\hline
Spatial resolution near wall & $\Delta x_{\mathrm{wall}}$ & 63.148 [nm] \\
\hline
\end{tabular}
\end{table}

In order to accurately capture the atom–field interaction physics, it is essential to construct an appropriate set of BA–MA modes, sampled over frequency and their corresponding degeneracy space. When the cavity quality factor becomes very high, the full width at half maximum (FWHM) of the resonance frequency becomes extremely narrow. If one samples the BA–MA field modes uniformly over frequency in such a case, a very fine frequency resolution $\Delta \omega$ is required to properly capture the resonance frequency. This inevitably leads to oversampling of the BA–MA modes across a broad frequency range, significantly increasing the computational load.

To address this issue, an adaptive frequency refinement technique is employed. Specifically, a classical FEM approach is first used, in which a point current source is applied at the TLS position. The resulting field values inside the cavity are extracted at each frequency to compute the spectral amplitude. Based on the obtained spectral amplitude, the adaptive frequency refinement technique adjusts the frequency resolution: finer sampling is applied near the resonance frequency — where the spectral amplitude exhibits rapid variation with respect to frequency — while coarser sampling is used in frequency regions farther from the resonance, where the spectral amplitude varies more gradually.

When the atom–field coupling strength exceeds the cavity decay rate, (partial) Rabi oscillations can be observed. On the other hand, when the cavity decay rate dominates over the coupling strength, the system exhibits spontaneous emission behavior instead. 
Therefore, simulations are conducted for two sets of cavity wall conductivity values, denoted as Set 1 and Set 2. 
Set 1 consists of the following conductivity values:
\[
\sigma_w = 1.0 \times 10^{11},\ 1.29 \times 10^8,\ 5.07 \times 10^7,\ 2.52 \times 10^7,\ 1.255 \times 10^7,\ 4.943 \times 10^6,\ 1.19 \times 10^6~[\text{S/m}]
\]
Set 2 consists of the following conductivity values:
\[
\sigma_w = 4.864 \times 10^5,\ 2.0345 \times 10^5,\ 6.2 \times 10^4,\ 0~[\text{S/m}]
\]

Table~\ref{tab:decay_rate_set1_nex2} and Table~\ref{tab:decay_rate_set2_nex2} present the computed resonance frequency $\omega_c$ (i.e., the frequency at which the spectral amplitude is maximized) and the corresponding decay rate $\kappa = \omega_c / \text{FWHM}$, obtained by applying the classical FEM approach for different cavity wall conductivities $\sigma_w$ for the set 1 and the set 2, respectively. A fixed non-uniform mesh is used for all simulations. For each value of $\sigma_w$, the ratio of the skin depth to the cavity discretization size ($\text{skin depth} / \Delta x_{\mathrm{cav}}$) is also listed. To ensure simulation accuracy, the mesh is designed such that $\text{skin depth} / \Delta x_{\mathrm{cav}}$ is greater than 10 when $\sigma_w$ is largest.
Note that the spontaneous emission rate $\Gamma$ in the presence of the conducting walls was evaluated by using the FEM simulations.
\begin{table}[h]
\centering
\caption{Conductance, cavity frequency $\omega_c$, decay rate, and $\delta_{skin}$ values for the set 1 (atom-field coupling strength $>$ decay rate)}
\label{tab:decay_rate_set1_nex2}
\begin{tabular}{>{\centering\arraybackslash}m{3.5cm} 
                >{\centering\arraybackslash}m{4cm} 
                >{\centering\arraybackslash}m{3.5cm} 
                >{\centering\arraybackslash}m{4cm}}
\toprule
$\sigma_{w}$ & $\omega_c$ & $\kappa$ & $\delta_{skin}/{\Delta x_{wall}}$ \\
\midrule
$1.000 \times 10^{11}$     & 49.9995          & $0.0005$               & $1.00 \times 10^{1}$ \\
$1.290 \times 10^{8}$      & 49.9934          & 0.0201                 & $2.79 \times 10^{2}$ \\
$5.070 \times 10^{7}$      & 49.9935          & 0.0500                  & $4.45 \times 10^{2}$ \\
$2.520 \times 10^{7}$      & 49.9930          & 0.1000                  & $6.31 \times 10^{2}$ \\
$1.255 \times 10^{7}$     & 49.9930          & 0.2000                  & $8.94 \times 10^{2}$ \\
$4.943 \times 10^{6}$     & 49.9925          & 0.4995                  & $1.42 \times 10^{3}$ \\
$1.190 \times 10^{6}$      & 49.9765          & 2.0003                  & $2.90 \times 10^{3}$ \\
\bottomrule
\end{tabular}
\end{table}

\begin{table}[h]
\centering
\caption{Conductance, cavity frequency $\omega_c$, decay rate, and $\delta_{skin}$ values for the set 1 (atom-field coupling strength $<$ decay rate)}
\label{tab:decay_rate_set2_nex2}
\begin{tabular}{>{\centering\arraybackslash}m{3.5cm} 
                >{\centering\arraybackslash}m{4cm} 
                >{\centering\arraybackslash}m{3.5cm} 
                >{\centering\arraybackslash}m{4cm}}
\toprule
$\sigma_{w}$ & $\omega_c$ & $\kappa$ & $\Gamma$ \\
\midrule
$4.8640 \times 10^{5}$     & 49.9000          & 4.5375               & $2.000 \times 10^{0}$ \\
$2.0345 \times 10^{5}$      & 49.4500          & 9.6750                 & $1.000 \times 10^{-1}$ \\
$6.2000 \times 10^{4}$      & n/a          & n/a                  & $5.000 \times 10^{-1}$ \\
0       & n/a          & n/a                  & $2.809 \times 10^{-1}$ \\
\bottomrule
\end{tabular}
\end{table}

For Set 1, the BA–MA field modes were sampled within a frequency band centered at $\omega_a$, using a fractional bandwidth of 20\%, as described previously. The adaptive sampling technique resulted in a total of $198$ frequency sampling points.
For the BA field modes, since this is a one-dimensional case, there are two degenerate BA modes corresponding to the incident plane waves propagating from the left and right, respectively.
For the MA field modes, due to the one-dimensional nature of the system, the three-dimensional orientation of the point noise current source is neglected. As a result, there are $N_{x_{\mathrm{cav}}}$ degenerate MA field modes, where $N_{x_{\mathrm{cav}}}$ denotes the number of grid points within the cavity wall region.

It is assumed that the two-level system (TLS) is initially excited, while all BA-MA field modes are in their ground states.  
Since the Jaynes–Cummings model conserves the total number of quanta, the total excitation remains constant over time.  
Given a single excitation initially, the quantum state of the system at any time $t$ will also contain a single excitation, which may transfer from the atom to the field modes.

Thus, the general form of the single-excitation quantum state can be written as
\begin{flalign}
\ket{\psi(t)} = C_{e,0}(t)\ket{e,\left\{0\right\}} + \sum_{l=1}^{N_f}C_{g,1_{l}}(t)\ket{g,1_l},
\label{eqn:single-quanta_qs}
\end{flalign}
where $l$ denotes the index of the discrete BA-MA field modes.  
Substituting Eq.~\eqref{eqn:single-quanta_qs} into the Schrödinger equation yields a set of $1 + N_f$ coupled differential equations for the probability amplitudes $C_{e,0}(t)$ and $C_{g,1_l}(t)$.  
These equations are numerically solved using the fourth-order Runge–Kutta method.

Once the probability amplitudes are obtained, we can compute observables such as the atomic population and the one-photon electric field amplitude.  
The atomic population is given by
\begin{flalign}
\mel{\psi(t)}{\hat{\sigma}_{-}\hat{\sigma}_{+}}{\psi(t)} = \left|C_{e,0}(t)\right|^2,
\end{flalign}
while the one-photon electric field amplitude is defined as
\begin{flalign}
\expval{\hat{E}_z(x,t)}_{sq}
\triangleq
\mel{g,\left\{0\right\}}{\hat{E}_z(x)}{\psi(t)}
+
\mel{\psi(t)}{\hat{E}_z(x)}{g,\left\{0\right\}}.
\end{flalign}
By expanding the expectation value, we obtain
\begin{flalign}
\expval{\hat{E}_z(x,t)}_{sq}
=
\sum_{l=1}^{N_f}
E_{l}(x) C_{g,1_l}(t)
+
\left[E_{l}(x) C_{g,1_l}(t)\right]^*.
\end{flalign}

Fig.~\ref{fig:ap_set_1} and Fig. \ref{fig:ap_set_2} show the atomic population dynamics for parameter sets 1 and 2, respectively.  
Each figure compares the results from the full BA-MA mode consideration (MMJC-BAMA) and the reduced model considering MA modes only (MMJC-MA), shown on the left and right panels.

In Fig.~\ref{fig:ap_set_1}, which corresponds to high-conductivity cavity walls, the atomic dynamics closely approach ideal Rabi oscillations as conductivity increases.  
This behavior is expected since highly conductive walls effectively isolate the interior of the cavity, leading to dominant interaction between the TLS and a high-$Q$ single cavity mode.  
As the conductivity decreases, the quality factor drops, and the Rabi oscillation becomes damped due to energy leakage.  
The envelope of this decay matches the expected cavity loss rate.  
A redshift in the oscillation frequency is also observed, stemming from the reduced effective coupling strength due to energy dissipation, which decreases the overall Rabi frequency.

When comparing MMJC-BAMA with MMJC-MA in set 1, we find negligible differences.  
This is because the high wall conductivity still suppresses BA field mode contributions, and the TLS dynamics are governed mainly by the MA field modes.

In contrast, Fig.~\ref{fig:ap_set_2} illustrates the results for low-conductivity walls.  
In this regime, the coupling strength is reduced and the spontaneous emission rate dominates over the coherent interaction.  
The lossy cavity modifies the local density of states, affecting the spontaneous emission rate.  
As conductivity decreases toward zero, the atomic population converges to that of free space.  
In this case, a clear discrepancy arises between MMJC-BAMA and MMJC-MA, since the BA field modes can now significantly interact with the TLS due to poor isolation by the cavity walls.  
Therefore, the reduced MMJC-MA model cannot fully capture the spontaneous emission dynamics in this regime.

Fig.~\ref{fig:MMJC_BAMA_MA_field_plot} displays the one-photon field amplitude $\expval{\hat{E}_z(x,t)}_{sq}$.  
For highly conductive walls, the field is well confined within the cavity and exhibits single-mode oscillations characteristic of ideal Rabi dynamics.  
As conductivity decreases, the field leaks outside the cavity, transitioning toward free-space spontaneous emission.  
In this regime, the MMJC-MA model fails to reproduce the correct field distribution, as it neglects the contribution of BA field modes which become increasingly important.

These numerical results demonstrate that the MMJC-BAMA formulation can capture the full non-Markovian dynamics of dissipative cavity quantum electrodynamics from first principles.  
The model successfully retrieves both the perfect Rabi oscillation and the free-space spontaneous emission behavior in their respective limits of high and low cavity wall conductivity.  
In particular, the inclusion of BA field modes becomes essential when the cavity walls are weakly conducting.

\begin{figure}[t]
\centering
\includegraphics[width=4in]{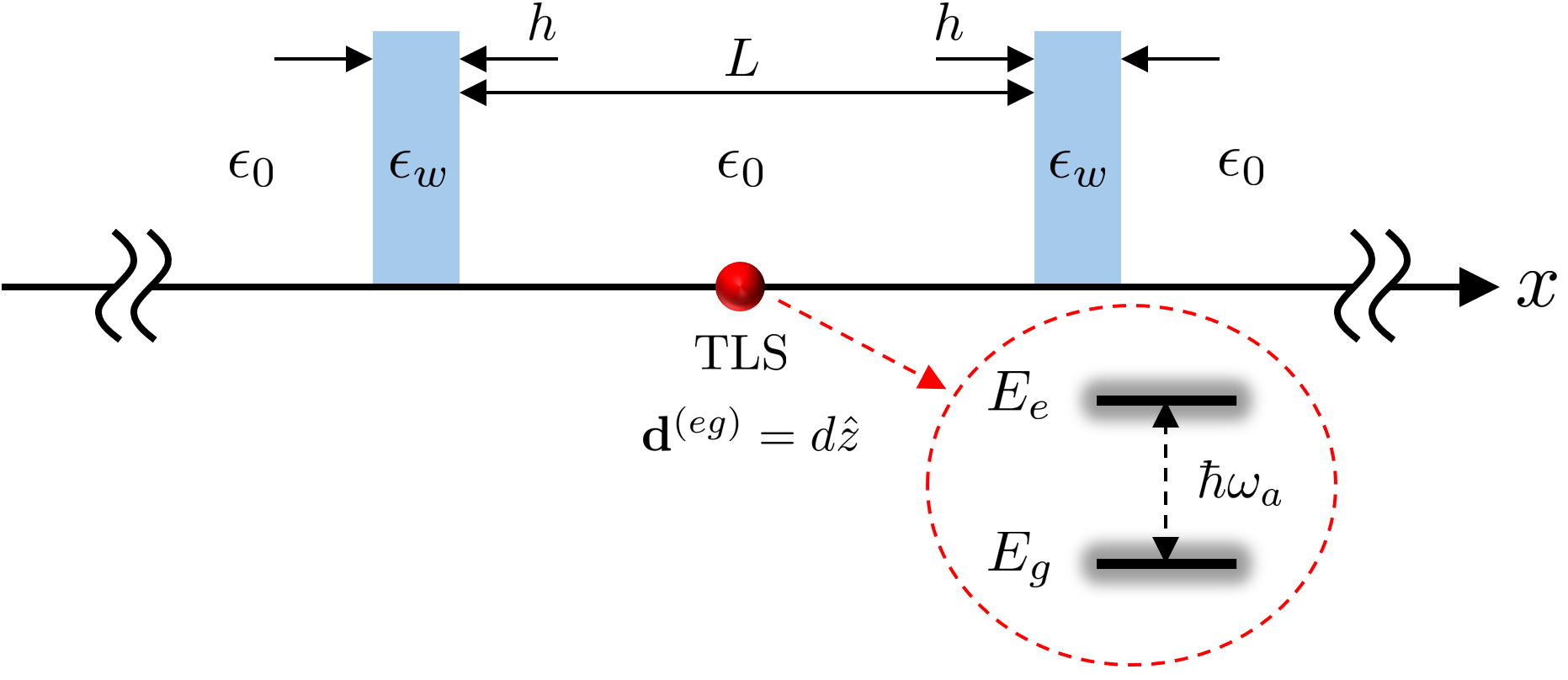}
\caption{Schematic of a TLS inside a one-dimensional cavity made of conducting walls with the Ohmic loss $\sigma_w$.}
\label{fig:lossy_cavity_schematic}
\end{figure}

\begin{figure}[t]
  \centering
  \begin{subcaptionbox}{MMJC-BAMA\label{fig_LC_AP_MMJC_BAMA_HL}}
    {\includegraphics[width=.465\textwidth]{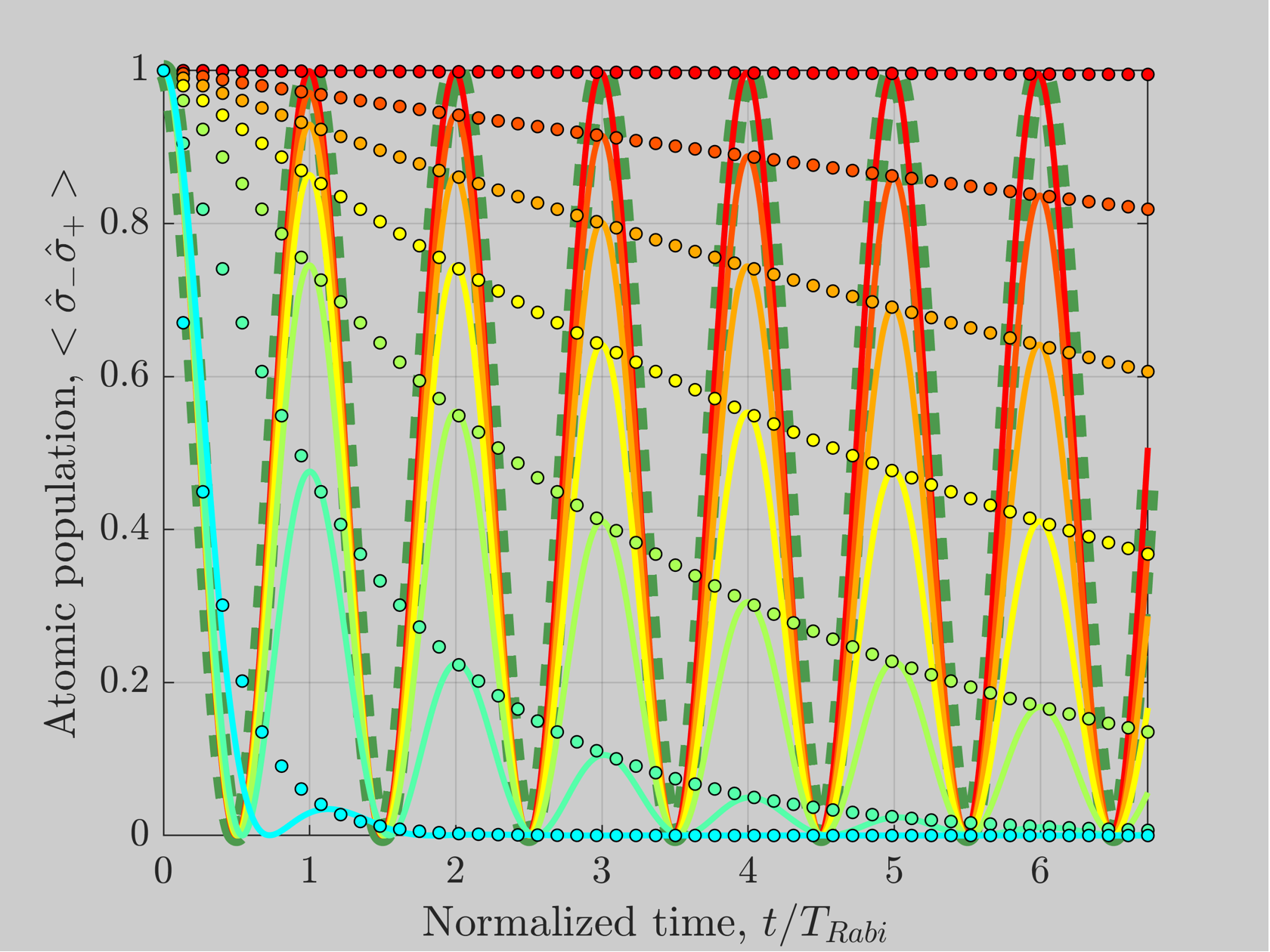}}
  \end{subcaptionbox}
  \begin{subcaptionbox}{MMJC-MA\label{fig_LC_AP_MMJC_MA_HL}}
    {\includegraphics[width=.465\textwidth]{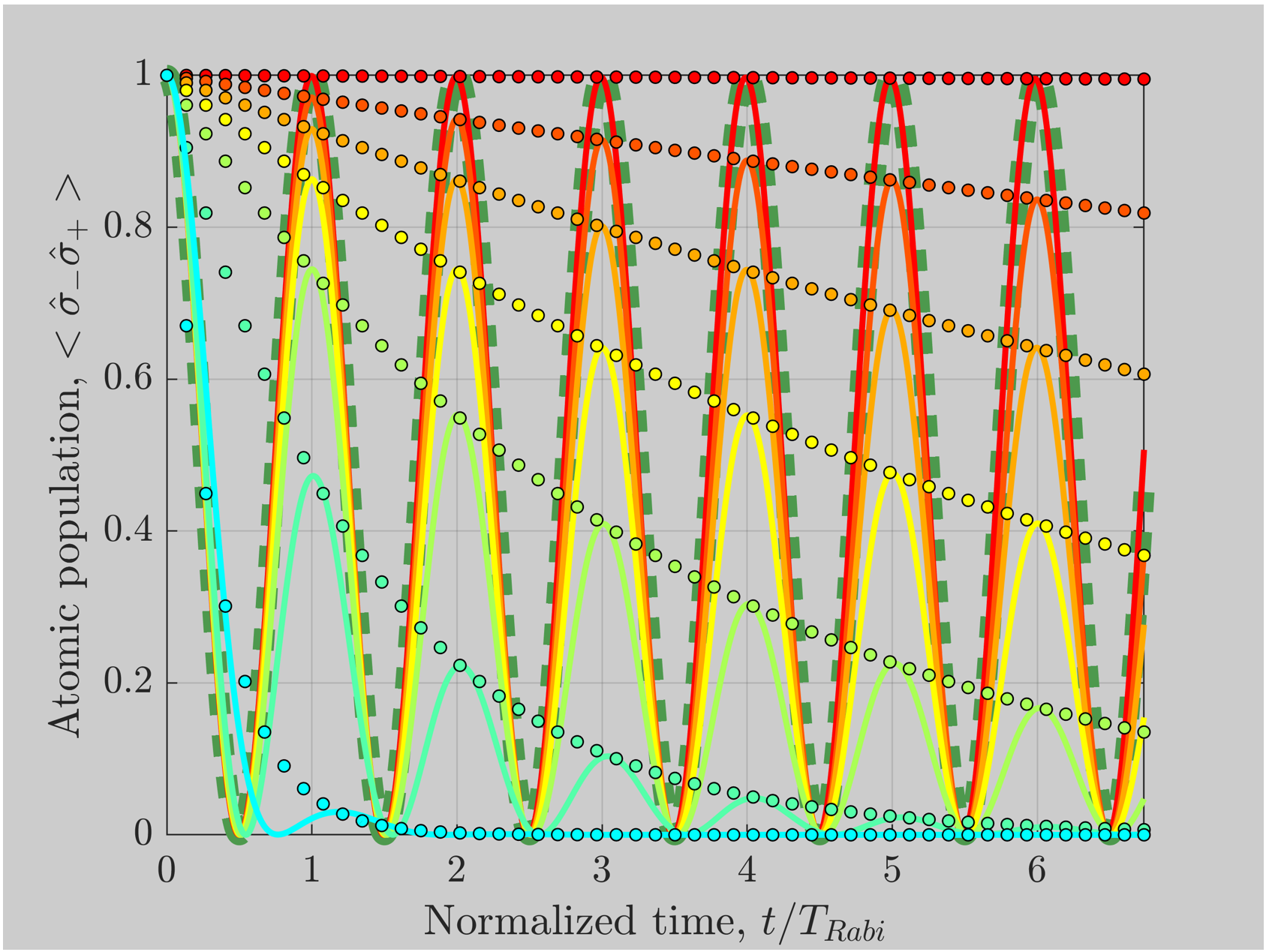}}
  \end{subcaptionbox}
  \caption{Atomic population $\expval{\hat{\sigma}_{-}\hat{\sigma}_{+}}$ for a TLS inside a one-dimensional lossy cavity with high Ohmic loss. Simulation results are obtained using the MMJC model combined with (a) both BA and MA fields, and (b) MA fields only. The curves are colored from red to orange, yellow, green, and cyan, corresponding to decreasing conductivity, corresponding to the cases with conductivities $\sigma_w = 10^{11}$, $1.28 \times 10^{8}$, $5.07 \times 10^{7}$, $2.52 \times 10^{7}$, $1.255 \times 10^{7}$, $4.943 \times 10^{6}$, and $1.19 \times 10^{6}$, respectively.  Solid lines indicate results from the MMJC-BAMA or MMJC-MA simulations, while the circle markers represent exponential decay functions with decay rates obtained separately from FEM simulations for each cavity configuration.}
  \label{fig:ap_set_1}
\end{figure}

\begin{figure}[t]
  \centering
  \begin{subcaptionbox}{MMJC-BAMA\label{fig_LC_AP_MMJC_BAMA_LL}}[0.45\textwidth]
    {\includegraphics[width=.465\textwidth]{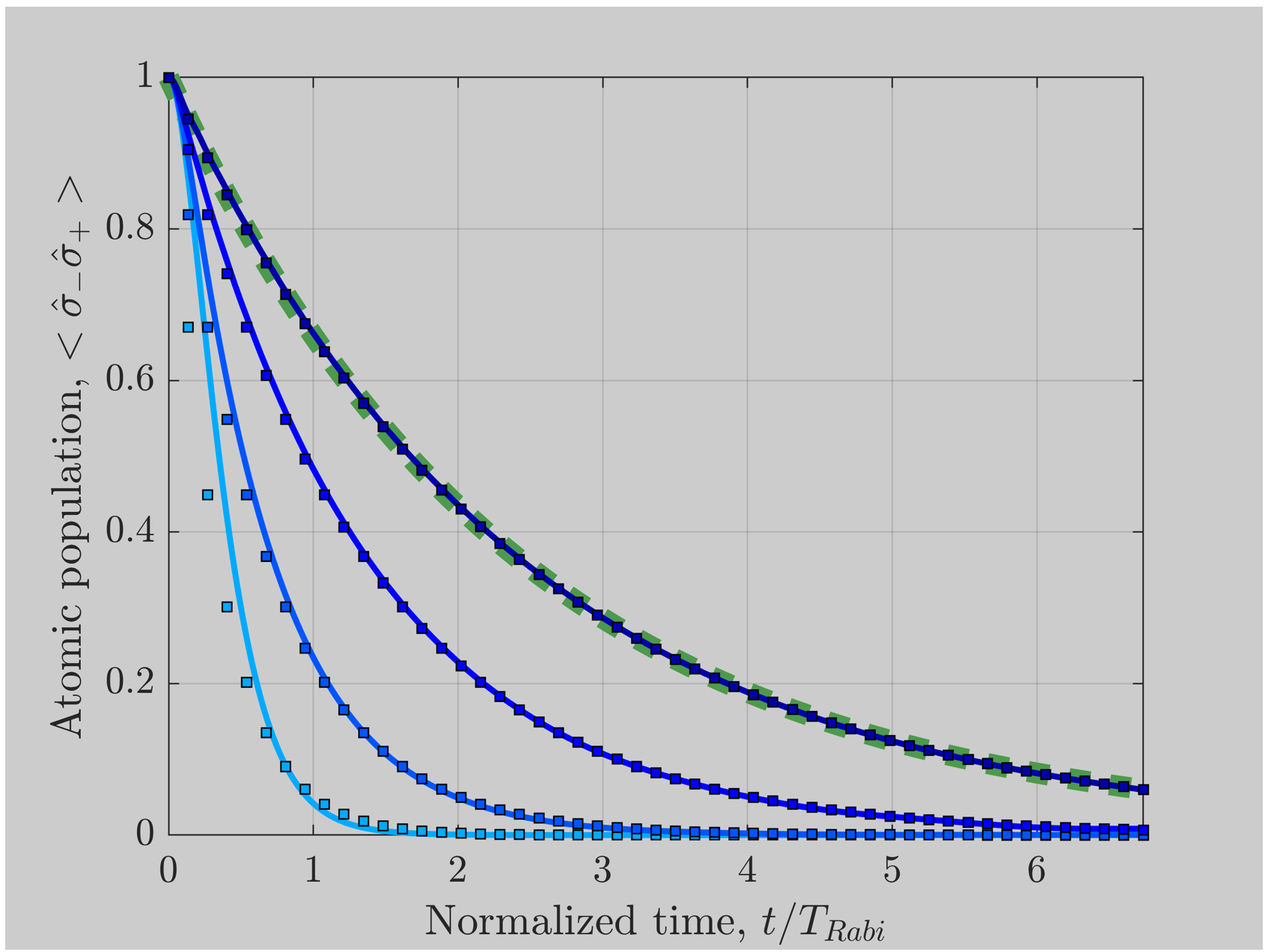}}
  \end{subcaptionbox}
  \begin{subcaptionbox}{MMJC-MA\label{fig_LC_AP_MMJC_MA_LL}}[0.45\textwidth]
    {\includegraphics[width=.465\textwidth]{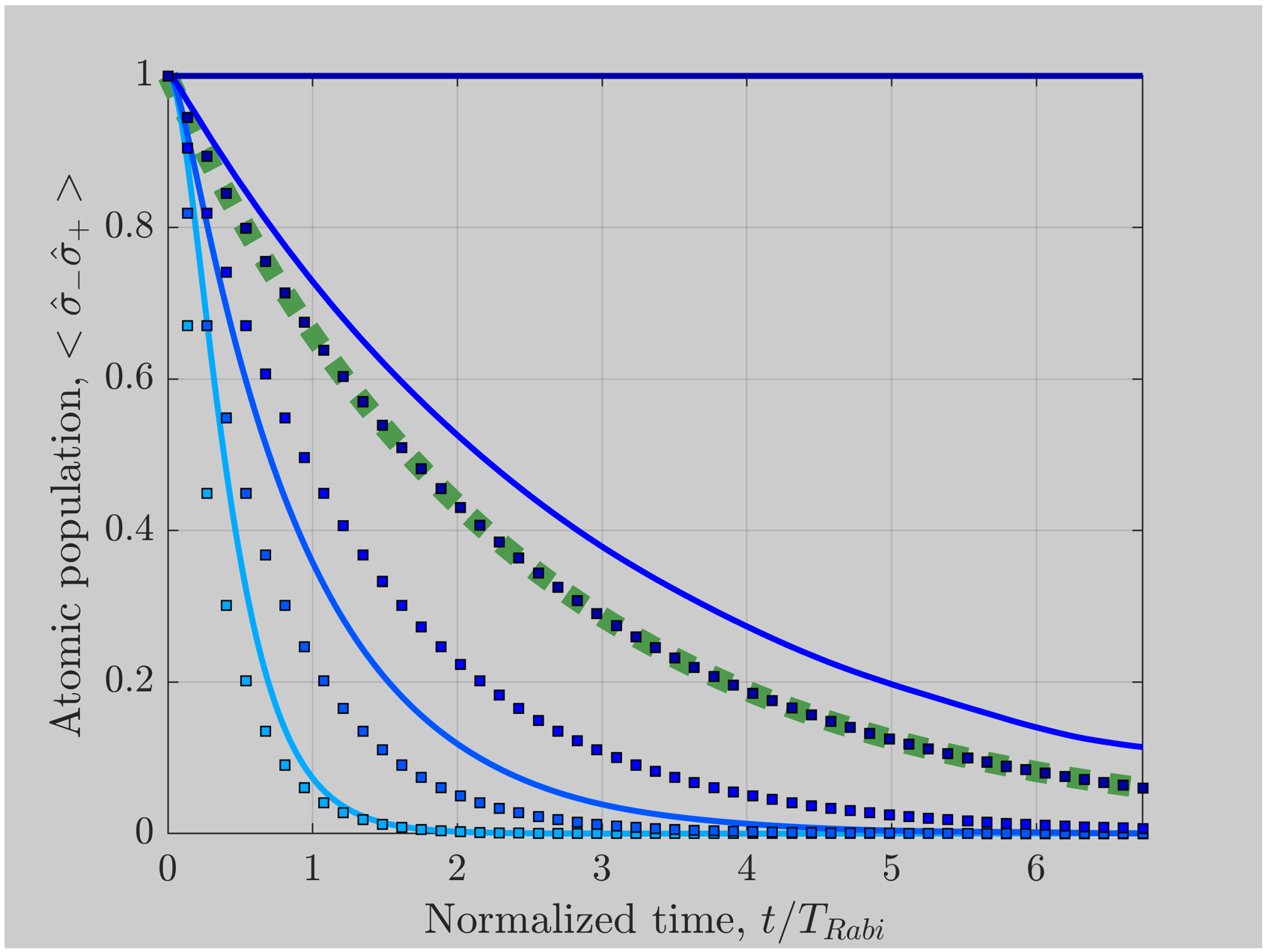}}
  \end{subcaptionbox}
  \caption{Atomic population $\expval{\hat{\sigma}_{-}\hat{\sigma}_{+}}$ for a TLS inside a one-dimensional lossy cavity with low Ohmic loss. Simulation results are obtained using the MMJC model combined with (a) both BA and MA fields, and (b) MA fields only. The curves colored from light sky blue to deep blue correspond to the cases with conductivities $\sigma_w = 4.864\times 10^{5}$, $2.0345 \times 10^{5}$, $6.2 \times 10^{4}$, and $0$, respectively.  Solid lines indicate results from the MMJC-BAMA or MMJC-MA simulations, while the circle markers represent exponential decay functions with the spontaneous emission rate obtained separately from FEM simulations for each cavity configuration.}
  \label{fig:ap_set_2}
\end{figure}

\begin{figure}[h]
  \centering
  \begin{subcaptionbox}{MMJC-BAMA ($\sigma_w = 10^{11}$)\label{fig:subfigA}}[0.45\textwidth]
    {\includegraphics[width=.3\textwidth]{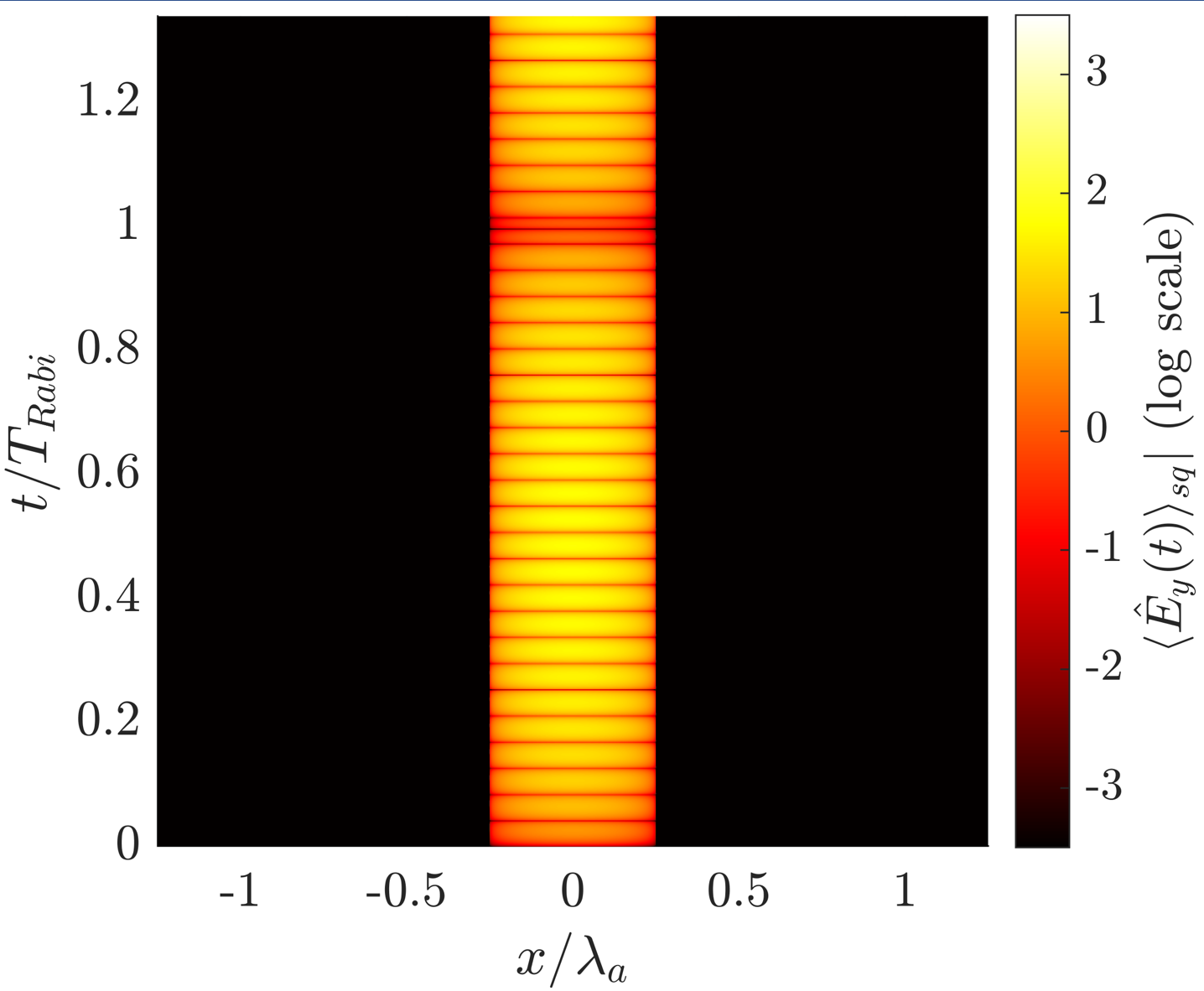}}
  \end{subcaptionbox}
  \begin{subcaptionbox}{MMJC-MA ($\sigma_w = 10^{11}$)\label{fig:subfigB}}[0.45\textwidth]
    {\includegraphics[width=.3\textwidth]{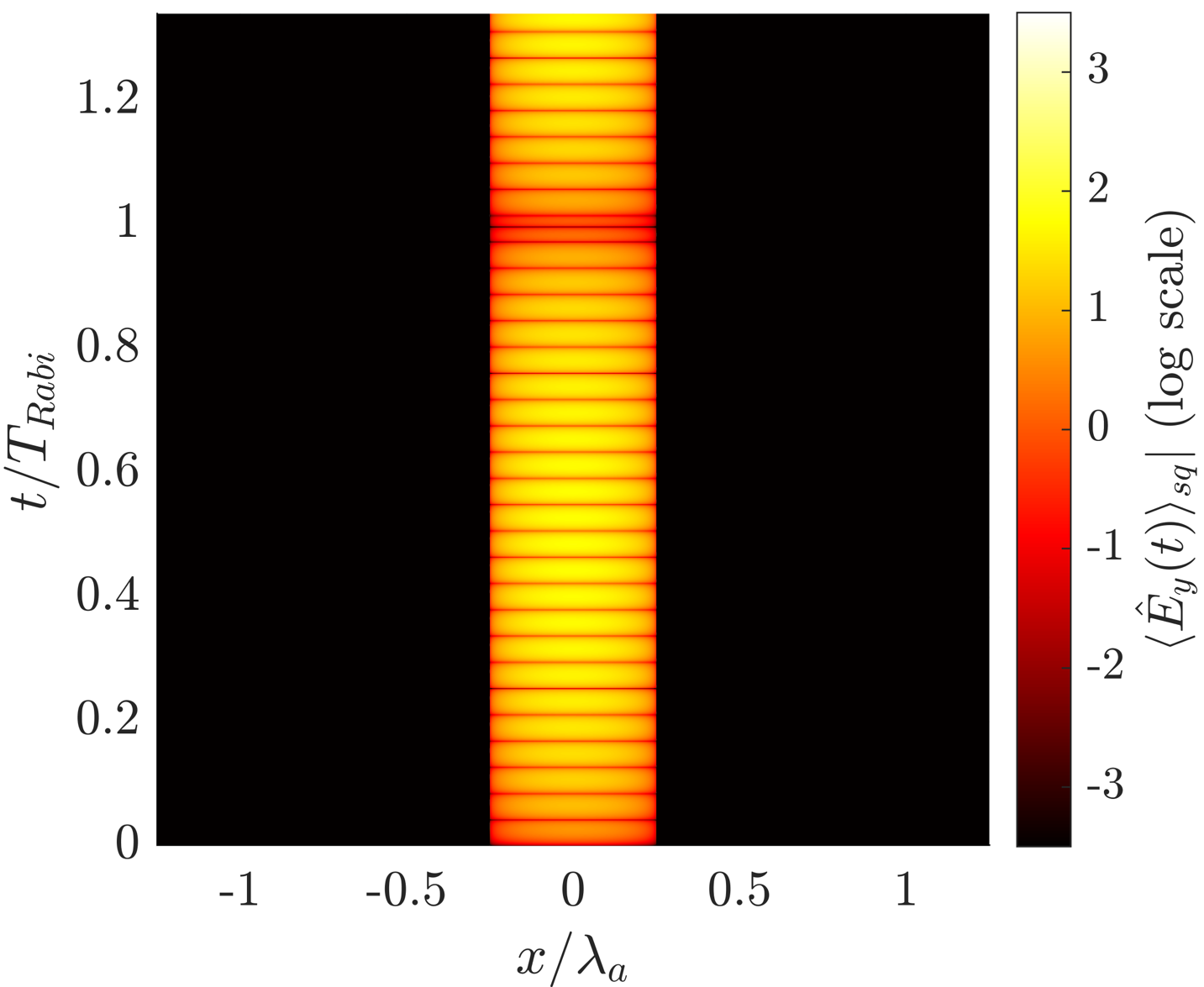}}
  \end{subcaptionbox}
  \\
  \begin{subcaptionbox}{MMJC-BAMA ($\sigma_w = 1.255\times 10^{7}$)\label{fig:subfigA}}[0.45\textwidth]
    {\includegraphics[width=.3\textwidth]{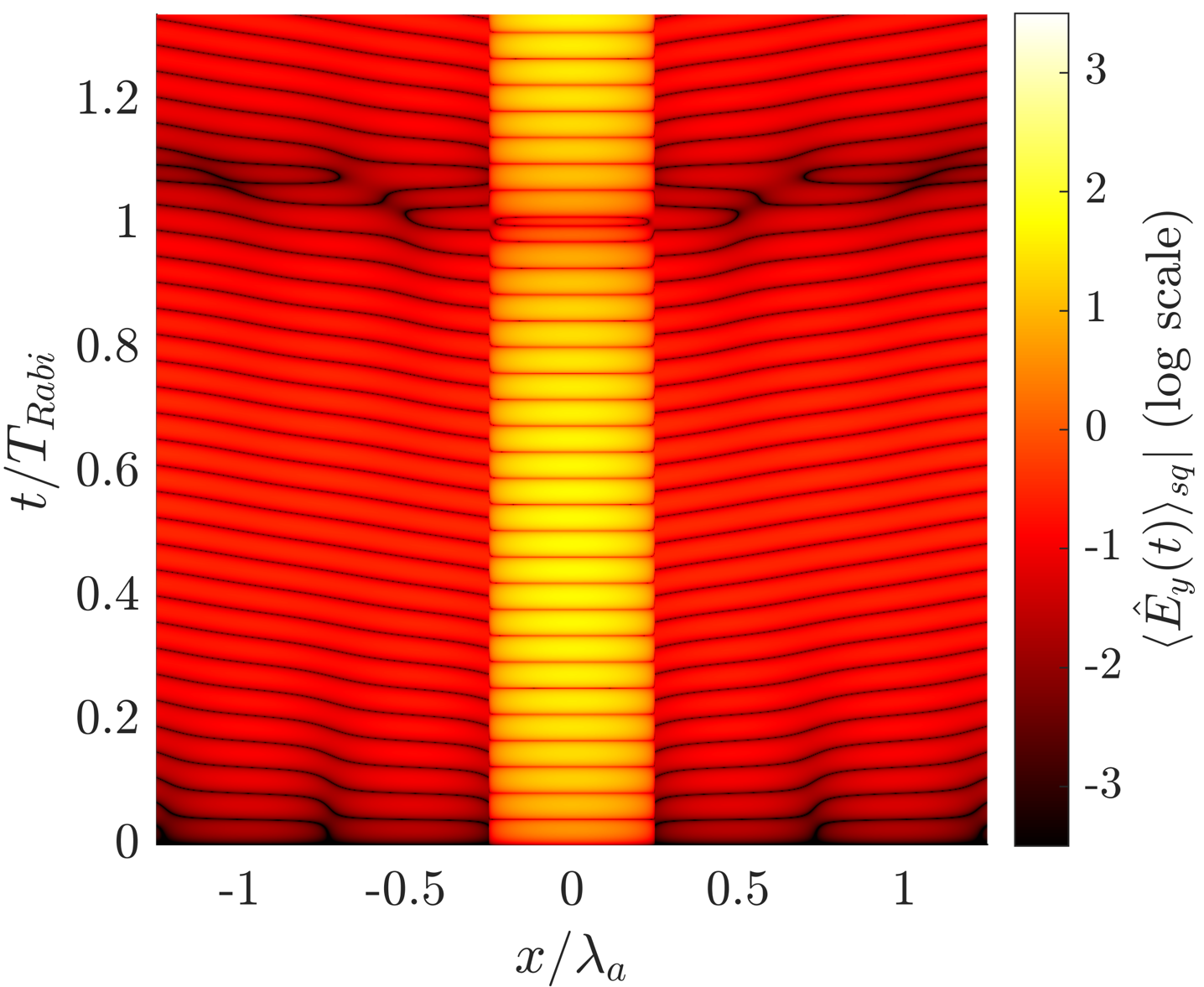}}
  \end{subcaptionbox}
  \begin{subcaptionbox}{MMJC-MA ($\sigma_w = 1.255\times 10^{7}$)\label{fig:subfigB}}[0.45\textwidth]
    {\includegraphics[width=.3\textwidth]{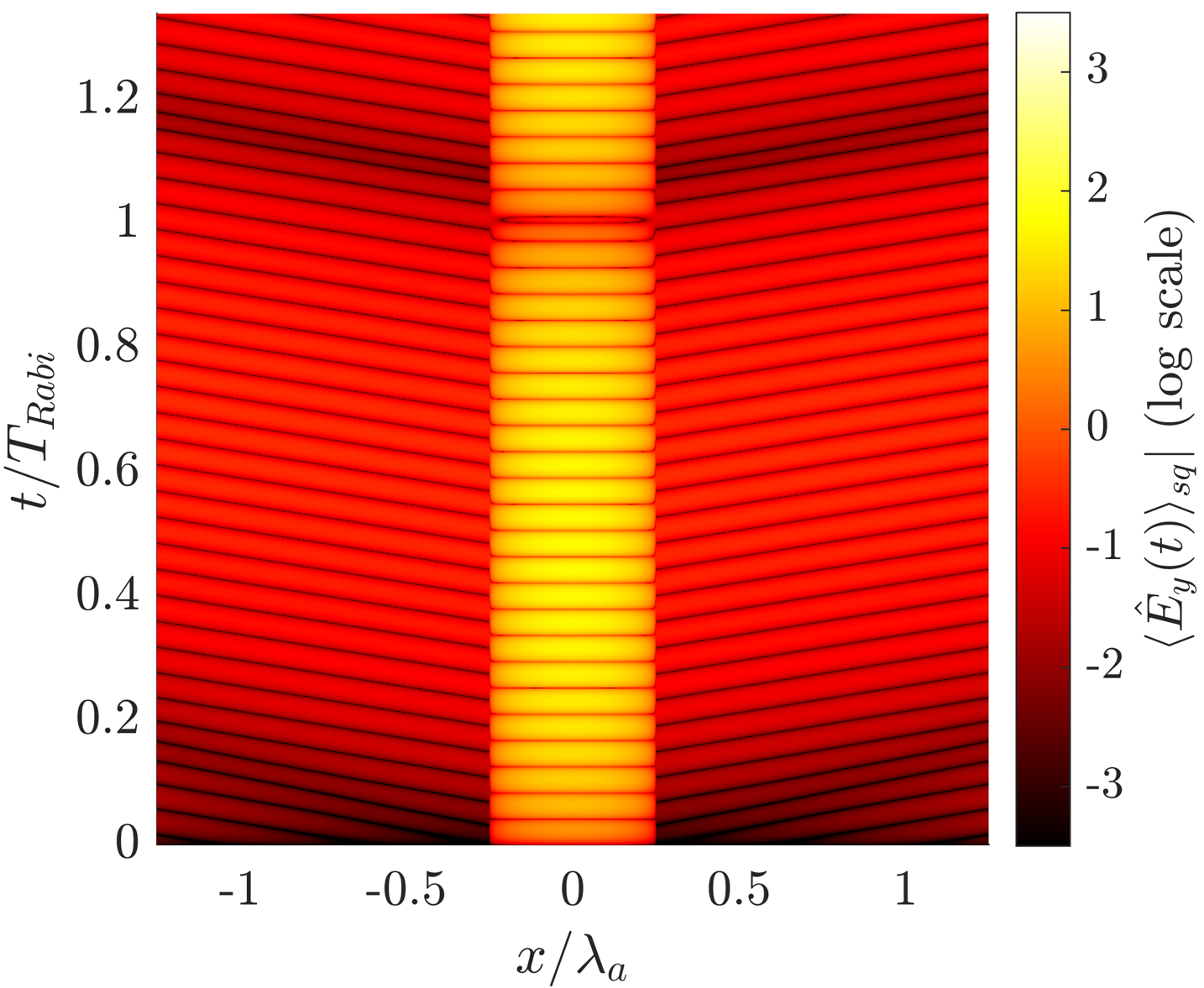}}
  \end{subcaptionbox}
  \\
  \begin{subcaptionbox}{MMJC-BAMA ($\sigma_w = 6.2\times 10^{4}$)\label{fig:subfigA}}[0.45\textwidth]
    {\includegraphics[width=.3\textwidth]{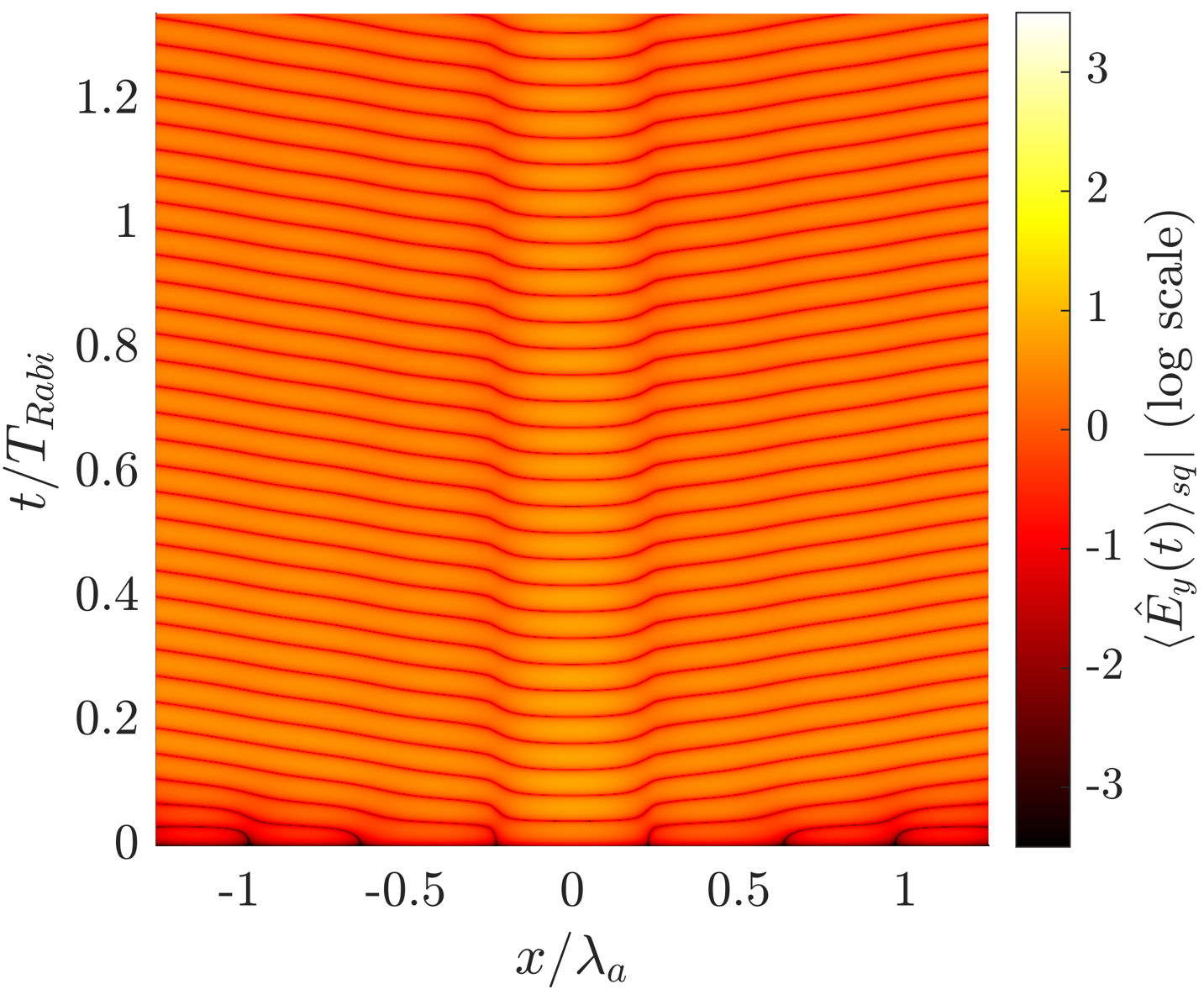}}
  \end{subcaptionbox}
  \begin{subcaptionbox}{MMJC-MA ($\sigma_w = 6.2\times 10^{4}$)\label{fig:subfigB}}[0.45\textwidth]
    {\includegraphics[width=.3\textwidth]{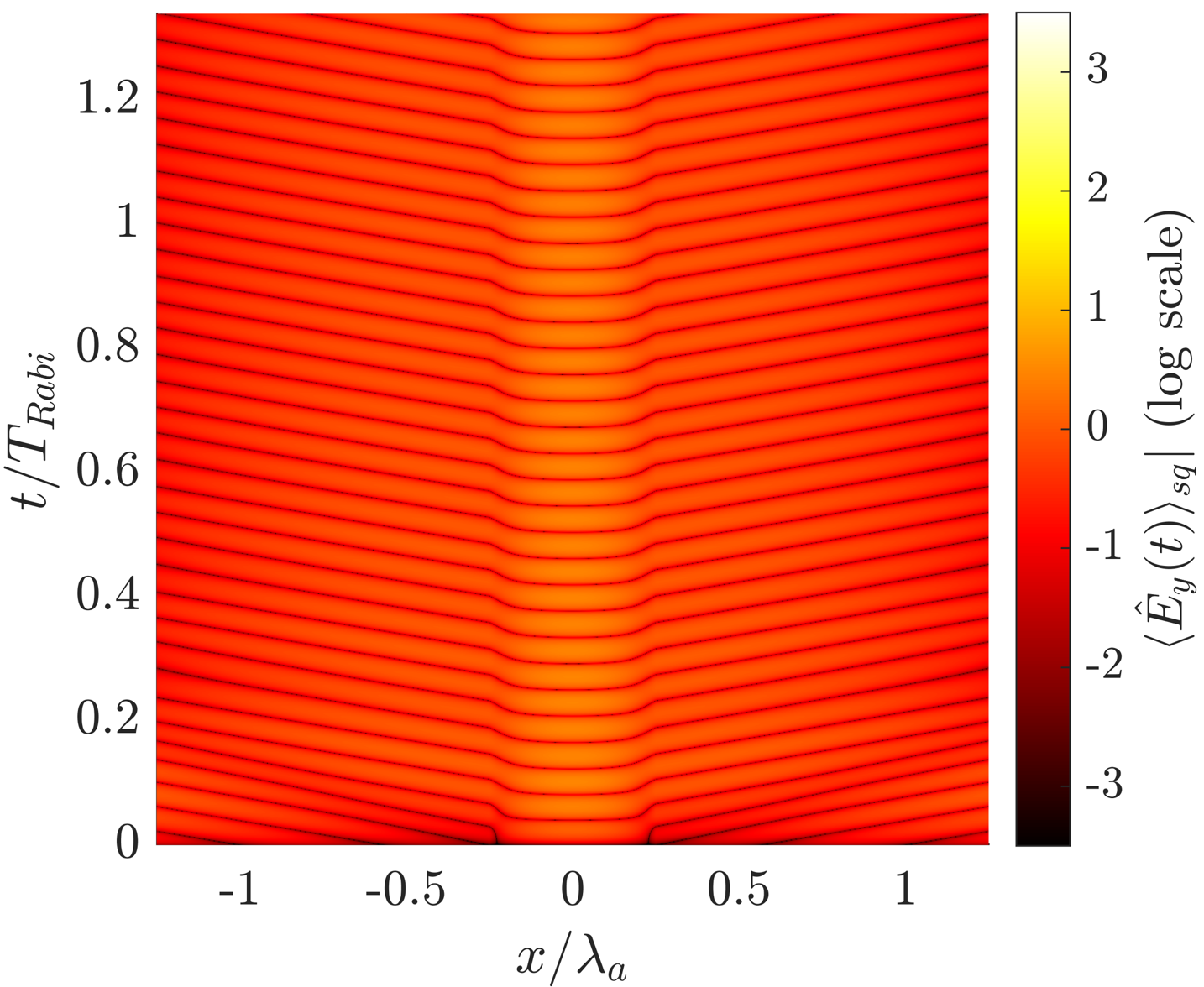}}
  \end{subcaptionbox}
  \caption{Field plots}
  \label{fig:MMJC_BAMA_MA_field_plot}
\end{figure}

\subsection{Super-radiance of multiple TLSs}
In this example, we extend numerical simulation to the multi mode Tavis-Cummings (MMTC) model. For specific, we validate whether our formalism can captures the collective atomic behavior, especially the super-radiance of multiple TLSs.  
The super-radiance is a collective atomic phenomenon in which $N$ ordered atoms interact with light collectively and coherently. This phenomenon occurs when the distance between atoms is much smaller than the atomic wavelength ($\lambda_a$). As the atoms get closer, the radiated fields become in-phase and interfere constructively that is clearly different from that of independent atom.

\begin{figure}[t]
\centering
\includegraphics[width=4in]{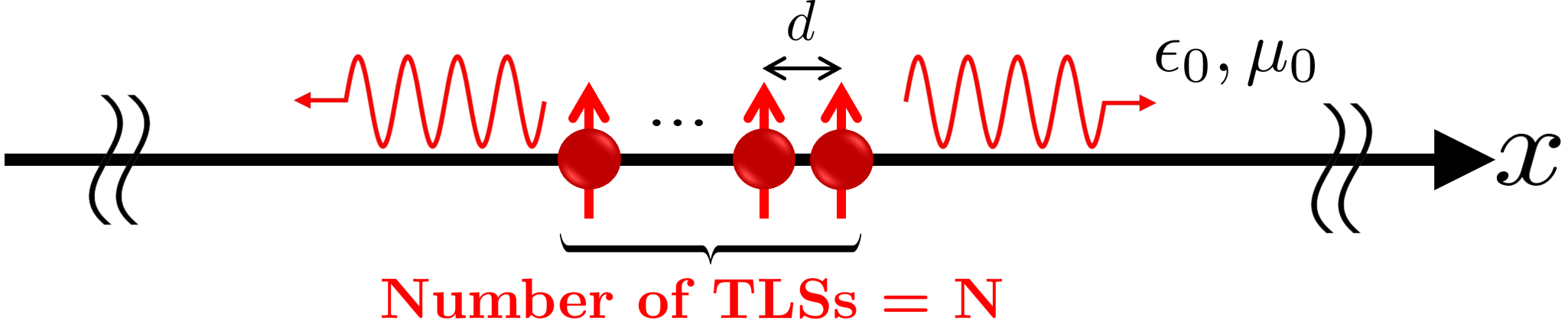}
\caption{Schematic of super-radiance.}
\label{fig:superradiance_sch}
\end{figure}

Let us consider the N identical TLSs with atomic frequency $\omega_a$. As illustrated in Fig. \ref{fig:superradiance_sch}, the TLSs are linearly arranged in free space with uniform atomic spacing ($d$). We consider one-dimensional simulation setup with PMLs implemented at both boundaries to model open environment. The simulation parameters are similar to those in Example 1, where the simulation domain without PML is 50 $\lambda_a$, $d_{eg}$ is 0.1C$\cdot$m. The atomic array is located at the center of the domain.

Unlike Dicke model for super-radiance, the MMTC employs rotating wave approximation, which ensures conservation of the total number of quanta.

\begin{figure}[t]
  \centering
  \begin{subcaptionbox}{  $d/\lambda_a\in \{0.01,0.04,0.06\}$ \label{fig_superradiance_fig1}}[0.45\textwidth]
    {\includegraphics[width=.465\textwidth]{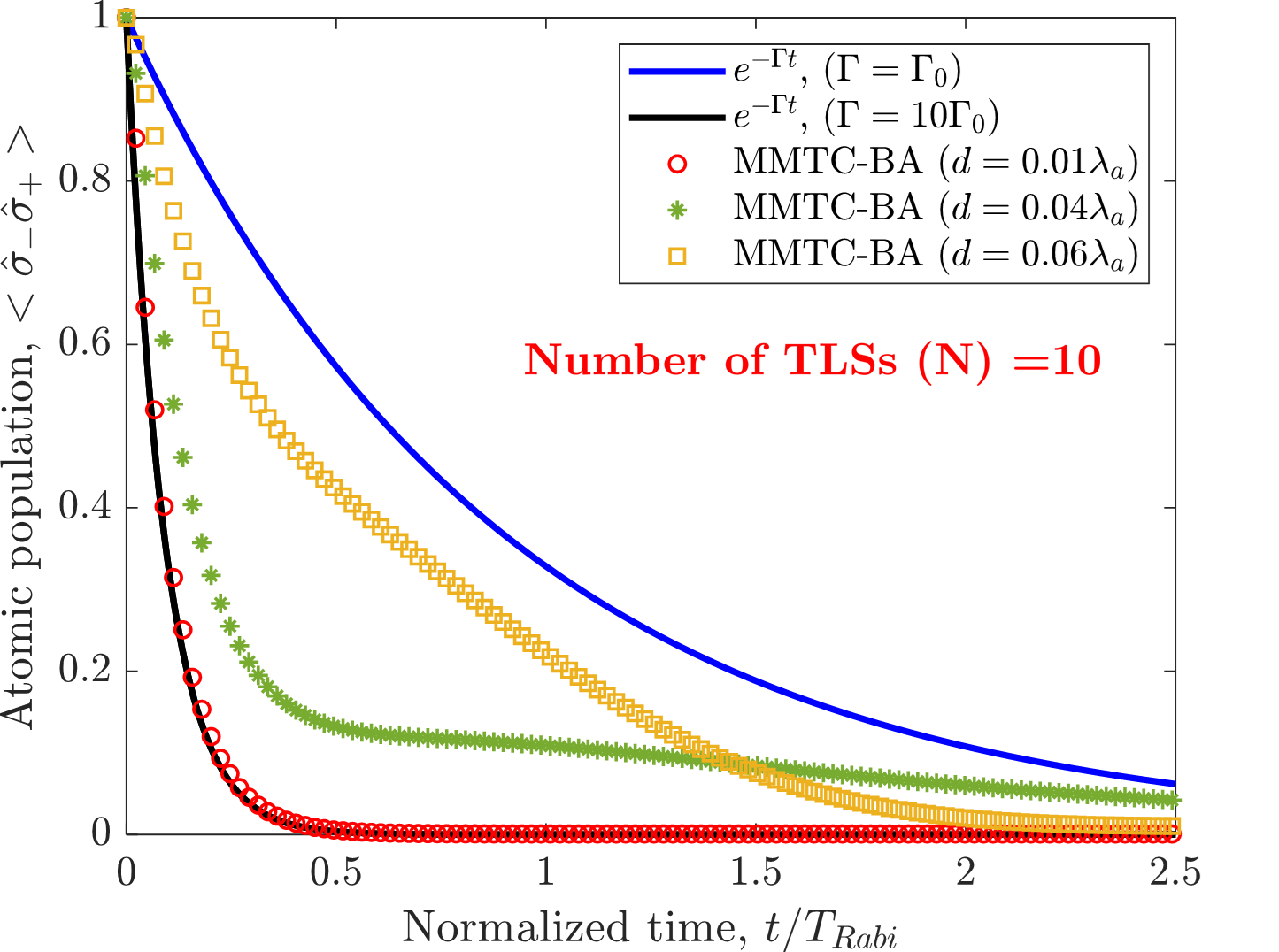}}
  \end{subcaptionbox}
  \begin{subcaptionbox}{$N\in\{2,4,15\}$\label{superradiance_fig2}}[0.45\textwidth]
    {\includegraphics[width=.465\textwidth]{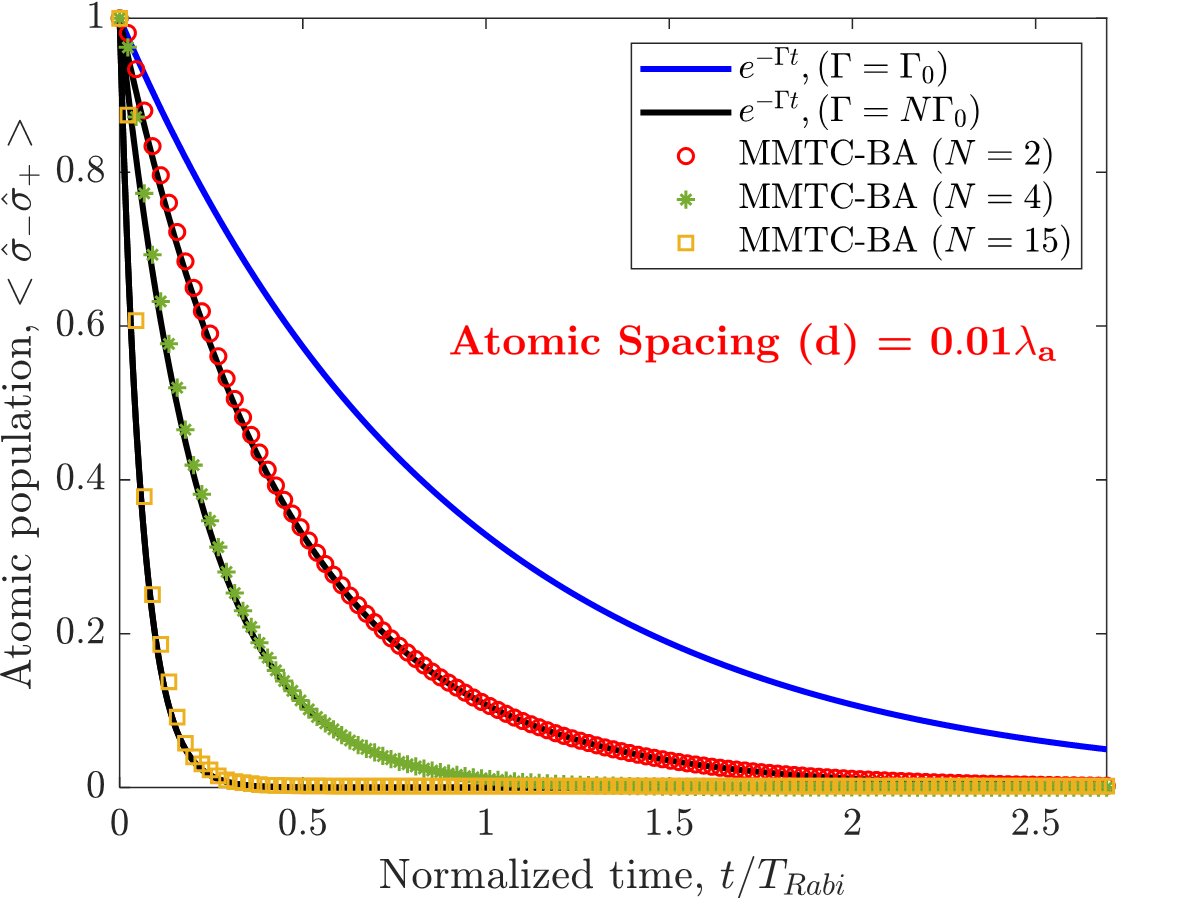}}
  \end{subcaptionbox}
  \caption{Super-radiance of a linearly arranged atom array in free space }
  \label{fig:superradiance}
\end{figure}

We consider several scenarios in which (i) the atomic spacing is varied with fixed number of atom, and conversly, (ii) the number of atom is varied with constant atomic spacing. Especially for atoms in free space, the supe-radiance can be evaluated by analyzing the time evolution of atomic population compared to that of single atom (i.e., exp($-\Gamma_0t$)). 
When the spacing between the TLSs ($d$) is much smaller than $\lambda_a$, the atomic population decays more faster and finally the atomic population decays at a rate approximately $N$ times greater than that of incoherent emission. In other words, the time evolution of the atomic population follows \(\exp(-N\Gamma_0 t)\). 
In Fig. \ref{fig:superradiance}, the time evolution of atomic population is illustrated for two scenarios. In Fig. \ref{fig:superradiance} (a), the atom decays more drastically as the spacing between adjacent atoms (d) decreases. When d becomes about 0.01$\lambda_a$, the time evolution of atomic population finally follows exp(-$N\Gamma_0t$) ($N=10$ in this case) which correspond to the super-radiant decay we expected. When $d$ is small enough, our formalism can capture the super-radiance phenomenon at different N ($N\in\{2,4,15\})$ as shown in Fig. \ref{fig:superradiance} (b). In other words, in cases where $d$ is sufficiently small, the results from our methods always well matched with theoretical exp(-$N\Gamma_0t$) value.

%% 20250317

\subsection{Entanglement sudden death of two TLSs}
Entanglement Sudden Death (ESD) refers to the phenomenon in which quantum entanglement between subsystems vanishes completely within a finite time due to interaction with the environment, even though the local decoherence of each subsystem may occur gradually \cite{Yu20024Finite,Yu2009Sudden}. 
Unlike classical decoherence processes where quantum properties decay asymptotically, ESD reveals that entanglement—a key quantum resource—can be lost abruptly. This effect has important implications for quantum information processing and quantum communication, as it highlights the fragility of entanglement under realistic, noisy conditions.

In this numerical example, we incorporate environmental and non-Markovian effects into the physics of entanglement sudden death (ESD) from first principles by considering BA-MA field modes in the Tavis–Cummings model. 
Following the approach in~\cite{Yu20024Finite}, we first assume that the two TLSs are initially prepared in the following mixed entangled state:
\begin{flalign}
\boldsymbol{\rho}(t)
=
\frac{1}{3}
\left[
\begin{matrix}
a(t) &  0 & 0 & 0 \\
0 & b(t) & z(t) & 0 \\
0 & z^{*}(t) & c(t) & 0 \\
0 & 0 & 0 & d(t)
\end{matrix}
\right]
\end{flalign}
for bases $\ket{e,e}$, $\ket{e,g}$, $\ket{g,e}$, and $\ket{g,g}$.
As the TLSs interact with the surrounding environment and the probability amplitudes of their eigenstates evolve over time, the degree of entanglement—quantified by the concurrence—can be calculated as follows~\cite{Yu20024Finite}.
\begin{flalign}
C(\boldsymbol{\rho}(t))
=
\frac{2}{3}\text{max}\left(0,\left|z(t)\right|-\sqrt{a(t)d(t)}\right)
\end{flalign}

\begin{figure}[t]
\centering
\includegraphics[width=4in]{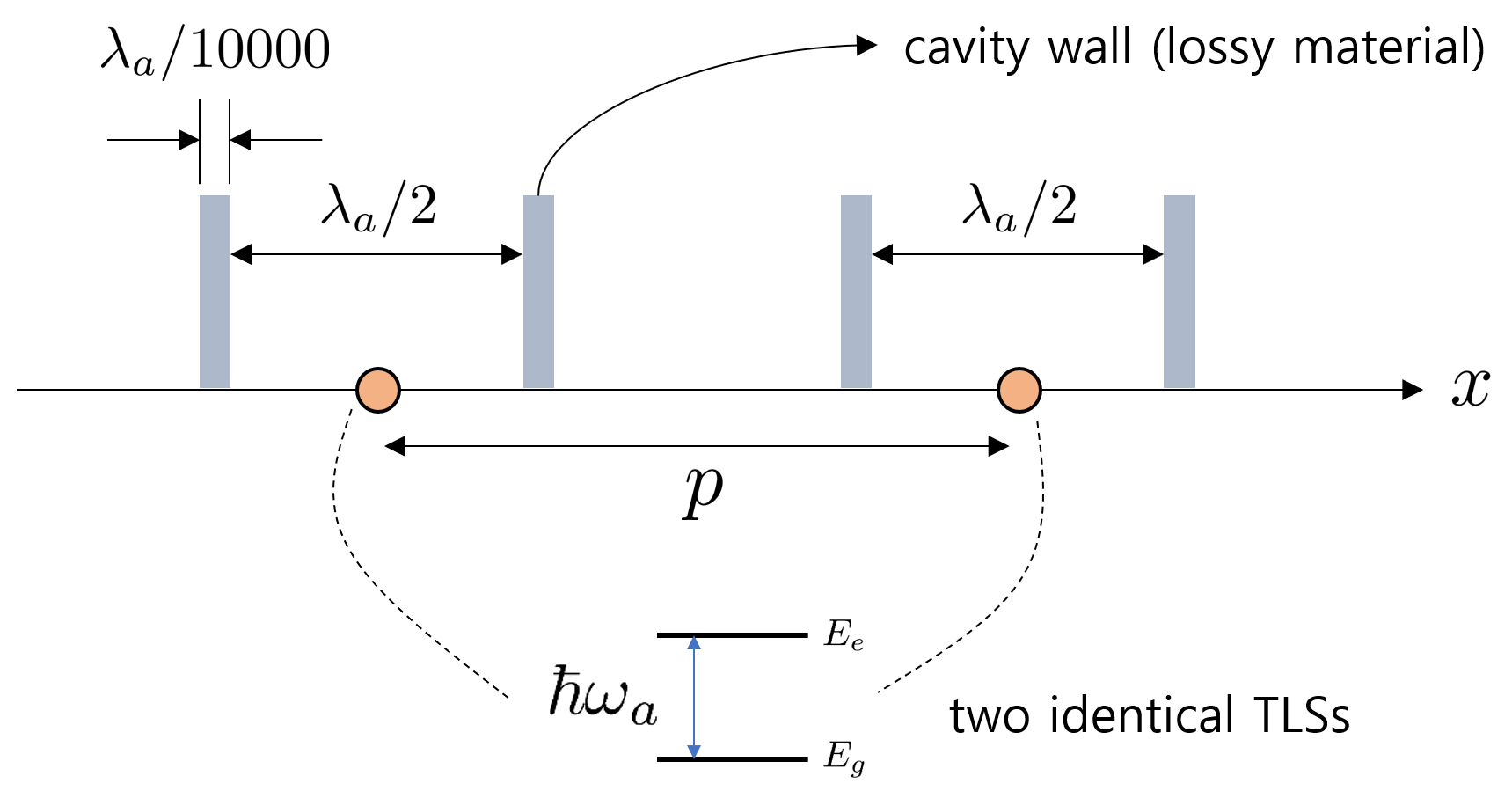}
\caption{The problem geometry of entanglement sudden death of two TLSs which are initially in a mixed entangled state.}
\label{fig:ESD_geom}
\end{figure}

As depicted in Fig. \ref{fig:ESD_geom}, in this calculation, we consider two types of environmental settings:
\begin{enumerate}
    \item First, the two TLSs are located in an open free-space region, separated by a distance $p$.
    \item Second, the two TLSs are placed at the centers of two separate cavities made of conductive dielectric walls with Ohmic losses. The distance between the two cavities is $p$.
\end{enumerate}
Through these two scenarios, we analyze how changes in the surrounding environment influence the physics of  ESD. In particular, in the second case, we investigate how the presence of lossy media affects ESD, and how increasing the cavity separation $p$ leads to non-Markovian effects that further impact ESD. Finally, we examine how the absence of the BA field contribution alters the ESD behavior in the second scenario.

Unlike the previous example, the initial quantum state of the two TLSs in this case includes 0-quanta, 1-quanta, and 2-quanta excitation states. In the Tavis–Cummings model, the total number of quanta is conserved over time. As a result, quantum states with different quanta numbers do not interact dynamically. 
Therefore, for each initial state belonging to a different quanta subspace, we solve three independent Tavis–Cummings models to obtain the time-dependent coefficients $a(t)$, $b(t)$, $z(t)$, and $d(t)$, which are then used to compute the concurrence.

Total number of eigenstates for 0-quanta, 1-quanta, and 2-quanta corresponds to
$N^{(0)}=1$, $N_{(1)} = N_a+N_f$, and $N_{(2)} = \binom{N_a}{2} + N_a \times N_f + \binom{N_a+N_f-1}{2}$.
Note that the total number of eigenstates for the 2-quanta case was obtained by considering three possible situations: (1) two quanta are contained in the two TLSs, (2) one quanta is contained in either TLs, and the other quanta is contained in one of BA-MA field mode, and (3) two quanta are contained in BA-MA field modes.
Thus, the total number of eigenstates grows exponentially when the quanta number increases.

In this numerical example, we fix $a(t) = 0.2$, set $b(t) = c(t) = z(t) = 1$, and define $d(t) = 1 - a(t)$. 
According to~\cite{Yu20024Finite}, the occurrence of ESD strongly depends on the initial quantum state. Within the Markovian approximation, ESD can only occur when $a(t)$ exceeds $1/3$.
In this example, we analyze the ESD phenomenon under non-Markovian conditions. The parameters of the TLSs are the same as in the previous example.

\begin{figure}[t]
\centering
\includegraphics[width=4in]{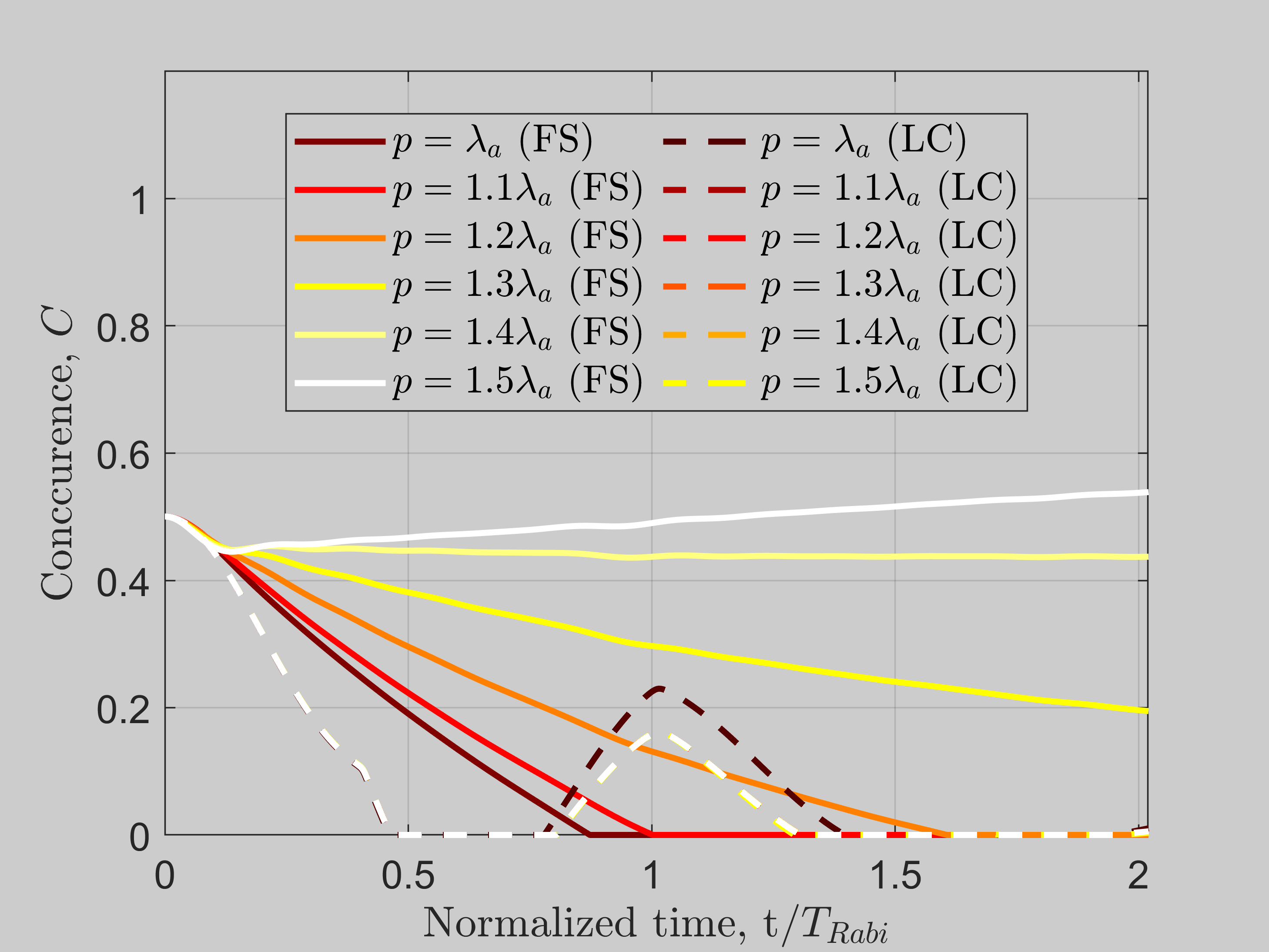}
\caption{The concurrence as a function of normalized time, $t/T_{\text{Rabi}}$, is shown for two scenarios: (1) two TLSs initially prepared in a mixed entangled state and separated by a distance $p$ in free space, and (2) each TLS placed at the center of a lossy cavity of size $L_{\text{cav}} = \lambda_a/2$, with wall thickness $L_{\text{wall}} = \lambda_a/10000$, and separated by a distance $p$. Note that FS and LC are abbreviations of free space and lossy cavity, respectively.}
\label{fig:ESD_result}
\end{figure}

The computed Conccurrence in time for various cases is presented in Fig. \ref{fig:ESD_result}.

When observing the results of the free space, the ESD indeed occurs at the finite time.
Note that in this case the spontaneous emission rate in the free space was 0.2812 (or, time for $e^{-1}$ corresopnds to $2.3949 T_{Rabi}$.)
More interestingly, despite the same value of $a(t)$, the ESD occurring times are different depending on the separation distance between the TLSs, which accounts for the non-Markovian physics.
In the extreme case, When the TLSs are separated by one and half wavelength, they fall into the subradiant regime; consequently, the conccurrence could be maintained almost as constant.
On the other hand, for the second scenario, the concurrence are periodically happen, i.e., rebirth of the ESD.

Unlike the first scenario, where any photon emitted by either TLS is irreversibly lost to free space, the cavities can also be modeled as undamped mirror-like structures or with very small decay rates. 
This situation results in a periodic sequence of perfect revivals of the atomic entanglement between TLSs A and B~\cite{Yu2009Sudden}. 
In our case, we consider a finite conductivity for the cavity walls; therefore, one can expect periodic rebirths of concurrence accompanied by gradual decay.

Observing the results in Fig.~\ref{fig:ESD_result}, we find periodic rebirths of the concurrence in the lossy cavity case. 
Unlike the first scenario, however, the concurrence behavior in the second scenario is only weakly affected by the distance between the two cavities. 
The generation and disappearance of entanglement is an informatics process, without any energy exchange between the two sites~\cite{Yu2009Sudden}. 
Therefore, entanglement can vanish or reappear within a finite time regardless of how far apart the two cavities are.

\section{Conclusion and future works}
We developed a novel numerical framework that incorporated the modified Langevin noise (M-LN) formalism into the multimode Jaynes– and Tavis–Cummings models, enabling a first-principles, non-Markovian analysis of atom–field interactions in dissipative electromagnetic (EM) environments. This approach accounted for both radiative losses and absorptive dissipation in general lossy or inhomogeneous causal media exhibiting dispersion and absorption.
The boundary- and medium-assisted (BA and MA) field modes—representing a continuum of EM modes in dissipative environments—were numerically constructed using the finite-element method (FEM). BA modes were extracted via plane-wave scattering problems, while MA modes were obtained by solving point-source radiation problems. To reduce computational cost, we employed an adaptive frequency refinement strategy and implemented coarse-graining over the degeneracy space of the BA and MA fields.
These numerically resolved BA and MA modes were integrated into the multimode Jaynes– and Tavis–Cummings models, enabling direct evaluation of atom–field coupling strengths. By appropriately truncating the Hilbert space and using a matrix-based Schrödinger equation solver, we obtained time-dependent probability amplitudes, from which expectation values of atomic populations, EM field observables, and entanglement metrics were computed.
The framework successfully captured non-Markovian atomic dynamics that are inaccessible to conventional quantum master equation approaches under the Markovian approximation. In particular, the method enabled the calculation of EM field observables such as single-photon field amplitudes and second-order correlation functions $g^{(2)}(\tau)$—quantities that are crucial for evaluating radiation patterns, coupling efficiency to optical fibers, and entanglement transfer to EM fields.
The proposed methodology was validated through four representative examples: (i) a two-level system near a perfect electric conductor (PEC) half-space; (ii) a dissipative cavity with limiting cases corresponding to free space and ideal Rabi oscillations; (iii) super-radiance in TLS arrays; and (iv) entanglement sudden death in dissipative cavities. These results demonstrated that the proposed framework can serve as a ground-truth numerical simulator for atom–field interactions in general dissipative EM environments.

This study opens several promising directions for future research:
\begin{itemize}
    \item \textbf{Development of computationally much more efficient algorithms:} 
    While the current framework offers high accuracy, it can be computationally demanding. Future work will focus on developing more efficient numerical algorithms—such as reduced-order modeling, parallelization strategies, and adaptive basis pruning techniques—to accelerate the simulation process while preserving physical fidelity.

    \item \textbf{Extension to higher-dimensional systems:}
    The present formulation was primarily demonstrated in lower-dimensional or symmetry-reduced settings. A natural extension is to apply the framework to full 3D geometries and multi-atom/multi-cavity systems, which will allow for more realistic modeling of quantum devices and nanophotonic environments.

    \item \textbf{Exploration of entanglement transfer and field observables:}
    Future studies will investigate the transfer of entanglement from atomic systems to photonic degrees of freedom, particularly in scenarios involving entanglement sudden death (ESD) and rebirth. The resulting electromagnetic field distributions and their correlation properties (e.g., $g^{(2)}(\tau)$) will be analyzed to better understand quantum information propagation in dissipative environments.
\end{itemize}

\clearpage
\appendix

\section{BA and MA fields in the one-dimensional space}
Since the numerical examples focused on one-dimensional cases, we elaborate here on the BA and MA fields corresponding to these one-dimensional scenarios.

\subsubsection{BA and MA fields}
The electric field operator has $z$ component only and is a function of $x$ such that the positive-frequency component of the monochromatic electric field can be expressed by
\begin{flalign}
\hat{E}^{(+)}_z(x,\omega)
=
{E}^{(+)}_{\text{(BA)},z}(x,\omega)
+
\hat{E}^{(+)}_{\text{(BA)},z}(x,\omega)
\end{flalign}
where
\begin{flalign}
\hat{E}^{(+)}_{\text{(BA)},z}(x,\omega)
&=
\sum_{\hat{k}_x= \pm}
{E}^{(+)}_{\text{(BA)},z}\left(x,\omega,\left\{\hat{k}_x\right\}\right)\hat{a}\left(\omega,\left\{\hat{k}_x\right\}\right),\\
\hat{E}^{(+)}_{\text{(MA)},z}(x,\omega)
&=
\int_{V_m}dx'
\hat{E}^{(+)}_{\text{(MA)},z}\left(x,\omega,\left\{x'\right\}\right)\hat{f}\left(\omega,\left\{x'\right\}\right).
\end{flalign}
Here, we used $\hat{k}_x$ to denote the degeneracy of the BA fields at $\omega$ in one-dimensional space for a single polarization, that is, propagating in the $+x$ or $-x$ directions.
The degeneracy of MA fields corresponds to a point source position $x'$ in a lossy medium defined in $V_m$, that is, $x'\in V_m$.
Similarly, we only consider a single orientation of point noise current sources, which is the $z$ direction.
Thus, degenerate BA and MA fields can be written more precisely by
\begin{flalign}
{E}^{(+)}_{\text{(BA)},z}\left(x,\omega,\left\{\hat{k}_x\right\}\right)
&=
\frac{i}{\sqrt{2\pi}} E_{\text{tot},y}\left(x,\omega,\left\{\hat{k}_x\right\}\right) \sqrt{\frac{\hbar\omega}{2}},\\
{E}^{(+)}_{\text{(MA)},z}(x,\omega,\left\{x'\right\})
&=
i k^2 G_E(x,x',\omega) \sqrt{\frac{\hbar \chi_{I}(x',\omega)}{\pi \epsilon_0}}
\end{flalign}
where
\begin{flalign}
\frac{d^2}{dx^2} E_{\text{tot},z}\left(x,\omega,\left\{\hat{k}_x\right\}\right) + k^2 \epsilon_r(x,\omega) E_{\text{tot},z}\left(x,\omega,\left\{\hat{k}_x\right\}\right) &= 0,
\\
\frac{d^2}{dx^2} G_E(x,x',\omega) + k^2 \epsilon_r(x,\omega) G_E(x,x',\omega)
&= - \delta(x-x').
\end{flalign}
Finally, one needs to solve one-dimensional plane-wave-scattering and point-source-radiation problems to find $E_{\text{tot},z}\left(x,\omega,\left\{\hat{k}_x\right\}\right)$ and $G_E(x,x',\omega)$, respectively.
More specifically,
\begin{flalign}
E_{\text{tot},z}\left(x,\omega,\left\{\hat{k}_x\right\}\right)
=
E_{\text{inc},z}\left(x,\omega,\left\{\hat{k}_x\right\}\right)
+
E_{\text{sca},z}\left(x,\omega,\left\{\hat{k}_x\right\}\right)
\end{flalign}
where
\begin{flalign}
\frac{d^2}{dx^2} E_{\text{sca},z}\left(x,\omega,\left\{\hat{k}_x\right\}\right) + k^2 \epsilon_r(x,\omega) E_{\text{sca},z}\left(x,\omega,\left\{\hat{k}_x\right\}\right)
=
-k^2 \chi(x,\omega) E_{\text{inc},z}\left(x,\omega,\left\{\hat{k}_x\right\}\right)
\end{flalign}
for a given incident plane wave
\begin{flalign}
E_{\text{inc},z}\left(x,\omega,\left\{\hat{k}_x\right\}\right)
=
e^{\pm i k x}
\end{flalign}
for $\hat{k}_x = \pm$ where $k=\omega/c$ denotes a wavenumber for $\omega$.

\subsection{FEM solution}
We make the use of the FEM formulation to find all degenerate BA and MA fields.
For BA fields, we expand the scattered component of the electric field in terms of the Whitney 0-forms by
\begin{flalign}
E_{\text{sca},z}\left(x,\omega,\left\{\hat{k}_x\right\}\right)
=
\sum_{i=1}^{N_0}
e^{(sca)}_i\left(\omega,\left\{\hat{k}_x\right\}\right)W^{(0)}_i(x).
\end{flalign}
The Whitney 0-form (also known as nodal element or barycentric coordinate) defined in $i$-node can be expressed 
\begin{flalign}
W^{(0)}_i(x) = 
\left\{
\begin{matrix}
\frac{x-x_{i-1}}{x_i-x_{i-1}} & \text{for } x_{i-1}\leq x\leq x_i \\
\frac{x_{i+1}-x}{x_{i+1}-x_{i}} & \text{for } x_{i}\leq x\leq x_{i+1} \\
0 & \text{elsewhere}
\end{matrix}
\right.
\end{flalign}

Then, one can transform the plane-wave-scattering problem into a finite-dimensional linear system by
\begin{flalign}
\left[\mathbf{S}\right]\cdot \mathbf{e}^{(sca)}
-
k^2 
\left[\mathbf{M}\right]\cdot \mathbf{e}^{(sca)}
=
\mathbf{j}
\end{flalign}
where
\begin{flalign}
\left[\mathbf{S}\right]_{i,j} 
&= 
\int_{V} dx \nabla W^{(0)}_{i}(x) \times \nabla W^{(0)}_{j}(x),\\
\left[\mathbf{M}\right]_{i,j} 
&= 
\int_{V} dx \epsilon_r(x,\omega) W^{(0)}_{i}(x) W^{(0)}_{j}(x),\\
\left[\mathbf{j}\right]_{i}
&=
\int_{V}dx E_{\text{inc},z}\left(x,\omega,\left\{\hat{k}_x\right\}\right) W^{(0)}_{i}(x),
\end{flalign}
and $V$ denote the solution domain.

To find degenerate MA fields, we solve for the point-source-radiation problem whose FEM counterpart corresponds to the following linear system:
\begin{flalign}
\left[\mathbf{S}\right]\cdot \mathbf{g}
-
k^2 
\left[\mathbf{M}\right]\cdot \mathbf{g}
=
\mathbf{j}
\end{flalign}
where 
\begin{flalign}
\left[\mathbf{j}\right]_i = 
\int_{V}dx 
\left[
-\delta(x-x') i k^2 \sqrt{\frac{\hbar \chi_I(x',\omega)}{\pi\epsilon_0}}
\right]
W^{(0)}_{i}(x)
=
-
i k^2 \sqrt{\frac{\hbar \chi_I(x',\omega)}{\pi\epsilon_0}}
W^{(0)}_{i}(x'),\\
\end{flalign}
such that
\begin{flalign}
E_{(\text{MA}),z}^{(+)}(x,\omega,\left\{x'\right\})
\approx
\sum_{i=1}^{N_0}
e^{(MA)}_i(\omega,\left\{x'\right\})
W^{(0)}_i(x).
\end{flalign}

\subsection{Mode volume for discrete BA and MA modes}
The mode volumes for BA and MA fields should be
\begin{flalign}
\mathcal{D}_l^{(f,BA)} &= {\Delta \omega},\\
\mathcal{D}_l^{(f,MA)} &= {\Delta \omega \Delta x'}.
\end{flalign}
where $\Delta \omega$ denotes the frequency differential, and $\Delta x'$ represents the spatial differential of the mesh within a lossy dielectric medium.
Again, both quantities may vary when employing adaptive frequency refinement and an irregular mesh.

\bibliographystyle{elsarticle-num} 
\bibliography{sample}

%% Text of bibliographic item

\end{document}